# Strain Effects on Oxygen Migration in Perovskites

## Updated January 26, 2016. See Update Note.

Tam Mayeshiba

>Materials Science Program

>University of Wisconsin-Madison, Madison, WI, 53706, USA

Dane Morgan (Corresponding Author)

>Department of Materials Science and Engineering,

>University of Wisconsin-Madison, Madison, WI, 53706, USA

>ddmorgan@wisc.edu

**Update Note, January 26, 2016:**
This draft was updated from the original version in order to:
1. Incorporate proof corrections that were made to the PCCP proof after submission, and which are not included in the previous Word document.
2. Correct the following error: The exponents of the Birch-Murnaghan equation in our Python-language equation of state code were erroneously given as 7/3, 5/3, and 2/3, which evaluated to 2, 1, and 0, respectively, instead of being given as 7.0/3.0, 5.0/3.0, and 2.0/3.0, evaluating to 2.333…, 1.666…, and 0.666…. These erroneously small exponents were applied to the ratio $V_0/V$ which was less than 1, producing $P(V_0,$ fixed $V)$ that was steeper than it should have been. Therefore, every pressure difference $P_{saddle}(V_{0,saddle},$ fixed $V) - P_{initial}(V_{0,initial},$ fixed $V)$ between a saddle point and initial point was reproduced at smaller intervals of $V_0$, or smaller fitted migration volumes, than was correct. Since the elastic strain model DMEPS scales linearly with migration volume (see Eq. 2 in the original paper), the resulting elastic strain model DMEPS were all smaller in magnitude than they should have been.
3. Correct minor data errors detected when the data was reorganized.

A short correction (1 page, with Figure 5R to correct Figure 5) was sent to PCCP in January 2016.

This updated version of the main paper and the updated version of the supporting information have not been peer-reviewed.

# Abstract

Fast oxygen transport materials are necessary for a range of technologies, including efficient and cost-effective solid oxide fuel cells, gas separation membranes, oxygen sensors, chemical looping devices, and memristors. Strain is often proposed as a method to enhance the performance of oxygen transport materials, but the magnitude of its effect and its underlying mechanisms are not well-understood, particularly in the widely-used perovskite-structured oxygen conductors.

This work reports on an *ab-initio* prediction of strain effects on migration energetics for nine perovskite systems of the form $LaBO_3$, where B = [Sc, Ti, V, Cr, Mn, Fe, Co, Ni, Ga]. Biaxial strain, as might be easily produced in epitaxial systems, is predicted to lead to approximately linear changes in migration energy. We find that tensile biaxial strain reduces the oxygen vacancy migration barrier across the systems studied by an average of 65 meV per percent strain for a single selected hop, with a low of 36 and a high of 86 meV decrease in migration barrier per percent strain across all systems. The estimated range for the change in migration barrier within each system is +/- 25 meV per percent strain when considering all hops. These results suggest that strain can significantly impact transport in these materials, e.g., a 2% tensile strain can increase the diffusion coefficient by about three orders of magnitude at 300 K (one order of magnitude at 500°C or 773 K) for one of the most strain-responsive materials calculated here ($LaCrO_3$).

We show that a simple elasticity model, which assumes only dilative or compressive strain in a cubic environment and a fixed migration volume, can qualitatively but not quantitatively model the strain dependence of the migration energy, suggesting that factors not captured by continuum elasticity play a significant role in the strain response.

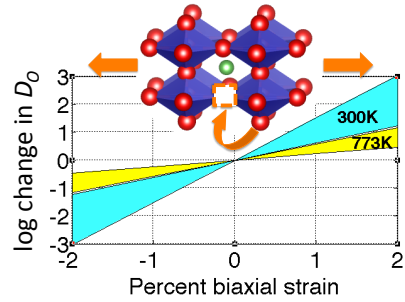

Computational results show that a 2% biaxial tensile strain may increase oxygen ion conduction, both in- and out-of-plane, by up to approximately three orders of magnitude at 300K in the most strain-sensitive $LaBO_3$ perovskites, where B = [Sc, Ti, V, Cr, Mn, Fe, Co, Ni, Ga].

# Introduction

Fast oxygen transport materials are important for creating efficient and cost-effective solid oxide fuel cells (SOFCs), gas separation membranes, oxygen sensors, chemical looping devices, and memristors.[1-8] Besides changing the chemical makeup of a material or changing the operating conditions of the device (e.g., temperature), strain is another possible mechanism to enhance oxygen diffusion. In a broad review over both metals and oxides, Yildiz summarized that stress resulting from tensile strain on materials affects the energy landscape, particularly by weakening interatomic bond strengths, which then results in lower defect formation energy, dissociation barrier, charge transfer barrier, adsorption energy, and, of particular interest to this paper, oxygen migration barrier.[8]

Several experimental studies have shown that strain in oxygen-conducting oxides can lead to faster oxygen bulk or surface diffusion.[9-13] The strongest effect is cited for Yttria-stabilized $ZrO_2$ (YSZ) and $SrTiO_3$ (STO) interfaces at 8 orders of magnitude enhancement in ionic conductivity in a range around 100°C to 225°C. This effect was attributed to strain and interfacial disorder in the YSZ/STO interface.[9] However, combined experimental and computational studies of (YSZ or ceria, $CeO_2$)/STO interfaces now suggest elastic strain contributions of 2-4 orders of magnitude, and attribute the remaining orders of magnitude primarily to electronic conductivity, if they attribute them at all.[12-14]

A number of computational studies have also examined the effects of strain on oxygen migration, with most studies focusing on fluorite structures. To aid in this discussion, we introduce the term "DMEPS" for "Delta in (oxygen) Migration (barrier) Energy per Percent Strain," where a DMEPS value is the slope of a plot of migration barrier energy versus percent strain. DeSouza, Ramadan, and Hörner find a total of 0.5 eV (out-of-plane) to 0.6 eV (in-plane) reduction in migration barrier for the fluorite $CeO_2$ at 7% biaxial tensile strain, with a migration barrier vs. strain slope of about -100 meV (out-of-plane) to -50 meV (in-plane) change per percent (biaxial) strain (i.e., DMEPS values of -100 meV/% strain and -50 meV/% strain, respectively) for low strains.[14] Note that in-plane refers to hop vectors entirely in the plane of the biaxial strain and out-of-plane refers to hop vectors with a component normal to the plane of biaxial strain. These studies were done using classical fitted pair potentials and correspond to about four orders of magnitude in ionic conductivity enhancement for 4% biaxial tensile strain at 500 K.[14] Schichtel, Korte, Hesse, and Janek predict a maximum of 2.5 orders of magnitude in ionic conductivity enhancement (only in-plane calculated) for a YSZ/STO lattice mismatch of 7.37% at 573 K using elastic strain theory.[12] With each 100 meV decrease in migration barrier producing a 0.88 increase in magnitude of conductivity at 573K (see Supporting Information Section S9), this value corresponds to a DMEPS of -38 meV/% strain. Kushima and Yildiz predict a maximum enhancement of 3.8 orders of magnitude (in-plane) for 4% biaxial tensile strain in YSZ at 400K using density functional theory migration barrier inputs into a kinetic Monte Carlo simulation.[15] Although their change of migration energy with strain is not linear, if it were, then this effect would correspond to a DMEPS of -75 meV/% strain. A density-functional theory study by Yang, Cao, Ma, Zhou, Jiang, and Zhong find a DMEPS of -50 to -90 meV/% strain for a single hop (out-of-plane, backwards and forwards; the slope is calculated between -1% and +1% strain) in the $A^{2+}B^{4+}O_3$ perovskite $BaTiO_3$, although interestingly their aim is to find a way to decrease oxygen migration in order to preserve ferroelectric behavior, and they propose compressive strain as a mechanism for doing so.[16] This last study underscores the point that, while the focus of this paper is on increasing oxygen migration, its findings could also apply to decreasing oxygen migration, for example for reducing corrosion.[8]

A number of explanations have been offered for the source of the coupling of migration energies and strain. Similar to the idea that doping with different-sized cations introduces local stress fields and lattice expansion, leading to lower migration energies,[8, 17] Kushima and Yildiz identified the mechanisms of strain effects on migration energies as a competition between the "elastic stretching" of cation-oxygen bonds, which weakens those bonds and also creates a larger "migration space," thus decreasing migration barrier, versus the plastic deformation of the material as bonds are completely broken and new, strong bonds are formed, thus increasing migration barrier.[15] Chroneos, Yildiz, Tarancón, Parfitt, and Kilner stress what they term "mechano-chemical" coupling as a mechanism, in which lattice strain affects the cation-oxygen bond strength.[18] Schichtel et al.'s elastic strain model, which has been used to predict changes in migration barrier with strain in YSZ and CSZ (Yttrium- and Calcium-stabilized zirconia) on various substrates, includes the isotropic pressure effects of strain, but neglects the changing local bond strengths mentioned by Chroneos et al.[12] While these studies to date shed some light on strain effects, there are still many uncertainties, and almost all studies have focused on the fluorite structure, leaving the important class of perovskite oxide oxygen conductors largely unexplored.

As a class of technologically important oxygen-conducting materials, perovskites may, along with fluorites, also benefit from strain-enhanced transport. For example, in SOFCs, which commonly use perovskite and fluorite materials, the major efficiency losses at low or intermediate temperature occur in the cathode and electrolyte, and can be mitigated by improving the rate of the oxygen reduction reaction at the cathode, and the rate of oxygen transport in both the cathode and the electrolyte.[2] One experimental study of strain effects on oxygen surface exchange and oxygen transport has been carried out on the perovksite Strontium-doped $LaCoO_3$, with the oxygen tracer diffusion coefficient D* increasing by about 1 order of magnitude when going from a compressive strain to a tensile strain, with a total strain difference of about 2.9%.[10] With some very significant assumptions and approximations, including that the whole increase is due to changes in the migration energy (see Results and Discussion), this increase corresponds to a DMEPS of -64 meV/% strain for out-of-plane hops. Another study on perovskite thin films of $(La_{0.5}Sr_{0.5})CoO_3$ on $SrTiO_3$ argued that changes in oxygen content and ordering were enabled by enhanced cation migration. While this work did not address changes in oxygen migration, it did suggest that the cation migration barriers were significantly enhanced by strain in the sample.[19] Additional studies link tensile strain with enhanced oxygen vacancy formation,[20, 21] which will also contribute to enhanced oxygen diffusion, and potentially to enhanced catalytic activity.[8, 21, 22] The present study aims to survey a range of perovskite materials and develop an understanding of how strain couples to migration barriers, assess the ability to understand these strain effects in terms of a simple elasticity model, and provide guidance on which materials might respond most strongly to strain engineering of oxygen kinetics.

This computational study focuses on bulk conduction and therefore most immediately serves as a foundation to understand strain effects in a system with strain over bulk-like distances, e.g., as may occur in a thin film used in experiments or small devices. However, the bulk trends may also provide guidance for understanding strain effects occurring at interfaces,[11], including surfaces, oxide superstructures,[23] and oxide heterostructures,[24-28] where strain effects may contribute to enhanced performance at perovskite material interfaces.[20, 29]

Using SOFCs as an example, an increase in ionic conductivity of just two orders of magnitude can transform a substandard material into a useful one at intermediate temperatures, so changes on this scale are of significant interest.[2, 30] In terms of specific devices, micro-SOFCs are on a size scale where epitaxial strain might be used, for example in anode- or cathode-supported electrolyte growth for parallel-layer devices. For a single-chamber configuration,[31, 32], lattice mismatch with the electrolyte may support strained anode or cathode growth.

Motivated by the potential to engineer oxygen conductors with strain and the limited knowledge of strain effects on perovskites, we here study the effects of elastic strain on the oxygen ion conductivity of perovskite systems of the type $ABO_3$, where A=La and B is a metal ion. Perovksites closely related to this set, with La and other rare earths on the A-site, typically with A-site doping by Sr or other alkali-earth elements and often multiple B site metals, are being used or considered as fast oxygen conductors in SOFCs, oxygen separation membranes and sensors, and chemical looping applications.[1-5] Therefore, the ability to enhance these materials with strain could have significant technological impact.

## Results and Discussion

In order to check the coupling between strain and oxygen ion conductivity, we take the migration barrier $E_{mig}$ as a simplified measure of conductivity, as described by Chroneos et al. (see Supporting Information Section S9 for details).[18] We employed the Vienna *Ab-initio* Simulation Package (VASP)[33-36] for density functional theory (DFT) calculations, using the MAterials Simulation Toolkit (MAST)[37] to automate sets of calculation workflows. See the Methods section and Supporting Information for more information on the parameters and workflow used. The Supporting Information also includes detailed discussion on additional hops not shown in the main text, derivation of the parameters used in the elastic strain model, and error analysis.

For the unit cells cells used in these studies, there are multiple symmetry-distinct hops, which increase in number when strain is introduced. In order to keep the calculations tractable, we have studied only a few specific hops for all systems and strain states, and all hops for just a couple of systems. Figure 1 shows the migration barrier versus strain results for a consistent (i.e., the same hopping atom and vacancy sites for each B cation) in-plane hop across all systems. The slope of each line gives the DMEPS value for that system. Figure 2 shows schematically what is meant by an in-plane and out-of plane hop with respect to the strain axes. Positive percent strain is tensile strain and negative percent strain is compressive strain. All systems show a decrease in migration barrier with increasing tensile strain. The average DMEPS is -66 meV per percent strain, with low and high magnitudes for DMEPS of -36 and -89 meV/% strain.

Supporting Information Section S8 shows the following additional details. First, a consistent out-of-plane hop gives comparable DMEPS results to that of the selected in-plane hop across all systems. Second, the selected in-plane and out-of-plane hops are among the lower barrier hops across all systems at zero strain, and are therefore a reasonable representative choice to model diffusion behavior in the perovskites. However, there is no evidence that suggests that the magnitude of a zero-strain migration barrier is correlated with the magnitude of its DMEPS. Therefore, all hops may need to be considered when considering the range of DMEPS for a material. Finally, third, for the cases where we have calculated DMEPS for all hops in the system (a total of 96 hops, of which 12 are symmetry-independent, producing distinct migration barrier values), there is no clear distinction between the magnitudes of DMEPS for in-plane hops versus DMEPS for out-of-plane hops; DMEPS for in-plane hops are neither consistently higher nor consistently lower than those for out-of-plane hops. Therefore, this main text limits the tabulated values to a single consistent hop (in-plane is chosen arbitrarily) and adds an additional range of

+/- 25 meV/% strain to the uncertainty in the calculated DMEPS in order to reflect the range of all the hops possible in a given system. We also note that the effects of strain apply, apparently similarly, to both in-plane and out-of-plane hops.

Table 1 shows the values of the fitted DMEPS from Figure 1, along with their fitting errors. The scatter in some of the DMEPS comes from structural instability (e.g. polymorphs), magnetic instability, and convergence problems, which we discuss in Supporting Information, Section S10, along with our general methodology. Despite the data scatter, we are confident that the trend of decreasing migration barrier with increasing tensile strain (negative DMEPS) is reproducible and significant for the perovskite systems.

Although strain may result in faster interfacial conduction along the plane of the strained interface,[11] our results are for bulk single-crystal conduction and therefore these results represent changes in the properties of bulk strained materials.

Figure 3 compares the DMEPS values obtained in this work with those obtained in other works of which we are aware. Where a log change in diffusion coefficient was given in the literature, the given temperature and lattice mismatch or strain were used to convert the log change in diffusion coefficient into a DMEPS value, making the significant approximation that all changes in diffusion coefficient were due to changes in migration barrier (see the following discussion on Figure 4, as well as Supporting Information, Section S9). The average DMEPS from this study are generally similar to those that have been found previously for both fluorites and perovskites and with both calculations and experiments. While there is no reason that the perovskites should have the quantitatively same DMEPS as the fluorites, it is reasonable to expect some similarity, given that they are both relatively open-structured oxides, and this qualitative agreement provides some validation of our results.

Figure 4 shows a more quantitative comparison between our DMEPS calculated for $LaCoO_3$, which is -80 meV/% strain for the out-of-plane hop (see Supporting Information, Section S8), and a DMEPS for Sr-doped $LaCoO_3$ (LSC) perovskite derived from Kubicek et al.'s experimental LSC oxygen diffusion data along the out-of-plane direction,[10] which we estimate as -64 meV/% strain. This comparison is the closest comparison that can be made between our calculated systems and those studied experimentally to this point. Note that we derive the DMEPS from the experimental data of the LSC under the assumption that all changes are due to only changes in migration barrier (see Supporting Information, Section S9). This assumption is certainly an approximation given that an increase in Co reduction is observed in strained LSC in experiment,[38] and that the oxygen vacancy formation energy in undoped $LaCoO_3$ was calculated to decrease with tensile strain within 2% tensile strain,[39] both of which may indicate increased oxygen vacancy concentration with strain. Neglecting this vacancy increase may lead to an overestimation in our DMEPS derivation. We also note that no reduction in activation energies was measured by Kubicek, et al., although there was significant uncertainty in the measurements.[10] In addition, while we took the experimental 475°C strain range of -1.9% compressive to 1.0% tensile strain for our derivation,[10] any strain relaxation that occurred in the film may also modify the derived DMEPS. Despite these limitations on the comparison, it is encouraging that the DMEPS for our out-of-plane hop, and also the DMEPS for our in-plane hop, at -60 meV/% strain, are both in good agreement with the DMEPS estimated from experiment. The good agreement provides some support for the results of our calculations and supports the hypothesis that at least some of the oxygen transport changes seen in the experimental studies of Kubicek, et al. are in fact due to bulk changes in migration energies induced by strain.

Overall, our total range of DMEPS represent values significantly lower and higher than previously reported, which is likely due to the relatively large number of compounds we have studied. Our results demonstrate that, even within one crystal structure, changing B-site cations can lead to a very wide range of oxygen migration strain response.

To explain the observed decrease of migration barrier with increasing percent biaxial strain, we consider the elastic model of Schichtel et al.,[12] which is given as Equation 1. First we convert this equation into a form where we can compare the predicted DMEPS value from this equation to that obtained from our calculations. Equation1 uses the Young's modulus $Y$, Poisson's ratio $v$, and applied biaxial strain $\epsilon_{12}$ to calculate a pressure $p$ due to strain. We take this pressure $p$ to be equivalent to a change in pressure $\Delta p$ compared to an assumed zero pressure at zero strain ($\Delta p = p - 0$). Dividing both sides by $\epsilon_{12}$ gives a slope of change in pressure with percent strain. Multiplying both sides by migration volume $V_{mig}$ transforms the slope of change in pressure with percent strain into a slope of change in migration free energy $G_{mig}$ per percent strain, which we then approximate by the change in migration enthalpy $H_{mig}$ per percent strain, as shown in Equation 2. Details of these derivations and calculations are in the Supporting Information, Section S12). For each strain case, our constant-volume migration barrier energies $E_{mig}$ are directly comparable to constant-pressure migration enthalpies (see Supporting Information, Section S9a), and therefore we also use DMEPS as the term for the elastic model slope of change in migration enthalpy per percent strain. Now the DMEPS predicted from Schichtel et al.'s elastic model and our ab initio calculation can be compared directly.

| | |
|---|---|
| $$p = -\frac{2}{3}\left(\frac{Y}{1-\nu}\right)\epsilon_{12}$$ | Eq. 1 |
| $$\frac{\Delta p * V_{mig}}{\epsilon_{12}} = -\frac{2}{3}\left(\frac{Y}{1-\nu}\right)V_{mig} = \frac{\Delta G_{mig}}{\epsilon_{12}} \approx \frac{\Delta H_{mig}}{\epsilon_{12}}$$ | Eq. 2 |

Figure 5 shows the elastic model DMEPS in meV/% strain as a function of the DMEPS we calculated directly from ab-initio DFT methods. From this figure, we see that the DMEPS predicted by the strain model qualitatively follow the same trends as our ab-initio DMEPS. The strain model DMEPS differ from the DFT-calculated DMEPS by an average of 16 +/- 13 meV/% strain (where the uncertainty represents one standard deviation of the error from the mean), with a maximum error of 40 meV/% strain and a root-mean-squared error of 20 meV/% strain, over all in-plane and out-of-plane hops shown in Figure 5. An equivalent but distinct approach to assessing the strain model is to use the ab-initio calculated DMEPS and the strain model to estimate vacancy migration volume and then compare this volume to the directly ab-initio calculated vacancy migration volume (see Supporting Information, Section S12d). This perspective shows that the effective migration volumes predicted by the strain model are similar to those calculated directly using the ab-initio methods, with some outliers. The root-mean-squared error for predicting migration volumes is 2 Å$^3$. The source(s) of discrepancies between the strain model results and the ab-initio data are not clear at this stage. The error is not as simple as a strain-dependent migration volume, as that would lead to a strong non-linear dependence in the ab-initio calculated DMEPS, whose constituent points are generally fairly linear. It might also be proposed that the presence of the vacancies alters the elastic properties, as has been predicted in some cases [19], as all the elastic constants are determined for the pristine material with no vacancies. Checks with select systems have shown that using elastic properties in the presence of vacancies can shift the predicted elastic-model DMEPS by some -10 to -20 meV/% strain, which in some cases may lead to better agreement, but still leaves significant deviation between the model and the fit results (see Supporting Information, Section S12b). We hypothesize that anisotropic effects account for the main deviations from the strain model, as the strain model uses a single isotropic pressure value and a single isotropic migration volume value. Other effects could include anharmonic effects causing deviations from linear elasticity and numerical noise in the calculations, particularly regarding the migration volumes.

Additional analysis, January 2016:

> Since the elastic strain model uses the zero-strain migration volume, its representation of the DFT DMEPS is expected to be most accurate near zero strain and less accurate for larger strains. Table E1 shows all slopes refit using only data between -1% and 1% strain. Figure E1 shows the agreement between the elastic strain model DMEPS and the low-strain DFT DMEPS. Comparing Figure E1 and Figure 5R shows a small improvement when using the low-strain data, resulting in an average difference of 14 +/- 12 meV/% strain with a maximum difference of 47 meV/% strain and a root-mean-squared error of 18 meV/% strain.

> The $LaCrO_3$ system may best describe the limit of the present elastic strain model's accuracy for representing DFT-fit DMEPS. Its well-behaved elastic constant fits lead to small error bars in the elastic model DMEPS (Table S12.2R). Its DFT migration barrier data, which follows a smooth and consistent slope along the -2% to 2% range of strain, leads to small error bars in the DFT DMEPS, whether the DFT DMEPS is calculated between -2% and 2% strain or between -1% and 1% strain (Figure 1, Figure S8.1R). Nevertheless, both the in-plane hop and the out-of-plane hop in Figure 5R show a disagreement between the elastic model DMEPS and the DFT DMEPS of approximately 20 meV/% strain. The range of disagreement is approximately 30 meV/% strain when all hops are considered (Figure S8.9R).

> Using a defected bulk modulus to calculate Young's modulus for the prefactor does not reduce the disagreement between the elastic model DMEPS and the DFT-fit DMEPS for both the in-plane hop and the out-of-plane hop in $LaCrO_3$ (see Figure S12.2R). Using a volume-relaxed migration volume rather than the Birch-Murnaghan fit migration volume also does not reduce the disagreement for both points (see Figure S12.5R).

Having examined each variable in the elastic model, we conclude that while the elastic model DMEPS are in qualitative agreement with the DFT-fit DMEPS, the error in the best fit cases is still approximately 30 meV/% strain for a single system considering all hops, approximately 20 meV/% strain for a single hop, and is more accurate at low strains for systems where the strain response is non-linear over a larger strain range.

Future modifications to the elastic model to yield better accuracy might include the use of anisotropic migration volumes and full elastic constant tensors. However, the inclusion of so much calculated data to the elastic model may disrupt the simplicity and generality of the approach. We reiterate that on the whole, the isotropic elastic model appears to qualitatively describe the response of oxygen migration barriers in perovskites with respect to biaxial strain.

## Methods

To calculate $E_{mig}$, we used density functional theory (DFT) as implemented in the Vienna Ab-initio Simulation Package (VASP)[33] with the climbing nudged elastic band method (CNEB) with 3 images.[40, 41] The pseudopotentials used were generated using the generalized gradient approximation and the projector-augmented-wave method[42, 43] with the Perdew-Wang 91 exchange correlation functional.[44-46] Valence electrons are listed in parentheses. The standard La pseudopotential was used ($5s^25p^65d^16s^2$). For the transition metals, the available pseudopotential with the most unfrozen electrons was used to assure the best possible accuracy: Sc_sv ($3s^23p^63d^14s^2$), Ti_sv ($3s^23p^63d^24s^2$), V_sv ($3s^23p^63d^34s^2$), Cr_pv ($3p^63d^44s^2$), Mn_pv ($3p^63d^54s^2$), Fe_sv ($3s^23p^63d^64s^2$), Co ($3d^74s^2$), Ni_pv ($3p^63d^84s^2$), and Ga_d ($3d^{10}4s^24p^1$). The soft oxygen pseudopotential was used.

A starting supercell of 2x2x2 formula units (40 atoms) was used, with 4x4x4 kpoints in a Monkhorst-Pack scheme.[47] Migration barriers were converged to within 20 meV relative to the choice of kpoint mesh. All calculations are started with the B-site cations in a ferromagnetic configuration. The same two migration directions were calculated for each system; one with an oxygen atom traveling within the plane of the strain, and one with an oxygen atom traveling out of the plane of the strain. Systems were strained equally along lattice parameters *a* and *b* by at least a grid of 0, ±1%, and ±2% of the original lattice parameters, with positive percentages as tensile strain and negative percentages as compressive strain, and a fit was performed in each strain case to find the equilibrium lattice parameter perpendicular to the plane (*c* lattice parameter). All CNEB calculations were performed at constant volume with internal relaxation. Calculation workflows were automated using the MAterials Simulation Toolkit (MAST), which is under development at the University of Wisconsin-Madison.[37]

We create a vacancy by removing from the supercell both an oxygen atom with its six electrons and an additional two electrons (see a discussion in Supporting Information, Section S7). This procedure is the computational equivalent of substituting lower-valence dopant atoms on A-sites or B-sites somewhere else in the crystal beyond the boundaries of the supercell. The advantage of this method is that it avoids the interaction between oppositely-charged defects by creating a single oxygen vacancy in the supercell without using dopant atoms.

For details on cutoff energy, smearing, CNEB parameters, CNEB workflow, simplifications involved, strain parameters, and charge compensation methods, see Supporting Information.

## Conclusions

We calculate that epitaxial tensile strain can reduce the migration barrier in perovskites with an overall average of about -65 meV per percent strain, with low and high DMEPS magnitudes of -36 and -86 meV/% strain, for a consistent in-plane hop. Assuming no other factors play a role, this decrease in barrier implies an overall average increase in ionic conductivity of about 2.5 orders of magnitude for a 2% biaxial tensile strain at 300K for some of the most responsive materials calculated here (LaCrO$_3$).

The amount of change depends strongly on the individual material and its DMEPS. On average, a 2% strain is not enough to transform a poor ionic conductor into a good ionic conductor, e.g. the strain effect on migration barrier is not enough to transform LaTiO$_3$ and LaVO$_3$ with high calculated migration barriers into the same class of high-performing oxygen conductors as La[Mn, Fe, Sc, Ga]O$_3$ with their lower migration barriers. However, the effect may be enough to extend the temperature range of a good high-temperature ionic conductor into a significantly lower temperature region. For example, the ionic conductivity of ionic conductors yttria-stabilized zirconia (YSZ), yttria-stabilized bismuth oxide (YSB), and LaGaO$_3$ with Sr, Mg, and Co doping (LSGMC) decline by about 2.4, 0.7, and 0.9 orders of magnitude between 750°C and 500°C,[2] which take them out of the usable range for SOFCs by 500°C; a single order of magnitude change could potentially be recouped using strain, and allow these materials to be used at 500°C. However, we note that maintaining significant strains in bulk materials over long times and at high-temperature, as occur in SOFCs, is extremely challenging. Therefore, the most likely role for strain to enhance SOFC systems is in thin-film devices or at interfaces. Overall, the strain effect on migration barrier in the perovskites studied here suggests that strain should be considered as a method for producing or enhancing perovskite fast oxygen-ion conductors for applications such as low or intermediate temperature SOFCs, gas separation membranes, chemical looping devices, and memristors.

The origin of the strain effects on migration energy is still somewhat unclear. We find that the effects of strain on migration barrier can be captured qualitatively by a simple migration volume elasticity model. The error in the elastic model DMEPS predictions using an ab-initio calculated migration volume is 16 +/- 13 meV/% strain compared to the DMEPS fit to DFT-calculated data.

Additional aspects of the migrating oxygen beyond its migration volume, perhaps associated with local distortions during migration, appear to be playing a significant role in the strain response of anion migration in these perovskite systems.

Another interesting challenge raised by this work is to understand the origin of the variability of strain response across different B cations, which we here demonstrate to range in DMEPS by over 50 meV/% strain across the B-site cations for a consistent hop. Our efforts at correlations with B-site cation size, d-electron filling, and other plausible descriptors did not yield any robust correlations. Understanding the origin of this variability in terms of controllable parameters could yield novel materials that are engineered to respond most effectively to strain.

## Acknowledgments

We would like to acknowledge the NSF Graduate Fellowship Program under Grant No. DGE-0718123 for partial funding of T. Mayeshiba. We would also like to thank the Professor Emeritus Raymond G. and Anne W. Herb Endowment for Physics, the UW-Madison Graduate Engineering Research Scholars Program, and the Robert E. Cech materials science scholarship for additional support of T. Mayeshiba. Computing resources in this work benefitted from the use of the Extreme Science and Engineering Discovery Environment (XSEDE), which is supported by National Science Foundation grant number OCI-1053575, and from the computing resources and assistance of the UW-Madison Center For High Throughput Computing (CHTC) in the Department of Computer Sciences. The CHTC is supported by UW-Madison and the Wisconsin Alumni Research Foundation, and is an active member of the Open Science Grid, which is supported by the National Science Foundation and the U.S. Department of Energy's

Office of Science. Support for D. Morgan, conference travel funds for D. Morgan and T. Mayeshiba, and the MAST tools applied in this work were provided by the NSF Software Infrastructure for Sustained Innovation (SI$^2$) award No. 1148011. We thank M. Gadre for helpful discussions about stress and strain.## Supporting Information

Supporting information is available and describes the following:

S1. Strain notation

S2. The 2x2x2 supercell and atomic positions

S3. Our orthorhombic-to-cubic assumption

S4. Pseudopotentials, electron smearing, and climbing nudged-elastic band calculations

S5. GGA versus GGA+U

S6. Ferromagnetic, high-spin starting configuration

S7. Charge compensation

S8. Jump directions, including all barriers for B=Mn and B=Cr

S9. Migration energy and its relationship to ionic conductivity

    S9a. Relating $H_{mig}$ at Constant Pressure and $E_{mig}$ at Constant Volume, for Unstrained and Strained Cases

    S9b. Approximating the Defected Volume with the Undefected Volume

S10. Straining supercells and the strained workflow

S11. Fitting and error analysis

S12. The elastic strain model

    S12a. Finding Poisson's ratio

    S12b. Finding bulk modulus

    S12c. Finding Young's modulus

S12d. Finding migration volume

# Figures and Tables

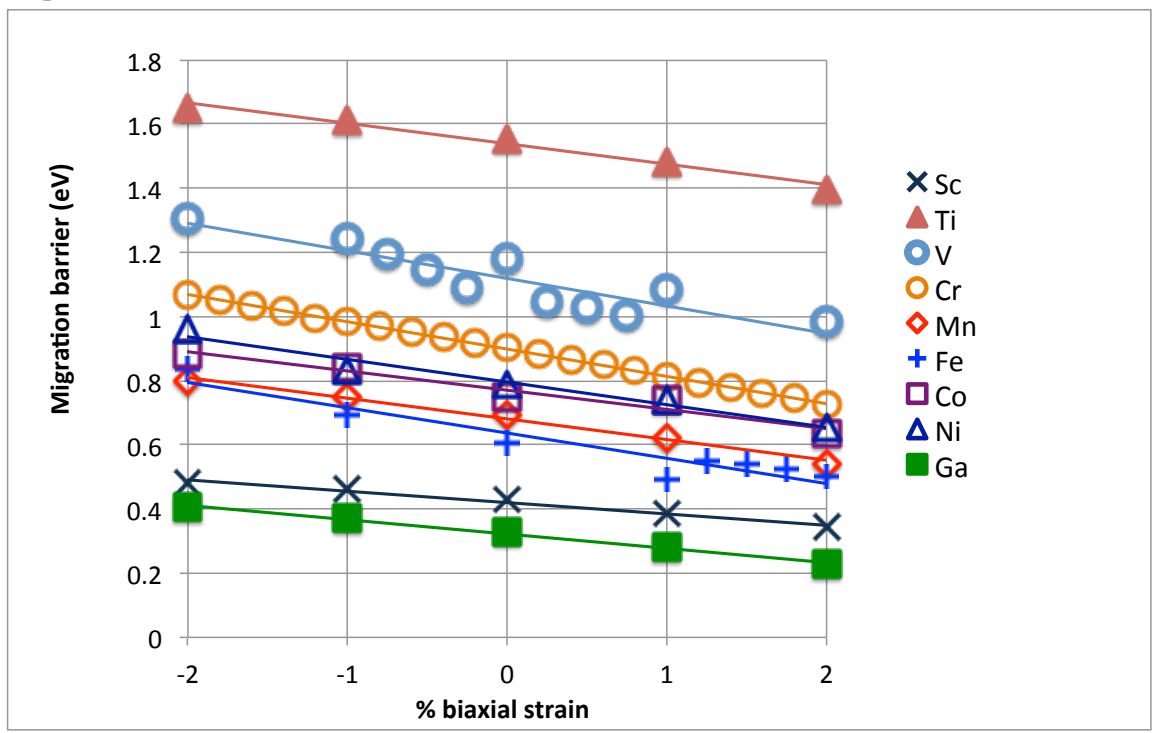

Figure 1. Migration barrier versus strain for a selected in-plane hop, which is shown schematically in Figure 2. See ESI, Section S2, for the atomic positions used (o31 to o30). The slope of each line is referred to in-text as a "DMEPS" value, which stands for "Delta (change) in Migration Energy per Percent Strain."

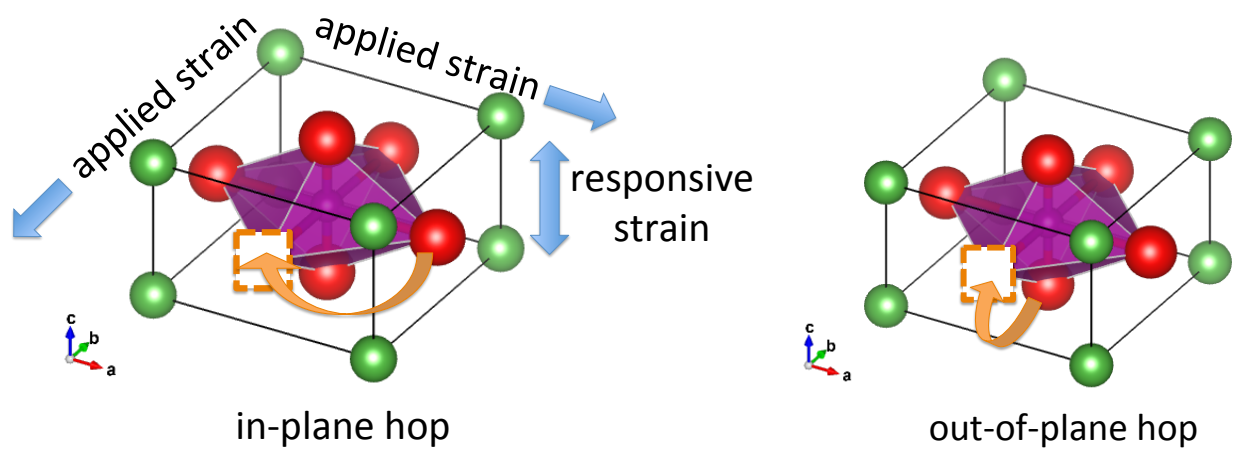

Figure 2. Schematic of an in-plane hop (left) and an out-of-plane hop (right) relative to the strain axes. The A-site cation is in green ("box corners"), the B-site cation is in purple (octahedral center), and the oxygen atoms are in red (octahedral vertices). The oxygen vacancy position is shown as an empty box with a dashed outline. This schematic shows an exaggerated example of applied biaxial tensile strain and its accompanying responsive compressive strain. For simplicity,

only a 1x1x1 segment of the actual 2x2x2 supercell is shown, and octahedral tilting is not depicted. See Supporting Information Section S2 for more detailed supercell information.

Table 1. DFT-fit DMEPS values and their errors. See ESI, Section S11, for error calculations. All numbers are in meV/% strain. While the total error range may suggest the possibility of positive DMEPS, we expect that the error estimates are not accurate in this region as we find no evidence for positive DMEPS and do not expect them for any similar perovskite systems.

| B-site cation | DMEPS fit to DFT (meV/% strain) | DMEPS fitting error (+/- meV/% strain) | Estimated bound for DMEPS over all hops (+/- meV/% strain) |
|---|---|---|---|
| Sc | -36 | 3 | 28 |
| Ti | -64 | 5 | 30 |
| V | -86 | 14 | 39 |
| Cr | -85 | 0.4 | 25 |
| Mn | -64 | 4 | 29 |
| Fe | -79 | 10 | 35 |
| Co | -60 | 8 | 33 |
| Ni | -70 | 8 | 33 |
| Ga | -44 | 2 | 27 |

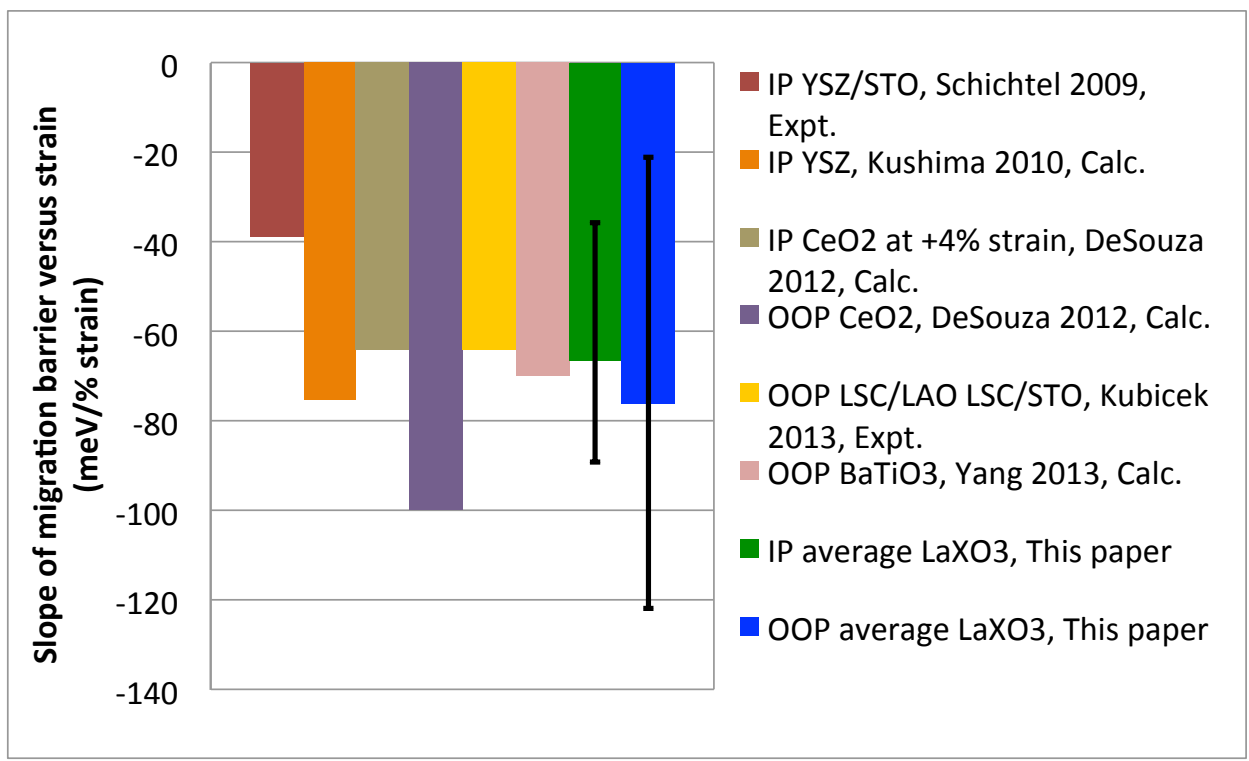

Figure 3. Comparison of literature values[3, 10, 12, 15] with values from this study. Error bars on the values from this study indicate the largest and smallest in-plane (IP) and out-of-plane (OOP) DMEPS values that were calculated for any system. DMEPS stands for "Delta (change) in Migration Energy per Percent Strain".

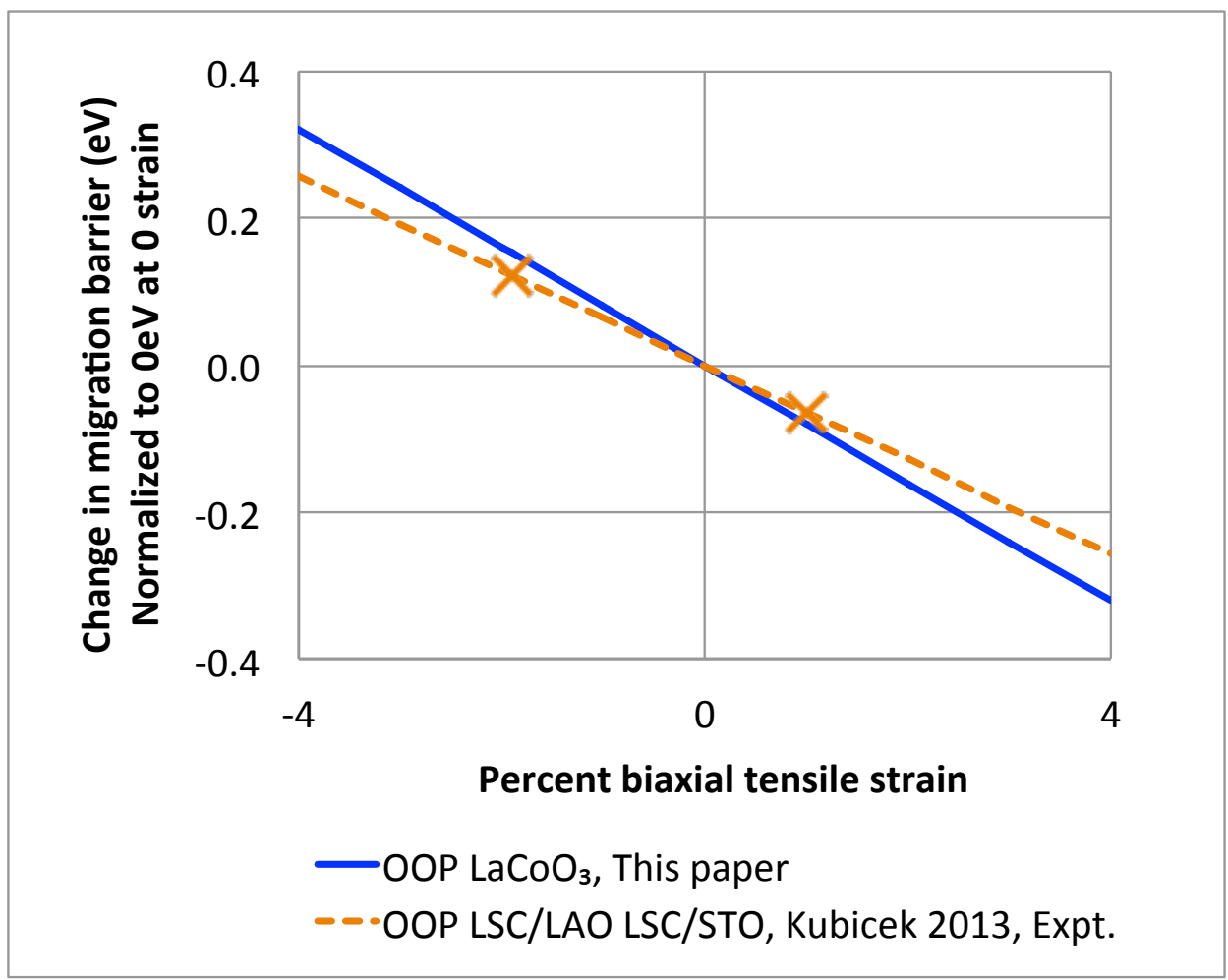

Figure 4. Comparison between our LaCoO$_3$ DMEPS for bulk oxygen diffusion and the Sr-doped LaCoO3 DMEPS for oxygen surface-to-depth diffusion of Kubicek et al.[10] DMEPS stands for "Delta (change) in Migration Energy per Percent Strain".

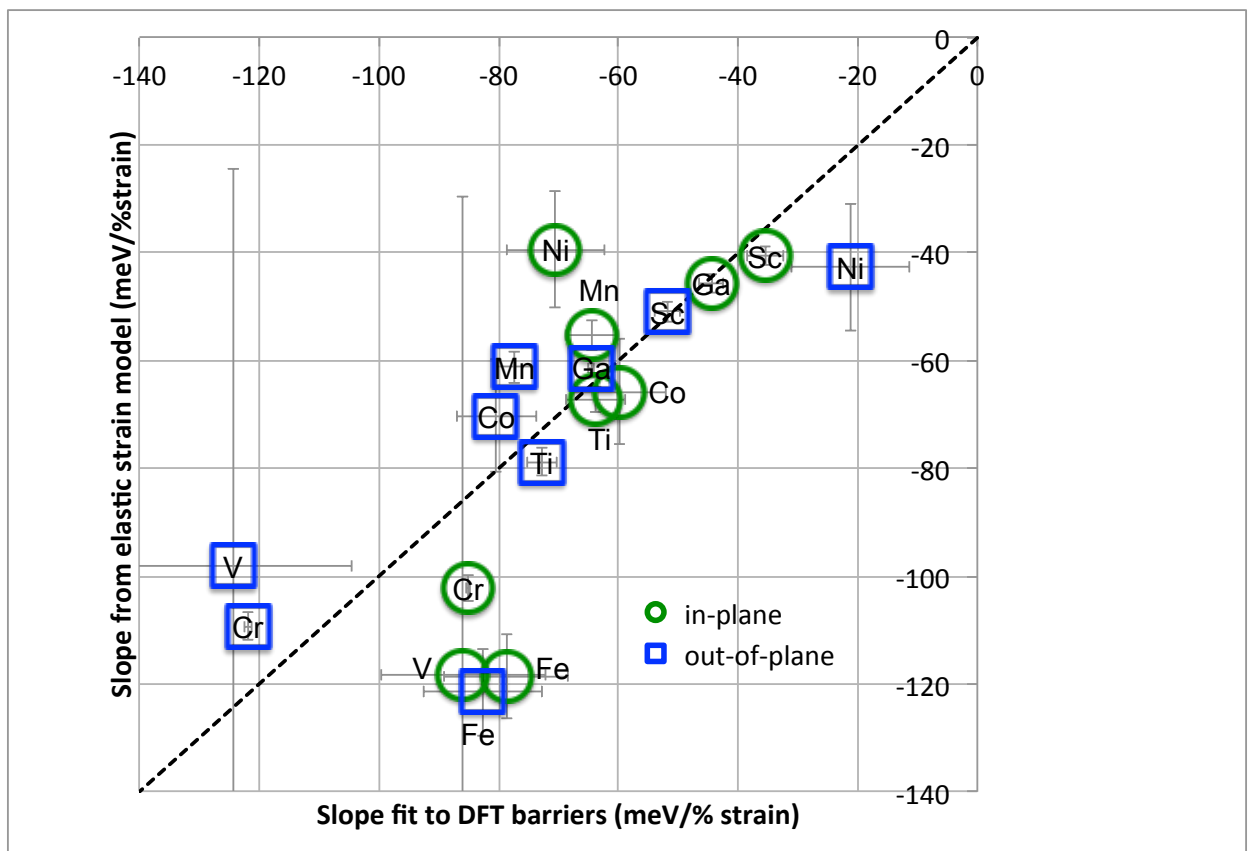

Figure 5. Elastic formula DMEPS versus fitted DMEPS for migration barrier as a function of strain (dashed line indicates perfect match). Data point is the center of each symbol. Error bars are based on uncertainties in fitting elastic constants and are discussed in the Supporting Information, Section S12. DMEPS stands for "Delta (change) in Migration Energy per Percent Strain". All error bars are symmetric.

Table E1. DFT-fit DMEPS using only biaxial strains between -1% and 1%.

| B-site | In-plane DMEPS fit to DFT (meV/% strain) | In-plane DMEPS fitting error (+/- meV/% strain) | Out-of-plane DMEPS fit to DFT (meV/% strain) | Out-of-plane DMEPS fitting error (+/- meV/% strain) |
|---|---|---|---|---|
| Sc | -37 | 2 | -54 | 0.4 |
| Ti | -65 | 4 | -75 | 1 |
| V  | -99 | 26 | -131 | 39 |
| Cr | -86 | 1 | -124 | 1 |
| Mn | -65 | 4 | -72 | 3 |
| Fe | -99 | 8 | -113 | 6 |
| Co | -51 | 23 | -100 | 11 |
| Ni | -47 | 2 | 4 | 0.5 |
| Ga | -45 | 2 | -65 | 0.5 |

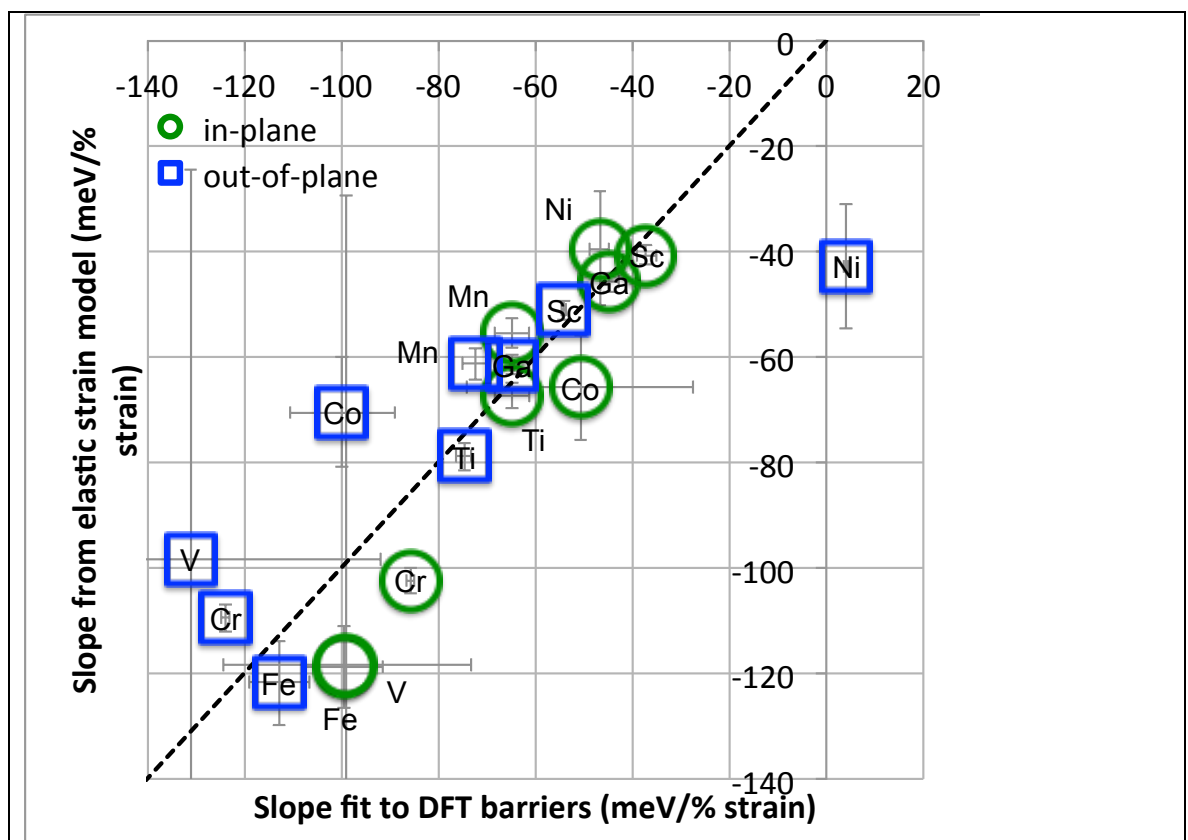

Figure E1. Elastic model slopes versus DFT-fit slopes from Table E1, using only biaxial strain data between -1% and 1%. The B-site cation Nickel point has a positive DFT-fit slope between -1% and 1%, producing the outlying position shown here, but its overall slope with strain is actually negative (see Figure S8.1R). The points for in-plane B-site cations Iron and Vanadium occupy a nearly identical spot.

**Strain Effects on Oxygen Migration in Perovskites – SUPPORTING INFORMATION**

**Updated January 26, 2016. See Update Note in main paper.**

Tam Mayeshiba

    Materials Science Program

    University of Wisconsin-Madison, Madison, WI, 53706, USA

Dane Morgan (Corresponding Author)

    Department of Materials Science and Engineering,

    University of Wisconsin-Madison, Madison, WI, 53706, USA

    ddmorgan@wisc.edu

# Contents

## S1. Strain Notations

Positive strain is taken as tensile strain. Negative strain is taken as compressive strain. Strain may be given in percentages or equivalently in multipliers, e.g. +2% strain is the same as a fractional multiplier of 1.02, and -2% strain is the same as a fractional multiplier of 0.98. For some raw data uploaded with the Supporting Information, strain is given in fractional multipliers multiplied by 1000, in order to allow for fine strain gridding without using periods in directory names, e.g. 1020 indicates +2% strain, 980 indicates -2% strain, and 1002 indicates +0.2% strain, or a multiplier of 1.002.

## S2. 2x2x2 Supercell

The general configuration of atoms for the 2x2x2 supercell is given in Figure S2.1, which was rendered using VESTA.[1] Atomic radii rather than crystal radii are used in this and other figures in order to prevent the cations from being obscured by the larger anions. The A-site cations are in green, numbers 1 through 8, the B-site cations are in purple, numbers 9 through 16, and the oxygen anions are in red, numbers 17 through 40. Additional oxygen atoms from neighboring repeated supercells are shown to complete the octahedra. The positions shown in Figure S2.1 are from $LaMnO_3$ and vary among the different compositions, although the qualitative structures are the same.

## S3. Orthorhombic-to-Cubic Assumption:

The relaxed perfect supercells for the various perovskites were orthorhombic rather than perfectly cubic, with an average angle between any two lattice vectors of 90.1°, a standard deviation of 0.3°, and a range between 89.8° and 91.4°.

For all strains, we strain the perfect supercell along lattice vectors *a* and *b* and fit for the lowest-energy strain along lattice vector *c* (see Section S10). For each strain case in lattice vectors *a* and *b*, the fitting equation is a cubic equation of supercell energy as a function of strain along lattice vector *c* (details are in Section S10). The local minimum in energy of the fit function is located, and the corresponding value of the strain along lattice vector *c* is taken.

Adjusting lattice vector *c* to be orthogonal to lattice vectors *a* and *b* changes supercell energies by less than one-tenth of 1 meV, as shown in Table S3.1. Due to the small difference in supercell energies, and the simplicity of using orthogonal principal axes, in all subsequent mathematical treatments, we assume that strain percentages along lattice vectors *a*, *b*, and *c* are equivalent to strain percentages along the three principal orthogonal axes, *x*, *y*, and *z*.

## S4. Pseudopotentials, Electron Smearing, and Climbing Nudged Elastic Band (CNEB) Calculations:

The choice of PAW-GGA PW-91 pseudopotentials was based on a recommendation on the VASP website for oxides[2] and from previous work on perovskites.[3] The soft oxygen pseudopotential was used, having a comparable maximum cutoff energy to most of the transition metal pseudopotentials, and having been shown to be adequate for many oxides.[3] The cutoff energy for relaxations and static calculations was taken as 1.5 multiplied by the cutoff energy suggested for the highest cutoff energy pseudopotential in the structure.

Gaussian smearing was used for all relaxations because the structure of most of the compounds appeared semi-metallic in VASP and Gaussian smearing would produce "reasonable results in most cases,"[4] without being prohibited for either insulators or metals. A smearing width of 0.05 eV was used for all calculations. A conjugate gradient algorithm was used for the ionic relaxations of the bulk cells and endpoints. A quasi-Newton algorithm with a force scaling factor of 0.5 was used for the nudged elastic band calculations.

The migration energy for an oxygen vacancy was determined using the climbing nudged elastic band method (CNEB) with 3 images (excluding the endpoints).[5, 6] Three images were used to ensure that the migration profile was demonstrating a single maximum rather than a local maximum-local minimum-local maximum, or to determine a more accurate energy for the global maximum. The climbing NEB method as opposed to the regular NEB method was used to ensure that one of the images climbed to the maximum energy transition state.[5] A spring constant of -5 eV/Å$^2$ was used.

In order to accomplish the NEB calculations in a quick, automated manner, a set of Python[7] scripts was written which took in a set of parameters, such as those necessary for the INCAR and POSCAR files in VASP, and then automate the calculation process. The scripts set up and run the first calculation, wait for the calculation to complete, modify the INCAR files as necessary, and submit the next calculation in the set, until all steps in the workflow are completed. These tools are part of the MAterials Simulation Toolbox (MAST), which is under development at the University of Wisconsin-Madison.[8]

The calculation steps for no strain were as follows: 1) bulk relaxation to an energy convergence of 1 meV/atom between relaxations, 2) creation and internal relaxation of two endpoints from the bulk, 3) static calculations of both endpoints, with tetrahedral smearing with Blöchl corrections[9] for a more accurate energy calculation, although keeping the same non-Gamma kpoint mesh, 4) linear interpolation of three images with center of mass adjustment from endpoint static runs, and using static endpoints as the NEB endpoints, 5) CNEB calculation, 6) static recalculation of all images with tetrahedral smearing. The NEB images were found to have no symmetry detected by VASP so are not expected to have any problems with trapping in high-symmetry states.

## S5. GGA vs. GGA+U

Many researchers use DFT+U[10] methods to treat correlations in transition metal oxides, including perovskites.[3] Although many exceptions in the literature exist,[11, 12] transition metal oxides are often treated with DFT+U in order to compensate for the electron self-interaction and excessive delocalization of d-orbital electrons in the plain GGA. Adopting values of U that have been optimized by studying non-perovskites[13] is often reasonable for studying related properties in an individual perovskite.

However, this study does not attempt to use U corrections due to uncertainty in their values, frequent convergence problems with their use,[14-16] and the fact that the present work is not focused on redox energies where U seems to play a particularly important role.[17] In particular, we expect that many of the errors associated with not using U will cancel when considering activation energies.

Figure S5.1 shows calculated GGA and GGA+U barriers, with some cases showing up to 1 eV difference, using U-values given in Table S5.1 (B=Sc and B=Ga are omitted as they were expected to have no need for U correction as they have no $d$ valence electrons in the 3+ state). These results suggest that the overall barrier magnitude is dependent on U value. However, in general we find no correlation of strain effects with barrier magnitude (see Section S8), so the changes in barriers with U do not necessarily suggest significant changes in the strain effects.

Furthermore, it is reasonable to expect that significant cancellations between barriers at different strain states will remove most of this U dependence. Therefore, we expect that our finding of decreased migration barrier with increasing tensile strain also applies to GGA+U barriers. That said, further study with U corrections and hybrid methods are clearly warranted in the future.

## S6. Ferromagnetic, High-Spin Starting Configuration

Some of the $LaXO_3$ perovskites have antiferromagnetic (AFM) structures below a certain Néel temperature.[18] The AFM structures arise from superexchange effects, mediated by the oxygen between two B-site cations.[19] The A-site cations and the oxygen anions have no magnetic moment and therefore no magnetic structuring.

There was noticeable disagreement between experimentally reported magnetic moments and our calculated bulk magnetic moments for Ti, V, Fe, Co, and Ni (see Table S6.1). This discrepancy may be due to several factors, including:

- Antiferromagnetic structures in experiment not present in the calculations, which are all ferromagnetic (FM)
- Incomplete treatment of the orbital moments, as in $LaTiO_3$,[20] which might require the addition of the spin-orbit coupling parameter in VASP
- Excessive delocalization of d-orbital electrons, which could be at least partially corrected by using GGA+U

The magnetic moments changed noticeably between the bulk and the endpoints for B-site cations Ti, Fe, Co, and Ni, and between the endpoints and the middle NEB image (the highest energy image) for Sc, Ti, Cr, Mn, Co, and Ni. This apparent change in magnetic moment occurs whether the moment is fixed using the MAGMOM tag in the INCAR or not.

In general we model our systems as FM. This choice is motivated by the fact that for SOFCs, which is our primary motivation, these systems are used under conditions of high temperature, where they are paramagnetic. While paramagnetic order is generally not practical to model, its more metallic character is often better approximated by a ferromagnetic than AFM arrangement.[3]. Nonetheless, we made an attempt to consider the effect of using the experimental AFM structures in place of the FM structure on the migration barriers (see Table S6.2). This result suggests that below temperatures where magnetic ordering occurs significant alterations in barriers from our FM values are possible. However, the impact does not seem to be large enough to change a very high barrier (over 1.5 eV) to a low barrier (less than 0.5 eV) material, or vice versa.

The initial magnetic moment on an atom is set as 1μBohr for each A-site cation, 5μB for each B-site cation, and 1μB for each oxygen anion, in a ferromagnetic configuration. We observe that VASP is able to relax these high spins to high-spin (B=Mn), intermediate-spin, low-spin, and no/nearly no magnetization states (La, O, B=Sc, B=Ga) in a sensible way, so that the A=1μB, B=5μB, O=1μB starting configuration may be consistently applied across all systems.

Figure S6.1, with references in Table S6.3, shows that calculated migration barriers in the compensated case agree well with the results of high-temperature experiments. The compensated case should be more similar to the experimental doped systems than the uncompensated case. (Note that activation energy is commonly given in the literature as $E_a$ to describe a combination of enthalpies; $E_a$ as calculated from an Arrhenius plot for the vacancy diffusion coefficient $D_v$ or for the ionic conductivity of Sr-doped $LaM^{III}O_{3-\delta}$ is equivalent to migration barrier enthalpy $H_{mig}$.[21, 22] Section S9 discusses the applicability of comparing our $E_{mig}(V)$ with $H_{mig}(P)$, while more information on compensation is in Section S7.) These results suggest that our ferromagnetic approximation is reasonable for treating the paramagnetic systems. More care should be taken when interpreting this data for use at lower temperature where strong magnetic ordering occurs.

**S7. Charge Compensation: Electron-Removal Compensation versus Doping:**

In this section, we describe our approach to compensating the charges associated with an oxygen vacancy formation and evaluate the difference between our electron-removal compensation mechanism, where we remove one oxygen atom along with two electrons, and actually doping the supercell.

An oxygen vacancy in a perovskite means that there is one fewer oxygen atom that can receive electrons donated by the cations. In order to preserve the charge neutrality of the overall crystal, the cations in the crystal must give up two fewer electrons for every oxygen vacancy, or equivalently, one can think of the oxygen vacancy as donating two electrons to the system. The donated electrons typically reduce transition metal B site cations in the material, or are at least formally considered to do so. For pure $A^{3+}B^{3+}O_3$ perovskite, the excess electrons will generally reduce the B-site cations (which typically contain transition metals) from 3+ to 2+. However, most perovskite systems used for fast oxygen conduction have lower-valence dopant atoms on the A- or B-sites, such as $Sr^{2+}$, which create $B^{4+}$ cations. For these doped systems the donated electrons may reduce some B-site cations from 4+ to 3+. In general the doping oxidizes the system more than the oxygen vacancies reduce the system, although this may not hold for all systems and can depend on temperature and oxygen partial pressure. Therefore, the system is predominantly a 4+ and 3+ B-site cation mixture and most oxygen will diffuse in an environment of 4+ and 3+ cations. The exact environment around the diffusing oxygen could be extremely complex. However, it is likely from simple electrostatic arguments that the 3+ will be closer to the vacancy and most systems are predominantly 3+. Furthermore, since doping levels and species vary, they open up a very wide-range of possible local environments. To keep the calculations tractable and avoid complexities of dopants couplings we therefore generally work with cells without explicit dopants. Thus we perform all calculations for undoped systems and 3+ cations.

In order to maintain 3+ cations even in the presence of the extra electrons donated by the oxygen vacancy we create a vacancy by removing from the supercell both an oxygen atom with its six valence electrons and an additional two electrons. This procedure is the computational equivalent of substituting lower-valence dopant atoms on A-sites or B-sites somewhere else in the crystal beyond the boundaries of the supercell. The advantage of this method is that it avoids the interaction between oppositely-charged defects by creating a single oxygen vacancy in the supercell without using dopant atoms.

Although the avoidance of explicitly including dopants greatly simplifies the calculation, it is important to assess if this approximately approach causes significant errors in the results. We therefore also performed a series of migration energy calculations with explicit Strontium doping in order to assess if the values are similar to those of the compensated method. For the doped supercells, we remove an oxygen atom (the atom only, without removing any extra electrons) and also substitute two Strontium atoms at two of the Lanthanum A-sites. VASP relaxes these cations to a +2 and a +3 state, respectively. For the $La_{0.75}Sr_{0.25}BO_3$ series, cation positions 7 and 8 were chosen (see Figure S2.1). This positioning is arbitrary. Differences created by different positioning of the A-site defects are expected to be small compared to the margin of error associated with the overall effects of including vs. excluding dopants and other errors in due to finite size and inherent DFT limitations. For example, the migration barrier for the O29 to O30 hop in $La_{0.75}Sr_{0.25}MnO_3$ with Strontium atoms in positions 4 and 5 is 0.98 eV, while the migration barrier with Strontium atoms in positions 7 and 8 is 1.00 eV.

The migration barrier difference between the two methods of compensation is shown in Table S7.1 as the $LaBO_3$ electron-removal compensated barrier minus the $La_{0.75}Sr_{0.25}BO_3$ Sr-doped barrier for each B-site cation. The magnitude of the difference is on average 140 meV, with a standard deviation around this difference of +/- 69 meV. This range of differences is below the range assumed solely for jump directions (see Section S8), and is likely a result of the different geometry imposed by adding Sr atoms to the doped cell. However, the shift has a clear direction, and the effect of Sr can be more usefully thought of as raising the no-dopant simulation results by about $140 \pm 69$ meV. This value provides a relatively easy shift one can apply to relate the values from the two approaches if needed. Given that the shift is relatively constant we expect that the strain response from our compensated calculations and the explicit Sr doped calculations will be similar, yielding similar energy/% strain slopes.

Overall, as this Sr coupling is hard to model accurately, electron removal is chosen to perform the strained, compensated calculations. We believe this approach better represents the pure compensation effect independent of which dopants and dopant placements are used to perform the compensation. We also believe that omitting the dopant atoms gives a more universal picture of strain effects for a given B-site cation. Nevertheless, measuring the strain effects on a variety of explicitly doped supercells would be a valuable future addition to the complete dataset. In Table S7.2 and Figure S7.1 through Figure S7.6, we provide migration barrier data for strained supercells with three explicit Sr A-site doping configurations for B-site cations Sc, Cr, and Mn. Table S7.2 shows that, with the exception of in-plane hops for Mn, the slopes for the explicitly doped supercells fall within the +/- 25 meV/% strain range of slopes that we expect given all hops in a supercell (see Section S8). The difference in slopes for the in-plane doped Mn supercells, as well as their higher-than-average errors in fitting, can be attributed to the migration barrier values for compressive strained supercells which are higher than would be expected for an approximately linear trend. The underlying cause of these increased barriers requires further investigation. In any case, Table S7.2 and Figure S7.1 through Figure S7.6 show clearly that in both explicitly doped supercells and in electron-removal charge-compensated undoped supercells, oxygen migration barriers decrease with increasing tensile strain.

## S8. Jump Directions

The migration of two oxygen atoms in two particular directions are calculated for each of the systems. The two calculated migration barriers were chosen so that one hop is in-plane (O31 to O30; see Figure S2.1 for positions) and one hop is out-of-plane (O29 to O30). This choice is an approximation as there are in fact multiple symmetry-independent hops in the unit cell due to the non-cubic symmetry of the low-temperature perovskite phase and the symmetry breaking of the vacancy. The in-plane hop is described in the main text, while the out-of-plane hop is plotted in Figure S8.1, with fitted slope values in Table S8.1. The slope values for the selected consistent in-plane and out-of-plane hops are plotted together in Figure S8.2. This data suggests perhaps a slight trend for more negative slopes for out-of-plane hops but this effect is almost certainly just an artifact of the specific hops and systems chosen as different hops can have quite a wide spread of slopes (see discussion below in this section). Figure S8.2 also shows that there is no clear trend in the slopes with the atomic number of the B-site cation.

The total range of barriers for systems where we have calculated all of the symmetry distinct hops is less than 300 meV (Figure S8.3 and Figure S8.4). Table S8.2 shows additional data for eight hops in different systems, with a maximum range of 600 meV and an average range of 280 +/- 160 meV. However, each particular barrier from Figure S8.3 and Figure S8.4 follows the same qualitative trend of decreasing with tensile strain that is seen for the barrier in the main text and the selected out-of-plane barrier, with representative examples given in Figure S8.5 and Figure S8.6. The range of the slopes of migration barriers for symmetry distinct hops in meV/% strain is some 50 to 70 meV/% strain from Figure S8.7 and Figure S8.8. While this range is large, all slopes for all barriers are decreasing slopes (we omit a single $LaCrO_3$ barrier for octahedron 9, -2% strain which had an unusual and probably erroneous total magnetic moment). We also note that the value of the slope does not correspond to the magnitude of the zero-strain barrier, and neither out-of-plane nor in-plane slopes are consistently larger than the other. Furthermore, the range in slopes does not change the overall prediction pattern for elastic theory-based slope versus DFT-calculated slope (Figure 5 in the main text, compared with Figure S8.9). The DMEPS predicted by the strain model qualitatively follow the same trends as the *ab initio* DMEPS for all hops in $LaCrO_3$ and $LaMnO_3$, when considering the entire cluster of points in each system. The spread of differences between the elastic model slopes and the DFT-fit slopes over the entire cluster of points for each system indicates that the limits of reasonable accuracy for using a single hop in a system to predict DMEPS with the elastic strain model are some 30 meV/% strain.

Figure S8.3, Figure S8.4, Figure S8.7, and Figure S8.8 show that groups of similar barrier and slope patterns (central B-site cations 10, 12, 13, and 15 versus central B-site cations 9, 11, 14, and 16) correspond to groups of octahedra which are similarly tilted around the b-axis (although not identically tilted in all respects), as can be seen from Figure S2.1. One additional observation that may be of interest for further study is that the same in-plane hops that have the largest difference between DFT DMEPS and elastic model DMEPS in LaMnO$_3$ actually have some of the smallest differences in LaCrO$_3$.

## S9. Migration Barrier $E_{mig}$ and Relationship to Ionic Conductivity:

In this paper we focus on migration energies, $E_{mig}$, but often wish to relate the to measured conductivities, particularly ionic conductivities. The connection between these quantities is discussed here.

Ionic conduction in perovskites is dominated by the movement of oxygen anions,[23, 24] while electronic conduction comes either from B-site cation electrons,[25] or from hole conduction at high oxygen partial pressures.[24] Significant effort is made to dope the perovskite A- and B-sites in order to produce the desired amounts of electronic and ionic conduction. For example, the doping of 2+ A-site cations (call them species M) in the place of 3+ A-site cations produces 2+-charged oxygen vacancies, as shown in Equation S9.1 and Equation S9.2:[26]

| | |
|---|---|
| $$A_2O_3 + B_2O_3 = 2ABO_3$$ | (S9.1) |
| $$2MO(A_2O_3) = 2M'_A + V_O^{\cdot\cdot} + 2O_O^x$$ | (S9.2) |

Assuming that oxygen ions are the only mobile ionic species in the material, conductivity is given by Equation S9.3:[23]

| | |
|---|---|
| $$\sigma_{total} = \sigma_{electronic} + \sigma_{oxygen\ ions}$$ | (S9.3) |

The assumption that oxygen ions are the only diffusing ions is reasonable for the perovskite system, especially in the context of the major intended uses. For example, SOFC operation depends on the adsorption and separation of gaseous oxygen, its motion as oxygen anions through the device, and the eventual recombination of those oxygen anions with hydrogen on the fuel side into water. In contrast, cations do not enjoy a large concentration of cation vacancies tailored by doping (cation doping produces anion vacancies), nor do they have a similar chemical reaction pathway that encourages unidirectional motion and keeps a supply of cations available. Furthermore, cation migration barriers are typically ~2.5-3 eV, making them far less mobile than oxygen.[27, 28] Massive cation motion may also imply phase segregation or material failure in the context of SOFCs.

Looking only at ionic conductivity, the ionic conductivity may be described through Equation S9.4, where $\eta$ or $C$ is the concentration of each species of charge-carrying ion, $q$ is the charge on each ion, and $\mu$ is the mobility of each ion; assuming only the motion of oxygen ions, the summation only contains one term:[29]

$$\sigma_{ionic} = \sum_i \eta_i q_i \mu_i = C_i q_i \mu_i \quad (S9.4)$$

The Nernst-Einstein equation relates mobility to the diffusion coefficient $D$ and changes the conductivity expression to Equation S9.5, where $k$ is the Boltzmann constant and $T$ is temperature, and the subscript $O$ is used for oxygen anions, $O^{2-}$:[25]

$$\sigma_{ionic} = C_i q_i \mu_i = \frac{C_i q_i^2 D_i}{kT} = \frac{C_O q_O^2 D_O}{kT} \quad (S9.5)$$

When an oxygen ion moves, it swaps spots with an oxygen vacancy. The diffusion coefficient for oxygen ions, $D_O$, can be related to the diffusion coefficient for oxygen vacancies, $D_v$, through their relative concentrations $C$ in Equation S9.6:[22]

$$D_O = D_v \frac{C_v}{C_O} \quad (S9.6)$$

Substituting in the expression for $D_O$ and recognizing that the square of the 2- charge on an oxygen anion is equivalent to the square of the 2+ charge on an oxygen vacancy, or $q_O^2 = q_v^2$, gives Equation S9.7,

$$\sigma_{ionic} = \frac{C_O q_O^2 D_O}{kT} = \frac{C_O q_O^2 D_v C_v}{kT C_O} = \frac{C_v q_O^2 D_v}{kT} = \frac{C_v q_v^2 D_v}{kT} \quad (S9.7)$$

or, alternatively, Equation S9.8,

$$\sigma_{ionic} = C_v q_v \mu_v = \frac{C_v q_v^2 D_v}{kT} \quad (S9.8)$$

Vacancy concentration and vacancy diffusion are both thermally-activated and may often be at least approximately expressed as functions of temperature in Equation S9.9, Equation S9.10, and Equation S9.11, where $\gamma$ is a geometric factor, $a$ is the jump distance, and $v_0$ is the vibrational frequency of the moving ion, the subscript "mig" stands for migration, and the subscript "form" stands for vacancy formation:[25]

$$D_v = \gamma a^2 v_0 \exp\left(\frac{-G_{mig}}{kT}\right) \quad (S9.9)$$

$$D_v = \gamma a^2 v_0 \exp\left(\frac{TS_{mig}}{kT}\right) \exp\left(\frac{-H_{mig}}{kT}\right) \quad (S9.10)$$

$$D_v = D_{v0} \exp\left(\frac{-H_{mig}}{kT}\right) \quad (S9.11)$$

(Here we use $G_{mig}$, $E_{mig}$, etc. as the Gibbs free energy and internal energy of migration; often they are also termed $\Delta G_{mig}$ or $\Delta E_{mig}$ to signify the change in energy during migration. However, since we are treating slopes in migration energy with respect to strain, or changes in change-of-energy-during-migration, we use $E_{mig}$ for the migration barrier energy, and $\Delta E_{mig}$ for the change in that migration barrier quantity, as with respect to strain.)

We will assume that for a perovskite doped for some practical purpose, the vacancy concentration comes primarily from aliovalent doping rather than from thermal activation, with thermal activation playing only a small part for vacancy formation energies between 2.4 and 5 eV.[3, 30] For example, for commercial LSGM ($La_{0.80}Sr_{0.20}Ga_{0.80}Mg_{0.20}O_{3-x}$),[31] the nominal vacancy concentration is 1 vacancy per two Strontium substitutions, or 0.1 per formula unit. Given a calculated $LaGaO_3$ volume of 488 Å$^3$ for 8 formula units, this is 8*0.1 = 0.8 vacancies, or a vacancy concentration $C_v$ of (0.8/488Å$^3$). In order to obtain this vacancy concentration from a purely thermally-activated process with a relatively low $H_{vf}$ of 3 eV, the initial vacancy concentration $C_{v0}$ would have to be thousands of vacancies/Å$^3$ at temperatures of 1173K or lower. However, for some systems the vacancy concentration will be dominated by thermally generated vacancies and the strain response of the diffusion and ionic conductivity may be strongly influenced by changes in $H_{form}$ with strain. We do not include these effects in the present work but they are an important area for further study.

Substituting in the temperature-dependent expression for $D_v$ in Equation S9.11 to the formula for ionic conductivity in Equation S9.8 gives Equation S9.12:

$$\sigma_{ionic} = \frac{q_v^2}{kT} C_v D_{v0} \exp\left(\frac{-H_{mig}}{kT}\right) \quad (S9.12)$$

Across two strains at a given temperature, the ionic conductivities may be compared as in Equation S9.13. Assuming that $C_v$ at strain 1 is similar to $C_v$ at strain 2, we arrive at Equation S9.14:

$$\frac{\sigma_{ionic,\epsilon 1}}{\sigma_{ionic,\epsilon 2}} = \frac{\frac{q_v^2}{kT}C_{v,\epsilon 1}}{\frac{q_v^2}{kT}C_{v,\epsilon 2}} \times \frac{D_{v0,\epsilon 1}\exp\left(\frac{-H_{mig,\epsilon 1}}{kT}\right)}{D_{v0,\epsilon 2}\exp\left(\frac{-H_{mig,\epsilon 2}}{kT}\right)} \quad (S9.13)$$

$$\frac{\sigma_{ionic,\epsilon 1}}{\sigma_{ionic,\epsilon 2}} \approx \frac{D_{v0,\epsilon 1}}{D_{v0,\epsilon 2}}\exp\left(\frac{-H_{mig,\epsilon 1}+H_{mig,\epsilon 2}}{kT}\right) \quad (S9.14)$$

We recognize that $D_{v0}$ has contributions from geometric factors, a correlation term, and phonons. Phonons are likely to be only weakly dependent on strain,[32] the correlation term is constant for dilute vacancies, and the overall geometric factors for the cell should be similar for small strains. These assumptions yield Equation S9.15, giving the ratio of ionic conductivities at different strains. The assumption that $C_v$ is independent of strain is likely not true in general, and Equation S9.14 and Equation S9.15 should be taken as limiting cases which include only the strain effect through migration energies without contributions from changes in vacancy content. However, we note that for doped perovskites, as are often used in oxygen conducting applications, the vacancy concentration is largely controlled by dopant concentration and not defect formation enthalpies. Under these quite common circumstances we do expect the vacncy concentration to have only a weak dependence on strain.

$$\frac{\sigma_{ionic,\epsilon 1}}{\sigma_{ionic,\epsilon 2}} \approx \exp\left(\frac{-H_{mig,\epsilon 1}+H_{mig,\epsilon 2}}{kT}\right) \quad (S9.15)$$

At a given temperature, making the assumption that all other quantities stay equal, Equation S9.15 makes it straightforward to relate changes in $H_{mig}$ to changes in conductivity. To translate experimental or literature data giving trends in $\sigma_{ionic}$ into a comparable slope value of $\Delta H_{mig}$/%strain we simply invert Equation S9.15 and divide the changes in $H_{mig}$ by the changes in strain to produce a slope. We report and compare these slopes directly to our calculated $\Delta E_{mig}$/%strain slopes. We justify the equivalence of the experimental constant-pressure strained migration enthalpies $H_{mig}$ and our calculated constant-volume strained migration barriers $E_{mig}$ in Section S9a, Section S9b, and Section S9c below.

### S9a. Relating $H_{mig}$ at Constant Pressure and $E_{mig}$ at Constant Volume, for Unstrained and Strained Cases

The following derivation will show that what is calculated with DFT, which is $E_{mig}$ at constant volume per formula unit $v'$, is approximately equivalent to $H_{mig}$ at constant pressure $P'$.

Suppose a diffusion experiment is run at some pressure $P'$ and temperature $T$. Making the assumptions described in Section 9 above in order to relate conductivities, we extract out a $H_{mig}$ at pressure $P'$. This pressure $P'$ is consistent from the initial defected state through the transition state to the final defected state. The initial defected state has a volume of $v'$ per formula unit, or a total volume of $V_{ol}=Nv'=V'$, where N is the number of formula units. The transition state has a slightly different volume (in order to stay at pressure $P'$), which we define as $V_{tst}=Nv' + V_{mig}(P')=V' + V_{mig}(P')$, where $V_{mig}(P')$ is the volume change associated with the atomic migration at pressure $P'$.

Now suppose we have a migration barrier calculation whose initial defected state is set to that same pressure $P'$ with the characteristic per-formula-unit volume $v'$, or total volume $Nv'$, for the same number of formula units as in the experiment (although practically speaking this would not be the case). For better convergence in the CNEB calculation, we fix the volume rather than the pressure, so that the transition state remains at constant volume $Nv'$, but its pressure increases.

We define two internal energies as functions of different variables: internal energy $E$ as a function of volume $V$, and internal energy $U$ as function of pressure $P$. Note that here $U$ is not the finite temperature internal energy, but still the zero-temperature internal energy equivalent to $E$ but written as a function of pressure. Then we define the migration energy as the transition state energy, $E_{tst}$ or $U_{tst}$, minus the energy of the initial defected state when the oxygen (or vacancy) is on the lattice site, $E_{ol}$ or $U_{ol}$, in Equation S9a.1 and Equation S9a.2. Note that because we are in a finite size supercell which can impact our results we explicitly keep $N$ as an independent variable.

The transition state energy can be split into a bulk-like term and a term associated with the transitioning atom (here, a single oxygen) and its surrounding atomic relaxations. The bulk-like term is an energy component that scales linearly with the number of formula units $N$. The term associated with the transitioning atom and its surrounding atomic relaxations, on the other hand, does not depend on the number of formula units $N$ for large $N$. For very few formula units, this local energy term may be affected by the supercell size, but as the calculation supercell gets larger, the local energy term converges to an asymptotic value. This dependence yields S9a.3 for all $N$, and S9a.4 for $N$ large enough that $E_1$ has reached its asympotic value. We note that $\varepsilon_0$ is just the energy per formula unit of the host system, here the undefected perovksite.

| | |
|---|---|
| $E_{mig}(V, N) = E_{tst}(V, N) - E_{ol}(V, N)$ | (S9a.1) |
| $U_{mig}(P, N) = U_{tst}(P, N) - U_{ol}(P, N)$ | (S9a.2) |
| $E_{tst}(V, N) = N\varepsilon_0(v) + E_1(V, N)$ | (S9a.3) |

$$E_{tst}(v, N) = N\varepsilon_0(v) + E_1(v) \ (in\ large\ N\ limit) \quad (S9a.4)$$

As we will from here forward be considering only cases where *N* is large enough that Equation S9a.4 holds and is otherwise fixed, we will not explicitly write the dependence on *N* in the following equations except when it is needed for clarity. Using the definition of enthalpy, $H_{mig}(P')$ can be written as Equation S9a.5 Using the definition of migration energy, $H_{mig}(P')$ can be further split into Equation S9a.6.

Now we make two major substitutions, changing the energy definition in the enthalpy from an energy *U* as a function of pressure *P* to an energy *E* as function of volume *V*. First, for the on-lattice energy, the on-lattice pressure *P'* was defined as having a corresponding volume *V'*, so we substitute the energy $E_{ol}(V')$ for the energy $U_{ol}(P')$ *(since they are the same value)*, giving Equation S9a.7. Second, for the transition-state energy, the transition state pressure *P'* was defined as having corresponding transition state volume $V_{tst} = V' + V_{mig}(P')$. Therefore, we substitute the energy $E_{tst}(V' + V_{mig}(P'))$ for the energy $U_{tst}(P')$, to give Equation S9a.8.

$$H_{mig}(P') = U_{mig}(P') + P'V_{mig}(P') \quad (S9a.5)$$

$$H_{mig}(P') = U_{tst}(P') - U_{ol}(P') + P'V_{mig}(P') \quad (S9a.6)$$

$$H_{mig}(P') = U_{tst}(P') - E_{ol}(V') + P'V_{mig}(P') \quad (S9a.7)$$

$$H_{mig}(P') = E_{tst}\left(V' + V_{mig}(P')\right) - E_{ol}(V') + P'V_{mig}(P') \quad (S9a.8)$$

In Equation S9a.9, we Taylor expand the $E_{tst}$ term.

$$E_{tst}\left(V' + V_{mig}(P')\right) \quad \text{(S9a.9)}$$

$$= E_{tst}(V') + V_{mig}(P')\left.\frac{dE_{tst}(V)}{dV}\right|_{V'} + \frac{1}{2!}\left(V_{mig}(P')\right)^2 \left.\frac{d^2 E_{tst}(V)}{dV^2}\right|_{V'}$$

$$+ \cdots$$

Using the implication of Equation S9a.4 that $E_{tst}$ is a function of only $N$ and $v$, and noting that at fixed $N$, $\frac{d}{dV} = \frac{1}{N}\frac{d}{dv}$, Equation S9a.9 can be rewritten as Equation S9a.10.

$$E_{tst}\left(\frac{V' + V_{mig}(P')}{N}, N\right) \quad \text{(S9a.10)}$$

$$= E_{tst}(v', N) + V_{mig}(P')\left.\frac{dE_{tst}(v', N)}{Ndv'}\right|_{v'}$$

$$+ \frac{1}{2!}\left(V_{mig}(P')\right)^2 \left.\frac{d^2 E_{tst}(v', N)}{N^2 dv'^2}\right|_{v'} + \cdots$$

Now susbtituting in Equation S9a.4 for $E_{tst}$ and dropping terms of O(1/$N$) and higher powers of 1/$N$ gives Equation S9a.11.

$$E_{tst}\left(\frac{V' + V_{mig}(P')}{N}, N\right) = E_{tst}(v', N) + V_{mig}(P')\left.\frac{d\varepsilon_0(v')}{dv'}\right|_{v'} \quad \text{(S9a.11)}$$

$$= E_{tst}(v', N) - P'V_{mig}(P')$$

Here we have defined used the fact that, as $\varepsilon_0$ is the bulk system energy per formula unit, its volume per formula unit derivative is just the negative pressure. Equation S9a.11 can be rewritten in terms of our total volumes as Equation S9a.12.

$$E_{tst}\left(V' + V_{mig}(P')\right) = E_{tst}(V') - P'V_{mig}(P') \qquad \text{(S9a.12)}$$

Substituting Equation S9a.12 into Equation S9a.8 yields Equation S9a.13, Equation S9a.14, and Equation S9a.15..

$$H_{mig}(P') = E_{tst}(V') - P'V_{mig}(P') - E_{ol}(V') + P'V_{mig}(P') \qquad \text{(S9a.13)}$$

$$H_{mig}(P') = E_{tst}(v', N) - E_{ol}(V') \qquad \text{(S9a.14)}$$

$$H_{mig}(P') = E_{mig}(V') \qquad \text{(S9a.15)}$$

We reiterate that the above only holds for large enough *N* that we can use Equation S9a.4 and drop terms of O(1/*N*) in the Taylor expansion in Equation S9a.10. With this result, we see that constant-volume migration barrier energies for a large enough supercell size can be used to approximate the values for constant-pressure enthalpies, assuming low temperature.

Note that the above derivation directly applies only for an isotropic system. However, in the general case one may have a migrating atom with an isotropic migration volume tensor and under mixed boundary coundiation, with some components at fixed strain and others at fixed stress. One can again ask the question whether the correct migration free energy (at low temperature) under the mixed boundary conditions is well represented by a fixed volume (and with fixed lattice vectors) ab inito calculated energy difference. The above derivation readily generalizes to this more complex situation, as each $\sigma_{ij}\epsilon_{ij}$ component of the migration energetics can be treated independently. If the migration is at fixed $\epsilon_{ij}$ then the calculation is being done at directly comparable boundary conditions. If the migration is at fixed $\sigma_{ij}$ then the calculation is approximately correct through an argument parallel to that above for the specific *ij* component.

### S9b. Approximating the Defected Volume with the Undefected Volume

In Section S9a we assumed that the starting pressure and volume for the calculations were $P'$ and $V'$, respectively, which is the volume for the defected supercell. However, the starting volume in our calculations is really $V_0$, the volume of the undefected supercell, or rather $\overline{V_0}$ with strain, where we use the bar over the variable to represent its value in our strained calculations. Therefore, the starting pressure is some $\overline{P_0}$ which is then modified by the effects of introducing a vacancy, and at volume $\overline{V_0}$ rather than $\overline{V'}$, the starting pressure is not exactly $\overline{P'}$. That said, with large enough cells (enough formula units $N$), these differences in volume and pressure diminish, as there is only one vacancy being introduced in a background of many formula units $N$, so the derivations in Section S9a still hold.

Through all of Section S9a and Section S9b we mention "large enough" formula units $N$. Our supercell sizes at an undefected supercell of 40 atoms or $N=8$ are not large enough for the approximations to become equalities within the limits of precision. However, we show in Table S12.1 (Section S12) that our values using migration energies are comparable with expected strain model values using isotropic pressures and volumes, even with our various assumptions, approximations, and distinctions (e.g. in starting pressures and volumes), and even at our relatively small $N=8$.

## S10. Straining Supercells
Strains along lattice vectors $a$ and $b$ were taken at ±1% and ±2% of the original lattice parameters, with positive values as tensile strain and negative values as compressive strain (for example, tensile strain of +2% would strain the lattice parameter to 1.02 times the original lattice parameter and a compressive strain of -2% would strain the lattice parameter to 0.98 times the original lattice parameter). Lattice mismatch strains of up to ~7 percent have been reported, although thin films may not allow as much strain, and strain may produce segregated phases.[33-35]

For each strain case (with lattice vectors *a* and *b* having equal fixed strains), the response in lattice parameter *c*, and therefore also the strained volume, was found by fitting a cubic function to the energies of a series of bulk calculations with different lattice vector *c* strains. At least seven lattice vector *c* strains were calculated to produce data points for the fit, in 1% steps in a range about 3% above and 3% below the estimated lattice vector *c* strain value, which from experience was known to be somewhat smaller than that given by a volume-conserving response. Each strain case consisted of a low kpoint-mesh (2x2x2 M) initial internal optimization, a 4x4x4 M kpoint-mesh internal optimization, and a static calculation, all at fixed volume (once the strain had been applied) and cell shape. Additional lattice vector *c* points were calculated as neccessary, for example, in order to distinguish between two distinct magnetic moment curves with B-cations Fe, Co, and Ni, and to choose only those points on a curve consistent with the magnetic moment trending near the volumetrically-likely lattice vector *c* strain (see Figure S10.1). To further explore the trends of c axis with strain, the lattice vector *c* strains were plotted against the lattice vector *a* and *b* strains. Fine-gridding of $LaCrO_3$ showed that such a plot is a smooth curve, fitting well to a quadratic function (see Figure S10.2). The fitted minimum-energy lattice vector *c* strain for each biaxial strain case was subsequently used, along with the strained lattice vectors *a* and *b*, to fix the volume and cell shape in all further calculations for that strain case and chemical system. Note that 0% strain was also subjected to the same fitting treatment, and we found that in most cases, 0% strain in lattice vectors *a* and *b* did not, in fact, correspond to 0% strain in lattice vector *c* but rather to a magnitude of 0.3% strain or lower, indicating small convergence errors in the original groundstate calculations.

We note that the total volume of a system could vary by as much as 5% among different strain states due to different equilibrium $c$ lattice parameter values.

Each undefected strain state bulk was allowed to relax only internally, and then the endpoint and NEB calculations proceeded as outlined in Section S2 through Section S7.

Occasionally, for systems B=V, Fe, Ni, where we were interested in checking apparent deviations from a linear slope in $E_{mig}$ with strain, more fine-gridded strains in $a$ and $b$ lattice parameters were evaluated, using a lattice parameter $c$ based on the quadratic fit of $c$ vs. strain in $a$. Also, some systems' defected endpoints and NEB images were started over using the fractional coordinates of the same system at a different strain to explore for metastable solutions. In these cases, the lowest-energy activated state energy, which corresponds to the lowest-energy barrier (since the endpoints remained the same), was taken to calculate the barrier. The finely gridded B=Cr system shows that for a well-behaved system, migration barrier versus strain falls along a smooth line (see B=Cr case in Figure 1 in the main text).

Outlier points occasionally exhibited some sort of polymorphic distortion, e.g. when comparing middle images among strain cases, an O-B-O bond angle for $LaVO_3$ suddenly changed its sign at the zero-strain case and then back again, rather than changing gradually with increasing tensile strain. An NEB calculation for $LaVO_3$ restarting with randomly perturbed images found a very close energy (within 0.005 eV per 8 formula units) defected structure with a completely different and rotated octahedron tilt configuration, although the oxygen vacancy remained in the same place. The Generalized Gradient Approximation (GGA) method used in this work (see Section S5) may also lead to unstable magnetic moments; for example, our GGA $LaFeO_3$, $LaNiO_3$, and $LaCoO_3$ systems showed different magnetic moments under different strain states (with a fluctuation of 0.35, 0.3, and 0.4 µB per Fe, Ni, and Co). Fixing magnetic moments could be arbitrary (for example, the magnetic moment for Ni in LaNiO$_3$ in GGA could vary between 0 and 0.4 µB per Ni across all strain cases, up to 0.3 µB per Ni within a single strain case, and up to 0.3 µB per Ni within a single NEB), and changes in magnetic moment profiles of NEBs from one strain case to another or between two NEBs of the same strain cases may also be a function of polymorphism. Total convergence among polymorphs and magnetic moments appears to be quite challenging and was not necessary to illustrate the clear overall trends we observe here, so we do not discuss them further in this paper. Where multiple migration barrier data was taken for a system using a given hop and at a given strain, the lowest migration barrier from initial to final endpoint is used.

## S11. Fitting and Error Analysis

For linear and quadratic fits, error analysis was derived from Hocking[36] as follows:

Our input variables are assumed to have no error, as they are set deliberately to a certain value (e.g. epitaxial strain, or system volume). These inputs correspond to the *x*-vector in Hocking's treatment, column vector [$x_1$; $x_2$; ...; $x_N$]. Our output variables, like migration barrier and pressure, correspond to the *y*-vector.

For a fit where the output axis intercept will be one of the fitting coefficients (e.g. there will be a +constant term at the end), we use X-matrices for linear and quadratic fits, as in Equation S11.1:

$$X = [\vec{J}\vec{x}] \text{ or } [\vec{J}\vec{x}\overrightarrow{x^2}] \quad (S11.1)$$

where vector **J** is a column vector of ones. In this notation $\bar{a}\bar{b}$ represents an N×2 matrix with the first column consisting of the N×1 vector $\bar{a}$ and the second column consisting of the N×1 vector $\bar{b}$.

For a fit where the output axis intercept (e.g. y-intercept) is set to zero, we use the following X-matrices for linear and quadratic fits, as in Equation S11.2:

$$X = [\vec{x}] \text{ or } [\vec{x}\overrightarrow{x^2}] \tag{S11.2}$$

The coefficient matrix $\beta$, with the lowest-order coefficient appearing as the top-most element, is found in Equation S11.3:

$$\hat{\beta} = (X^T X)^{-1} X^T \vec{y} \tag{S11.3}$$

The "hat" matrix $H$ is given in Equation S11.4:

$$H = X(X^T X)^{-1} X^T \tag{S11.4}$$

The estimated standard deviation, squared, is given in Equation (S11.5):

$$s^2 = \frac{\sigma^2 \chi^2 (N-p)}{N-p} \approx \frac{RSS}{N-p} = \frac{RSS}{rank(I-H)} \tag{S11.5}$$

where I is an appropriately-sized identity matrix, $N$ is the sample size (number of input, output pairs), and *RSS* is the residual sum of squares, defined below in Equation (S11.6) and Equation (S11.7) the residual $r_i$ is the difference between the observed and estimated value of the output. Note that when using Python's numpy package, the method numpy.linalg.matrix_rank should be used to calculate *rank(I-H)*, rather than using numpy.rank.

$$RSS = \sum_{i=1}^{N} r_i^2 \quad \text{(S11.6)}$$

$$r_i = y_i - \hat{y}_i \quad \text{(S11.7)}$$

The variance and therefore standard error in the coefficients themselves is given in vector form in Equation (S11.8):

$$var[\hat{\beta}_i] = (std\ errors^2) = diag((X^T X)^{-1}) * s^2 \quad \text{(S11.8)}$$

Explicitly, for the DFT-fitted slopes, the standard error therefore works out to Equation S11.9, where the standard error in the slope, $s_{\beta 1}$, is given as a function of the residual sum of squares $\Sigma r_i^2$ and the sum of input squares, where $x_i$ are the input percent strains (e.g. "-2"):

$$s_{\widehat{\beta_1}} = \sqrt{\frac{\frac{1}{N-2}\sum_{i=1}^{N} r_i^2}{\sum_{i=1}^{N}(x_i - \bar{x})^2}} \quad \text{(S11.9)}$$

## S12. Elastic Strain Model

We make extensive comparison to the simple elastic model proposed by Schichtel et al.,[35] which relates strain and pressure as in Equation S12.1:

$$p = -\frac{2}{3}\left(\frac{Y}{1-\nu}\right)\epsilon_{12} \quad \text{(S12.1)}$$

Here $\epsilon_{12}$ is biaxial strain, $Y$ is the Young's modulus, $v$ is the Poisson's ratio, and $p$ is the resulting pressure. At zero strain, the pressure is zero, so $p$ can also be given as the change in pressure due to strain, where $\Delta p = p - 0$. Therefore, at fixed temperature and for an unchanging number of particles, a change in migration Gibbs free energy due to pressure goes as Equation S12.2 and Equation S12.3, assuming that migration volume $V_{mig}$ changes little at different strains (an assumption which is supported by our data in Table S12.1):

| | |
|---|---|
| $\Delta G_{mig} = V_{mig} \Delta p = -\frac{2}{3}\left(\frac{Y}{1-v}\right)\epsilon_{12} V_{mig}$ | (S12.2) |
| $\frac{\Delta G_{mig}}{\epsilon_{12}} = -\frac{2}{3}\left(\frac{Y}{1-v}\right) V_{mig}$ | (S12.3) |

Migration volume $V_{mig}$ increases by 1 Å$^3$ between -2% and +2% biaxial strain (see Table S12.1). Therefore, the transition from ESI Equation S12.2 to ESI Equation S12.3 is not as well defended as it previously was. However, Figure 1R and Figure S8.1R show fairly linear decreases of migration barrier with respect to strain for most systems. A linear decrease corresponds to a constant DMEPS. Using a single strain-independent $V_{mig}$ value produces a constant DMEPS (ESI Equation S12.3 and S12.4). Therefore, we continue with the assumption of a strain-independent $V_{mig}$ as calculated from the no-strain case for the strain range between -2% and +2% biaxial strain.

Rearranging Equation S12.2 gives $\Delta G_{mig}/\Delta p = V_{mig}$, for a fixed-pressure, constant temperature system for a given strain case. As in Section S9, our calculated $E_{mig}$, which are done at fixed volume at each strain case, can be thought of as equivalent to $G_{mig}$, giving Equation S12.4 for the predicted results from the Schichtel model for our calculated slopes.

$$\frac{\Delta E_{mig}}{\epsilon_{12}} = -\frac{2}{3}\left(\frac{Y}{1-\nu}\right)V_{mig} \tag{S12.4}$$

Table S12.1 shows, first, that the perfect strained cell pressures are similar to those expected from Equation S12.1 and, second, that were we to take a fixed strained pressure and a fit-calculated migration volume (Section S12d), we would arrive at a $\Delta G_{mig}$ value similar to that predicted by Equation S12.2 and also similar to our directly-calculated $\Delta E_{mig}$ values.

The comparison in the text shows that there are a number of quantitative discrepancies between predictions from this strain model and the migration energy slopes calculated directly with ab initio methods.

### S12a. Finding ν for Use in the Elastic Model

For a linearly elastic isotropic material, the strain energy density $U$ is given in Equation S12a.1, where the principal axes are denoted by subscripts 1, 2, and 3.[37]

$$U = \frac{1}{2}\lambda(\epsilon_1 + \epsilon_2 + \epsilon_3)^2 + G(\epsilon_1^2 + \epsilon_2^2 + \epsilon_3^2) \tag{S12a.1}$$

The elastic constants $\lambda$ and $G$ can be written in terms of the Poisson's ratio $\nu$ and Young's modulus $Y$ of the material (given as $E$ in the reference text), in Equation S12a.2 and Equation S12a.3, which may be substituted into Equation S12a.1 to give Equation S12a.4.

$$\lambda = \frac{\nu Y}{(1+\nu)(1-2\nu)} \tag{S12a.2}$$

$$G = \frac{Y}{2(1+\nu)} \tag{S12a.3}$$

$$U = \frac{\nu Y}{2(1+\nu)(1-2\nu)}(\epsilon_1 + \epsilon_2 + \epsilon_3)^2 + \frac{Y}{2(1+\nu)}(\epsilon_1^2 + \epsilon_2^2 + \epsilon_3^2) \quad \text{(S12a.4)}$$

Using $\epsilon_1 = \epsilon_2$ for biaxial strain, $\epsilon_3$ is obtained from the lattice vector $c$ fitting, and is not equal to zero, as the cases discussed here are for the thin-film plane stress case, rather than for a plane strain case.

The strain energy density $U$ is the change in energy due to strain, per unit volume when strained, and was defined as in Equation S12a.5:

$$U = \frac{E_{strained}^{bulk\ supercell} - E_{unstrained}^{bulk\ supercell}}{V_{strained}^{bulk\ supercell}} \quad \text{(S12a.5)}$$

In principle, it is possible to perform a nonlinear fit using the strain and strain energy data in order to find $\lambda$ and $G$ or $Y$ and $\nu$. However, in practice, such fitting produced unreasonable negative Poisson's ratios and was in general badly determined, with large ranges of constant pairs that had very similar root-mean-squared errors.

Instead, we use the plane-stress approximation to calculate a Poisson's ratio $\nu$ from our data, then use the Poisson's ratio and the bulk modulus to calculate the Young's modulus $Y$. From Barber, we obtain Equation S12a.6, Equation S12a.7, and Equation S12a.8 for the plane-stress case ($\sigma_{zz}=0$), where $E$ is the Young's modulus.[38] Given that $\epsilon_{xx}$ and $\epsilon_{yy}$ are identical for our cells, and taking the approximation that $\sigma_{xx}$ and $\sigma_{yy}$ for our near-cubic cells are also equal, we derive Equation S12a.9 through Equation S12a.15.

$$\epsilon_{zz} = \frac{-\nu}{E}(\sigma_{xx} + \sigma_{yy}) \quad \text{(S12a.6)}$$

| | |
|---|---|
| $$\epsilon_{xx} = \frac{\sigma_{xx}}{E} - \frac{\nu\sigma_{yy}}{E}$$ | (S12a.7) |
| $$\epsilon_{yy} = \frac{\sigma_{yy}}{E} - \frac{\nu\sigma_{xx}}{E}$$ | (S12a.8) |
| $$\epsilon_{xx} = \epsilon_{yy} = \frac{\sigma_{xx}}{E} - \frac{\nu\sigma_{yy}}{E} = \frac{\sigma_{xx}}{E} - \frac{\nu\sigma_{xx}}{E}$$ | (S12a.9) |
| $$\epsilon_{xx} = \frac{(1-\nu)\sigma_{xx}}{E}$$ | (S12a.10) |
| $$\sigma_{xx} = \frac{E\epsilon_{xx}}{(1-\nu)}$$ | (S12a.11) |
| $$\epsilon_{zz} = \frac{-2\nu\sigma_{xx}}{E}$$ | (S12a.12) |
| $$\epsilon_{zz} = \frac{-2\nu E\epsilon_{xx}}{E(1-\nu)}$$ | (S12a.13) |
| $$\frac{\epsilon_{zz}}{\epsilon_{xx}} = \frac{-2\nu}{(1-\nu)}$$ | (S12a.14) |
| $$\nu = \frac{\frac{\epsilon_{zz}}{\epsilon_{xx}}}{\left(\frac{\epsilon_{zz}}{\epsilon_{xx}} - 2\right)} = \frac{\frac{\epsilon_c}{\epsilon_a}}{\left(\frac{\epsilon_c}{\epsilon_a} - 2\right)}$$ | (S12a.15) |

Using the definition of engineering strain as ΔL/L,[39] and noting that our zero-strain fractions are 1 for lattice vectors *a* and *b*, but usually slightly over 1 for lattice vector *c* due to fitting (explained in Section S10, Straining supercells, e.g. 1.003), we take strains from Equation S12a.16 through Equation S12a.18, where "strain fraction" is the strain fraction multiplier, e.g. 1, 1.02, 0.99, etc.)

| | |
|---|---|
| $$\epsilon_a = \frac{(strain\ fraction\ a - 1) * a}{a} = (strain\ fraction\ a) - 1$$ | (S12a.16) |
| $$\epsilon_c = \frac{(strain\ fraction\ c - zero\ strain\ fraction\ c) * c}{c}$$ | (S12a.17) |
| $$\epsilon_c = strain\ fraction\ c - zero\ strain\ fraction\ c$$ | (S12a.18) |

These normalized $\epsilon_c$ values were fit against the $\epsilon_a$ values to produce a quadratic fit, Equation S12a.19. Because zero strain in *a* should produce zero strain in *c* by definition (and from the normalization in Equation S12a.18, the intercept *g* in Equation S12a.19 is set as zero. Substituting Equation S12a.19 into our Equation S12a.15 for Poisson's ratio produces Equation S12a.20.

| | |
|---|---|
| $$\epsilon_c = d\epsilon_a^2 + f\epsilon_a + g = d\epsilon_a^2 + f\epsilon_a$$ | (S12a.19) |
| $$v = \frac{\frac{\epsilon_c}{\epsilon_a}}{\left(\frac{\epsilon_c}{\epsilon_a} - 2\right)} = \frac{\epsilon_c}{\epsilon_c - 2\epsilon_a} = \frac{d\epsilon_a^2 + f\epsilon_a}{d\epsilon_a^2 + f\epsilon_a - 2\epsilon_a} = \frac{d\epsilon_a + f}{d\epsilon_a + (f - 2)}$$ | (S12a.20) |

To take *v* as a material constant defined at small strains, we take *v* in the limit as $\epsilon_a$ goes to zero and arrive at Equation S12a.21 (remember that here, *f* is just the first-order coefficient of the fit of $\epsilon_c$ as a function of $\epsilon_a$), with error defined in Equation S12a.22 and Equation S12a.23.

| | |
|---|---|
| $$v = \frac{f}{f - 2}$$ | (S12a.21) |
| $$\left(\frac{\sigma_v}{v}\right) = \sqrt{\left(\frac{\sigma_f}{f}\right)^2 + \left(\frac{\sigma_f}{f}\right)^2}$$ | (S12a.22) |
| $$\sigma_v = v\left(\frac{\sigma_f}{f}\right)\sqrt{2}$$ | (S12a.23) |

$\sigma_f$ is the first element of the matrix in Equation S12a.24 and the X matrix is defined in Equation S12a.25. (There is no *J*-vector of ones, since the intercept is set to (0,0).)

| | |
|---|---|
| $$\left[ diag((X^T X)^{-1}) * \frac{RSS \text{ of } \epsilon_c(\epsilon_a)}{rank(I - H \text{ of } \epsilon_c(\epsilon_a))} \right]$$ | (S12a.24) |
| $$X = \left[ \vec{\epsilon_a} \; \vec{\epsilon_a^2} \right]$$ | (S12a.25) |

The calculated Poisson's ratio values are given in Table S12.2 using data from -2% to 2% strain, inclusive. It is possible that the introduction of an oxygen vacancy may change the Poisson's ratio for the 2x2x2 simulation supercells. However, this effect has not been included in the present calculations. We use the Poisson's ratio calculated from perfect cells as a materials constant and do not include effects of the vacancy.

### S12b. Finding the Bulk Modulus (for *Y* and $V_{mig}$)

Finding the bulk modulus was necessary for calculating the Young's modulus *Y* and the migration volume $V_{mig}$. In order to calculate the bulk modulus $B_0$ and the derivative of the bulk modulus, $B_0$', for each system, nine pressure-volume pairs of the perfect bulk were calculated for each system. Each pressure-volume point was a static calculation at a different volume, where the volume was equally strained along each lattice vector *a*, *b*, and *c*, starting at 5% compressive strain and increasing in increments of 1% to 3% tensile strain. In the following fitting we found that the direct fit to a Birch-Murnaghan equation was numerically unstable and gave results very sensitive to the initial values chosen in the optimization. Therefore, we have first fit to a cubic equation and then used the results of that fit to obtain the parameters for the Birch-Murnaghan equation.

The nine points were easily fit to a well-matching cubic equation for each system, producing *V(P)* (see Figure S12.1). The coefficients of the cubic fit were then used to derive the bulk modulus, $B_0$, and its derivative, $B_0'$, as shown in Equation S12b.1 through Equation S12b.14:

| | |
|---|---|
| $$B_0 = -V \left(\frac{\partial P}{\partial V}\right)_{P=0}$$ | (S12b.1) |
| $$B = -V \left(\frac{\partial P}{\partial V}\right)$$ | (S12b.2) |
| $$B = \frac{-V}{\left(\frac{\partial V}{\partial P}\right)}$$ | (S12b.3) |
| $$V(P) = jP^3 + kP^2 + lP + m, \{j, k, l, m\} \in \mathbb{R}$$ | (S12b.4) |
| $$\left(\frac{\partial V}{\partial P}\right) = 3jP^2 + 2kP + l$$ | (S12b.5) |
| $$B = \frac{-V}{\left(\frac{\partial V}{\partial P}\right)} = \frac{-(jP^3 + kP^2 + lP + m)}{3jP^2 + 2kP + l}$$ | (S12b.6) |
| $$B_0 = \frac{-V}{\left(\frac{\partial V}{\partial P}\right)_{P=0}} = \frac{-(0+0+0+m)}{0+0+l} = \frac{-m}{l}$$ | (S12b.7) |
| $$B_0 = \frac{-m}{l}$$ | (S12b.8) |
| $$\sigma_{B_0} = B_0 \sqrt{\left(\frac{\sigma_m}{m}\right)^2 + \left(\frac{\sigma_l}{l}\right)^2}$$ | (S12b.9) |
| $$B_0' = \left(\frac{\partial B}{\partial P}\right)_{P=0} = \frac{\left(\frac{-\partial V}{\partial P}\right)\left(\frac{\partial V}{\partial P}\right) - \left(\frac{\partial^2 V}{\partial P^2}\right)(-V)}{\left(\frac{\partial V}{\partial P}\right)\left(\frac{\partial V}{\partial P}\right)} = -1 + \frac{V\left(\frac{\partial^2 V}{\partial P^2}\right)}{\left(\frac{\partial V}{\partial P}\right)\left(\frac{\partial V}{\partial P}\right)}\bigg|_{P=0}$$ | (S12b.10) |

| | |
|---|---|
| $$\left(\frac{\partial^2 V}{\partial P^2}\right) = 6jP + 2k$$ | (S12b.11) |
| $$B_0' = -1 + \frac{(jP^3 + kP^2 + lP + m)(6jP + 2k)}{(3jP^2 + 2kP + l)(3jP^2 + 2kP + l)}\bigg|_{P=0}$$ | (S12b.12) |
| $$B_0' = -1 + \frac{(0 + 0 + 0 + m)(0 + 2k)}{(0 + 0 + l)(0 + 0 + l)}$$ | (S12b.13) |
| $$B_0' = -1 + \frac{2mk}{l^2}$$ | (S12b.14) |

(Remember that *k, l,* and *m* are just real-valued coefficients.)

The error in the coefficients *j, k, l,* and *m* can be approximated as the fourth, third, second, and first elements, respectively, of the matrix in Equation S12b.15, with the X-matrix given in Equation (S12b.16). The calculated bulk moduli and their errors are given in Table S12.3.

| | |
|---|---|
| $$\left[diag((X^T X)^{-1}) * \frac{RSS\ of\ V(P)}{rank(I - H\ of\ V(P))}\right]$$ | (S12b.15) |
| $$X = \begin{bmatrix}\vec{J} & \vec{P} & \vec{P^2} & \vec{P^3}\end{bmatrix}$$ | (S12b.16) |

Although we calculated the bulk modulus for the undefected system, it is possible that it is altered by the specific state of the system during the calculation of the migration energies. As we wish to compare the strain model to the migration energy calculations such a change in bulk modulus could produce errors. In particular, vacancy concentration (presence or absence of the single vacancy), placement (initial or activated state), and compensation status (compensated using electron removal or uncompensated) can all have an effect on bulk modulus, as given in Table S12.4. The strain data is for compensated systems, so using bulk modulus in the presence of the compensated vacancy, the two check cases (B=Cr and B=Mn) indicate an increase in the bulk modulus and therefore in the Young's modulus *Y* (using the Poisson's ratio from Section S12a), prefactor, and magnitude of the calculated elastic model slope. The full elastic model could be redone with all bulk modulus calculations at the same vacancy concentration, placement, and compensation status as the migration barrier calculations, rather than using the undefected-cell bulk modulus. However, Figure S12.2 shows that including the effect of the vacancy on the bulk modulus, using the largest downward shift to the elastic model slope, shifts the elastic model slope to be steeper, which brings better agreement to the DFT slopes for some points but worse agreement for other points. Looking at Figure S12.2 and Table S12.4, if the in-plane and out-of-plane points for B-site cation Chromium could be made to shift in different directions using different bulk moduli, then including vacancy effects for the bulk moduli would be clearly indicated. However, the hops share the same endpoint, and the endpoint bulk modulus shifts both points toward steeper model slopes. The compensated NEB bulk moduli also shift both points toward steeper model slopes. Therefore, the utility of including vacancy effects on the bulk modulus is unclear, especially given the larger error in calculating the bulk modulus with vacancy effects, due to the increased standard error in the fit coefficients.

The main conclusion remains that the elastic model provides a good qualitative, if not quantitative, description of the change in migration barrier versus strain.

## S12c. Finding Y for Use in the Elastic Strain Model

Using our bulk modulus values, we calculate the Young's modulus $Y$ from Equation S12c.1.[37] The error is given in Equation S12c.2.

| | |
|---|---|
| $Y = 3B_0(1 - 2\nu) = 3B_0 - 6B_0\nu$ | (S12c.1) |
| $\sigma_Y^2 = 3^2 \sigma_{B_0}^2 + 6^2 (B_0\nu)^2 \left[\left(\frac{\sigma_{B_0}}{B_0}\right)^2 + \left(\frac{\sigma_\nu}{\nu}\right)^2\right]$ | (S12c.2) |

Therefore, our prefactor for the elastic model is defined in Equation S12c.3, with error treatment in Equation S12c.4:

| | |
|---|---|
| $prefactor = \dfrac{-2Y}{3(1 - \nu)}$ | (S12c.3) |

$$\text{error in prefactor} = \text{prefactor} * \sqrt{\left(\frac{\sigma_Y}{Y}\right)^2 + \left(\frac{\sigma_\nu}{\nu}\right)^2} \quad \text{(S12c.4)}$$

The calculated values for Young's modulus and the prefactor are given in Table S12.2.

### S12d. Finding Migration Molume for Use in the Elastic Strain Model:

The migration volume is defined as the change in volume of the system from the initial state to the activated state. Calculation of the volume of the fully relaxed activated state through direction optimization is not possible without some method to constrain the reaction coordinate degrees of freedom, as the system is unstable in the activated state and will relax to its initial or final state. We evaluated constrained relaxations, where only volume is allowed to relax, but felt that this approach may be inaccurate due to the many internal degrees of freedom that are constrained (this data is shown in Table S12.5). The CNEB method naturally constrains just the reaction coordinate in the activated state but is implemented at fixed volume, and therefore does not allow relaxation of the activated state volume during the CNEB calculation. To overcome this problem, for each system and each hop, at the initial state and at the activated state, we used the bulk modulus $B_0$, its derivative with respect to pressure, $B_0'$, the fixed volume $V$ (common to the shared initial state and both activated states), and the pressure $P$ at that fixed volume in the third-order Birch Murnaghan equation, Equation S12d.1, in order to calculate the expected zero-pressure volume $V_0$. The Birch-Murnaghan equation is reproduced below as Equation S12d.1:

$$P(V) = \frac{3B_0}{2}\left[\left(\frac{V_0}{V}\right)^{\frac{7}{3}} - \left(\frac{V_0}{V}\right)^{\frac{5}{3}}\right]\left\{1 + \frac{3}{4}(B_0' - 4)\left[\left(\frac{V_0}{V}\right)^{\frac{2}{3}} - 1\right]\right\} \quad \text{(S12d.1)}$$

We solved for $V_0$ by evaluating each prospective $V_0$ volume in the range of 300 to 600 cubic Angstroms (for our 2x2x2 supercell size) in steps of 0.1 cubic Angstroms, paired with the known V, and took the closest match to the observed P for each case. Then we calculate the migration volume as in Equation S12d.2, with error treatment in Equation S12d.3:

| | |
|---|---|
| $dV_{mig} = V_{0,activated} - V_{0,initial}$ | (S12d.2) |
| $\sigma_{dV_{mig}} = \sqrt{\sigma_{V_{activated}}{}^2 + \sigma_{V_{initial}}{}^2} = \sqrt{0.1^2 + 0.1^2} = 0.1\sqrt{2}$ | (S12d.3) |

Note that the values of $\sigma_{Vactivated}$ and $\sigma_{Vinitial}$ are taken as 0.1 Å$^3$, as this is the step size used in obtaining the values. The obtained $dV_{mig}$ values are listed in Table S12.5.

As a different way to evaluate the elastic strain model than comparing slopes, we can supply the DFT migration barrier slopes on the left-hand side of Equation S12.4 and calculate anticipated volumes, then compare those volumes against either our Birch-Murnaghan equation migration volumes, or our volume-only relaxation migration volumes. Figure S12.3 and Figure S12.4 show elastic-model anticipated volumes, using the Young's modulus, Poisson's ratio, and DFT-fit slopes in the elastic model Equation S12.4 in order to calculate a volume, compared to the Birch-Murnaghan calculated volumes and the volume-only relaxation calculated volumes, respectively. The root-mean-squared error for predicting migration volumes is 2 Å$^3$ for either migration volume method, but an inspection of Figure S12.3 and Figure S12.4 shows somewhat better qualitative agreement for Figure S12.3 with Birch-Murnaghan migration volumes, excluding outliers.

Figure S12.5 reproduces the elastic model analysis in the main text, but using volume-only relaxations for $V_{mig}$ rather than Birch-Murnaghan migration volumes. Figure S12.5 shows somewhat more scatter in the outlier points than Figure 5. For Figure S12.5, most points with DFT slopes between -50 and -80 meV/% strain seem slightly under-predicted by the elastic model and may benefit from some correction in bulk modulus due to vacancy effects, similar to those in Figure S12.2.

## R1. Full update list

This is the full update list corresponding to the Update Note in the main text. This list is provided to easily identify changes between the published version and the updated version. The changes have already been incorporated into the main text and into this supporting information document. Tables and figures listed below are replaced with their original names and no longer designated with R.

### R1.1. Revised Results for the Main Text

Below is a list of revisions for the main text. "S" indicates a table or figure in the ESI.
- Table 1R to replace Table 1 due to a typographical error in DMEPS for B-site cation Vanadium and consistent use of lowest migration barrier from initial to final endpoint when multiple migration barriers exist for a system at a strain (affects B-site cations Vanadium and Iron).
- Figure 1R to replace Figure 1 due to corrections in Table 1R.
- Figure 5R to replace Figure 5 due to corrected Table S12.5R.

### R1.2. Revised Discussion for the Main Text

Below is a list of revised discussion and observation points in order of appearance.

Abstract:
- The steepest DMEPS is changed to -86 meV/% strain and the average DMEPS is changed to -65 meV/% strain.

p. 2719:
- The DMEPS predicted by the strain model qualitatively follow the same trends as our *ab initio* DMEPS. *The DMEPS from the strain model are no longer uniformly smaller than the ab initio DMEPS.*
- The strain model DMEPS differ from the DFT-calculated DMEPS by an average of 16 +/- 13 meV/% strain (where the uncertainty represents one standard deviation of the error from the mean) with a maximum error of 40 meV/% strain and a root-mean-squared error of 20 meV/% strain.
- Effective migration volumes predicted by the strain model are no longer all larger than those calculated directly using the *ab initio* methods. The root-mean-squared error for predicting migration volumes is 2 Å$^3$.

Additional error analysis, Figure E1, and Table E1 added.

### R1.3. Revised Results for the ESI

Below is a list of revisions for the ESI.
- Table S6.3R to replace Table S6.3 due to:
    - A change in the column heading $E_{mig}$ into $H_{mig}$. Activation energy is commonly given in the literature as $E_a$ to describe a combination of enthalpies; $E_a$ as calculated from an Arrhenius plot for the vacancy diffusion coefficient $D_v$ or for the ionic conductivity of Sr-doped LaM$^{III}$O$_{3-\delta}$ is equivalent to migration barrier enthalpy $H_{mig}$.[21, 22] ESI Section S9 discusses the applicability of comparing our $E_{mig}(V)$ with $H_{mig}(P)$.
    - Additional explanation of sources used
    - Using data obtained directly from the referenced sources in Table 5 of Lybye et al. rather than the referenced values, and including more values
- Table S8.1R to replace Table S8.1 due to:
    - Correction of a spreadsheet error for the -0.75% strain point for out-of-plane B-site cation Vanadium
    - Consistent use of lowest migration barrier from initial to final endpoint when multiple migration barriers exist for a system at a strain (affects B-site cations Vanadium, Iron, and Nickel).
    - Minor difference in the error value for B-site cation Titanium due to rounding
- Table S12.1R to replace Table S12.1 due to corrections in migration volumes from the Birch-Murnaghan equation
    - Minor rounding corrections are also present in the table
- Table S12.2R to replace Table S12.2 due to:
    - Correction to B-site cation Vanadium Poisson's ratio (and subsequently the rest of the table columns) where strains over +2% are no longer used, for consistency with other systems.
    - Minor differences due to rounding for B-site cations Iron and Nickel
    - Also, the prefactor value is left negative to correspond to its definition in ESI Equation S12c.3.
- Table S12.4R to replace Table S12.4 due to the inclusion of a new data row for B-site cation Chromium. Previously, the hop direction had not been specified and was for the out-of-plane hop. We show that hop direction does not make a significant difference in bulk modulus or prefactor.
- Table S12.5R to replace Table S12.5 due to:
    - Corrections in migration volumes from the Birch-Murnaghan equation
    - Correction to B-site cation Vanadium out-of-plane Birch-Murnaghan migration volume, where the saddle image pressure used is now that of the zero-strain NEB calculation used in Table 8.1R and Figure S8.1R, instead of from a skipped calculation that was one of several done for zero strain.
    - Correction to B-site cation Vanadium out-of-plane volume-only relaxation, where the zero-strain NEB calculation used in Table 8.1R and Figure S8.1R is used, instead of from a skipped calculation that was one of several done for zero strain.

- - Minor correction in the value for B-site cation Scandium in-plane volume-only relaxation due to Excel rounding down instead of rounding up on 3.050
- Figure S6.1R to replace Figure S6.1 due to corrected Table S6.3R.
- Figure S8.1R to replace Figure S8.1 due to corrected Table S8.1R.
- Figure S8.2R to replace Figure S8.2 due to corrected Table S8.1R.
- Figure S8.9R to replace Figure S8.9 due to corrected Table S8.1R and corrected Table S12.5R.
- Figure S12.2R to replace Figure S12.2 due to:
  - Corrected Table S12.5R and corrections in migration volumes not shown explicitly in any table
  - Correction in caption to reference the correct figure
  - Also, points for other systems that were present in Figure S12.2 are omitted in Figure S12.2R for clarity.
- Figure S12.3R to replace Figure S12.3 due to corrected Table S8.1R and corrected Table S12.5R.
- Figure S12.4R to replace Figure S12.4 due to corrected Table S8.1R and corrected Table S12.5R.
- Figure S12.5R to replace Figure S12.5 due to corrected Table S8.1R and corrected Table S12.5R.

## R1.4. Revised Discussion for the ESI

Below is a list of revised discussion and observation points in order of appearance in the ESI.

ESI, p. 12:
- The DMEPS predicted by the strain model qualitatively follow the same trends as the *ab initio* DMEPS for all hops in $LaCrO_3$ and $LaMnO_3$, when considering the entire cluster of points in each system.
- The spread of differences between the elastic model slopes and the DFT-fit slopes over the entire cluster of points for each system indicates that the limits of reasonable accuracy for using a single hop in a system to predict DMEPS with the elastic strain model are some 30 meV/% strain.
- One additional observation is that the same in-plane hops that have the largest difference between DFT DMEPS and elastic model DMEPS in $LaMnO_3$ actually have some of the smallest differences in $LaCrO_3$.
- The elastic strain model no longer predicts very similar migration volumes for all hops.

ESI, p. 14: Figure 4 is referenced but should have been Figure 5, and is now Figure 5R.

ESI, p. 30: Migration volume $V_{mig}$ increases by 1 Å$^3$ between -2% and +2% biaxial strain (see Table S12.1). Therefore, the transition from ESI Equation S12.2 to ESI Equation S12.3 is not as well defended as it previously was. However, Figure 1R and Figure S8.1R show fairly linear decreases of migration barrier with respect to strain for most systems. A linear decrease corresponds to a constant DMEPS. Using a single strain-independent $V_{mig}$ value produces a constant DMEPS (ESI Equation S12.3 and S12.4). Therefore, we continue with the assumption of a strain-independent $V_{mig}$ as calculated from the no-strain case for the strain range between -2% and +2% biaxial strain.

ESI, p. 31:
- Note improved agreement between the energy change $PV_{mig}$, which is described by fixed strained pressure multiplied by fit-calculated migration volume, and the directly calculated $\Delta E_{mig}$.
- Note improved agreement between the energy change $\Delta G_{mig}$ expected from the Schichtel model and the directly calculated $\Delta E_{mig}$.

ESI, p. 37: Including the effect of the vacancy on the bulk modulus shifts the elastic model slope to be steeper, which brings better agreement for some points but worse agreement for other points. Looking at Figure S12.2R and Table S12.4R, if the in-plane and out-of-plane points for B-site cation Chromium could be made to shift in different directions using different bulk moduli, then including vacancy effects for the bulk moduli would be clearly indicated. However, the hops share the same endpoint, and the endpoint bulk modulus shifts both points toward steeper model slopes. The compensated NEB bulk moduli also shift both points toward steeper model slopes. Therefore, the utility of including vacancy effects on the bulk modulus is unclear, especially given the larger error in calculating the bulk modulus with vacancy effects, which is produced by the standard error in the fit coefficients themselves.

ESI, p. 40:
- The points on Figure S12.4R are no longer clustered closer to the guideline than those in Figure S12.3R. The root-mean-squared error for predicting migration volumes is 2 Å$^3$, as with using Birch-Murnaghan migration volumes, but an inspection of Figure S12.3R and Figure S12.4R shows somewhat better qualitative agreement for Figure S12.3R with Birch-Murnaghan migration volumes, excluding outliers.
- The points on Figure S12.5R are no longer clustered closer to the guideline than those in Figure 5R.
- For Figure S12.5R, most points with DFT slopes between -50 and -80 meV/% strain seem slightly under-predicted by the elastic model and may benefit from some correction in bulk modulus due to vacancy effects, similar to those in Figure S12.2R.

# Tables

## Table S3.1. Supercell energy comparison for orthogonal versus non-orthogonal lattice vector c.

Table S3.1. Supercell energy comparison for orthogonal versus non-orthogonal assumption for lattice vector $c$.

| | | Fitting equation coefficients for supercell energy (eV) | | | | Energy difference (eV), using original $c$ fraction * sin(angle) | |
|---|---|---|---|---|---|---|---|
| % strain, $a$ and $b$ | Fitted $c$ fraction, greatest magnitude | $(c\ \text{fraction})^3$ | $(c\ \text{fraction})^2$ | $c$ fraction | constant | 91.4° | 89.8° |
| -2 | 1.0268 (B=Mn) | 706.11 | -1869.1 | 1605.0 | -778.78 | 3E-05 | 1E-07 |
| 2 | 0.9838 (B=Fe) | -2231.90 | 6980.0 | -7253.2 | 2182.30 | -3E-06 | -7E-07 |

## Table S5.1. U-values for GGA+U

Table S5.1. U-values for GGA versus GGA+U barriers.[13, 40]

| B-site cation | U-J value for B-site, J=1 eV |
|---|---|
| Ti | 4.0 |
| V | 3.1 |
| Cr | 3.5 |
| Mn | 4.0 |
| Fe | 4.0 |
| Co | 3.3 |
| Ni | 6.4 |

## Table S6.1. Magnetic moment per B-site cation from FM configuration

Table S6.1. Magnetic moment per B-site cation in LaBO$_3$ perovskites, when relaxed from a high-spin ferromagnetic starting configuration.[a]

| B-site | Experimental (μB) | Bulk (μB) | Uncompensated | | Compensated | |
| --- | --- | --- | --- | --- | --- | --- |
| | | | Endpoint (μB) | Middle image (μB) | Endpoint (μB) | Middle image (μB) |
| Sc | 0 | 0.0 | 0.0 | 0.3 | 0.0 | 0.0 |
| Ti | 0.46[20] | 0.1 | 0.2 | 0.3 | 0.3 | 0.3 |
| V | 1.4[41] | 2.0 | 2.0 | 2.0 | 1.7 | 1.7 |
| Cr | 2.8±0.2[42] | 3.0 | 3.0 | 3.3 | 3.0 | 3.0 |
| Mn | 3.9±0.2[42] | 4.0 | 4.0 | 4.3 | 4.0 | 4.0 |
| Fe | 4.6±0.2[42] | 3.5 | 3.3 | 3.3 | 3.5 | 3.5 |
| Co | 2[b,43] | 1.6 | 1.4 | 1.3 | 1.7 | 1.8 |
| Ni | 1[c,44] | 0.3 | 0.1 | 0.3 | 0.4 | 0.2 |
| Ga | 0 | 0.0 | 0.0 | 0.0 | 0.0 | 0.0 |

[a] The full relaxation was followed by a static calculation. The calculated magnetic moment for the B-site cation in the bulk or endpoint is taken by dividing the total supercell magnetic moment by 8. This procedure re-attributes to the B-site cations the small moments which the VASP calculations sometimes put onto the La$^{3+}$ and O$^{2-}$ ions. The magnetic moment for the first endpoint was in all cases the same as the magnetic moment for the second endpoint, to the accuracy displayed here.

[b] According to Saitoh et al., LaCoO$_3$ is nonmagnetic at 0K to paramagnetic at 90K, with a purported transition from $t_{2g}^6$ to $t_{2g}^5 e_g^1$, which would give a moment of 2 μB per Co ion.

[c] According to Sreedhar et al., the configuration is $t_{2g}^6 e_g^1$.

## Table S6.2. Effect of AFM structure on LaXO3 calculations

Table S6.2. Effect of antiferromagnetic structure on LaXO3 calculations, using a 4x4x4M kpoint mesh.

| B-site | Exper. AFM | $E_{bulk, AFM} < E_{bulk, FM}$ [b] | $E_{mig, AFM} - E_{mig, FM}$ [a] (compensated) | $E_{mig, AFM} - E_{mig, FM}$ (uncompensated) | Néel Temp. (K)[45] |
|---|---|---|---|---|---|
| Sc | not magnetic[18] | N/A | N/A | N/A | (not listed) |
| Ti | G-type[46] | No | N/A | N/A | paramagnetic |
| V | C-type, cited[18] | -0.22 | -0.03 | 0.04 | 137 |
| V | G-type[47] | No | N/A | -0.11 | 137 |
| Cr | G-type[42] | -0.72 | -0.15 | 0.36 | 320 |
| Mn | A-type[42] | No | N/A | N/A | 100 |
| Mn | G-type (for consistency) | No | N/A | N/A | 100 |
| Fe | G-type[42] | -0.30 | -0.11 | -0.39 | 750 |
| Co | None (<90 K) to paramagnetic[43] | N/A | N/A | N/A | (not listed) |
| Ni | None (<15K) to paramagnetic[44] | N/A | N/A | N/A | paramagnetic |
| Ga | not magnetic[48] | N/A | N/A | N/A | (not listed) |

[a] Migration barriers were taken from 3-image CNEB calculations
[b] If the relaxed bulk energy calculated in VASP with an antiferromagnetic high-spin starting configuration was lower than the relaxed bulk energy calculated with a ferromagnetic starting configuration, then more AFM calculations were pursued.

## Table S6.3. Literature values for comparison with LaXO3 barriers.

Table S6.3. Literature values for Figure S6.1R. Activation energy as calculated from an Arrhenius plot for the vacancy diffusion coefficient $D_v$ or for the ionic conductivity $\sigma_i$ of Sr-doped $LaM^{III}O_{3-\delta}$ is equivalent to migration barrier enthalpy $H_{mig}$.[21, 22]

| Actual material | B-site cation | Temp. (°C) | $H_{mig}$ (eV) | Notes | Source |
|---|---|---|---|---|---|
| $La_{0.9}Sr_{0.1}Sc_{0.9}Mg_{0.1}O_{3-\delta}$ | Sc | 800 | 0.5 | Activation energy for $\sigma_i$ as assumed from $\sigma_{total}$ in reducing atmosphere (Lybye et al, p.98) [24] | [24] |
| $La_{0.9}Sr_{0.1}ScO_{3-\delta}$ | Sc | 730-980 | 0.71 | Activation energy for $\sigma_{total}$ in $N_2$, which is expected to be $\sigma_i$ (Nomura and Tanase, p. 234) [21] | [21] |
| $La_{0.9}Sr_{0.1}ScO_{3-\delta}$ | Sc | 330-480 | 0.47 | Activation energy for $\sigma_{total}$ in $N_2$, which is expected to be $\sigma_i$ (Nomura and Tanase, p. 234) [21] | [21] |
| $La_{0.7}Ca_{0.3}CrO_3$ | Cr | 900-1000 | 0.81 | Activation energy for $D_v$ | [49] |
| $La_{0.79}Sr_{0.20}MnO_{3-\delta}$ | Mn | 700-860 | 0.726 | Activation energy from chemical diffusion coefficient $\tilde{D}$ [a] | [50] |
| $La_{0.8}Sr_{0.2}MnO_3$ | Mn | 850-1000 | 1.47 | Activation energy for $D_v$ converted from $\tilde{D}$ (De Souza and Kilner, Fig. 8, line B, and Yasuda and Hishinuma, Fig. 10) [26, 51] | [26, 51] |
| $LaFeO_{3-\delta}$ | Fe | 900-1100 | 0.767 | Activation energy for $D_v$ | [22] |
| $La_{0.9}Sr_{0.1}FeO_{3-\delta}$ | Fe | 850-1100 | 0.819 | Activation energy for $D_v$ | [22] |
| $La_{0.75}Sr_{0.25}FeO_{3-\delta}$ | Fe | 900-1050 | 1.182 | Activation energy for $D_v$; authors note that large activation energy may come from inaccuracy in $D_O^*$ (Ishigaki et al. 1988, p. 184) [22] | [22] |
| $LaCoO_{3-\delta}$ | Co | 800-1000 | 0.798 | Activation energy for $D_v$ | [22] |
| $La_{0.9}Sr_{0.1}CoO_{3-\delta}$ | Co | 800-1000 | 0.819 | Activation energy for $D_v$ | [22] |
| $LaCoO_3$ | Co | 850-1000 | 0.781 | Activation energy for $D_v$ | [52] |
| $La_{0.9}Sr_{0.1}Ga_{0.9}Mg_{0.1}O_{3-\delta}$ | Ga | 200, 800 | 1.2 | Activation energy for $\sigma_i$ as assumed from $\sigma_{total}$ in reducing atmosphere (Lybye et al., p.98). [24] Temperature is given as 200°C on Lybye et al., p.98 and 800°C | [24] |

| | | | | in Lybye et al., Table 5. [24] | |
|---|---|---|---|---|---|
| La$_{0.9}$Sr$_{0.1}$Ga$_{0.9}$Mg$_{0.1}$O$_{3-\delta}$ | Ga | 1000 | 0.6 | Activation energy for $\sigma_{total}$, expected to be almost purely ionic (Lybye et al., p.99, p.101) [24] | [24] |
| La$_{0.9}$Sr$_{0.1}$GaO$_{3-\delta}$ | Ga | 730-980 | 0.6 | Activation energy for $\sigma_{total}$ in N$_2$, which is expected to be $\sigma_i$ (Nomura and Tanase, p. 234) [21] | [21] |
| La$_{0.9}$Sr$_{0.1}$GaO$_{3-\delta}$ | Ga | 430-580 | 0.81 | Activation energy for $\sigma_{total}$ in N$_2$, which is expected to be $\sigma_i$ (Nomura and Tanase, p. 234) [21] | [21] |

$^a$According to Equation 16 in Ishigaki et al., the chemical diffusion coefficient for vacancy mediated diffusion can be equated to the vacancy diffusion coefficient using $\widetilde{D} = -\frac{1}{2}\frac{1}{\left(\frac{\partial \ln C_v}{\partial \ln P(O_2)}\right)}D_v$.[22] For a similar system, the denominator of the second fraction approaches a constant (Yasuda and Hishinuma, Fig. 7)[51] at higher oxygen partial pressures. Therefore, for this case, $\widetilde{D} \propto D_v$, and on an Arrhenius plot, the activation energy of $\widetilde{D}$ would be equivalent to the activation energy of $D_v$.

## Table S7.1. Difference between electron-removal compensated migration barriers and doped migration barriers.

Table S7.1. Difference between electron-removal compensated migration barriers and doped migration barriers. This hop is from O29 to O30 (unstrained).

| B-site | Barrier for LaBO$_3$ compensated - barrier for LaBO$_3$ doped (eV) | With shift from mean of compensated minus doped barrier (eV) |
|---|---|---|
| Sc | -0.07 | 0.07 |
| Ti | -0.12 | 0.02 |
| V | -0.11 | 0.03 |
| Cr | -0.21 | -0.07 |
| Mn | -0.25 | -0.11 |
| Fe | -0.24 | -0.10 |
| Co | -0.10 | 0.04 |
| Ni | -0.07 | 0.07 |
| Ga | -0.08 | 0.06 |
| *Largest value* | *-0.25* | *-0.11* |
| *Mean value* | **-0.14** | *0.00* |
| *RMS value* | 0.16 | **0.07** |

## Table S7.2 In-plane and out-of-plane slopes for La$_{0.75}$Sr$_{0.25}$BO$_3$ supercells, compared with undoped slopes.

Table S7.2 In-plane and out-of-plane slopes for La$_{0.75}$Sr$_{0.25}$BO$_3$ supercells, compared with undoped slopes. In-plane hop is from oxygen position o31 to o30 (see Figure S2.1). Out-of-plane hop is from oxygen position o29 to o30. Cross-body diagonal dopant positions are a1 and a8. Cross-face diagonal dopant positions are a2 and a8. In-line dopant positions are a4 and a8.

|  | B-site cation | Dopant position: | Slope of migration barrier versus strain (meV/% strain) +/- fitting error | | | |
|---|---|---|---|---|---|---|
|  |  |  | Cross-body | Cross-face | In-line | No dopants (electron-removal compensated) |
| In-plane hop | Sc |  | -24 +/- 1 | -26 +/- 1 | -31 +/- 1 | -36 +/- 3 |
|  | Cr |  | -91 +/- 1 | -90 +/- 1 | -92 +/- 1 | -85 +/- 0 |
|  | Mn |  | -106 +/- 9 | -94 +/- 9 | -111 +/- 16 | -64 +/- 4 |
| Out-of-plane hop | Sc |  | -34 +/- 3 | -31 +/- 1 | -47 +/- 2 | -52 +/- 2 |
|  | Cr |  | -111 +/- 2 | -98 +/- 1 | -110 +/- 1 | -122 +/- 1 |
|  | Mn |  | -68 +/- 12 | -73 +/- 4 | -91 +/- 6 | -77 +/- 3 |

## Table S8.1. Out-of-plane slopes and slope error

Table S8.1. Out-of-plane slopes and slope error for Figure S8.1R.

| B-site cation | Out-of-plane slope fit to DFT (meV/% strain) | Out-of-plane slope error (meV/% strain) |
|---|---|---|
| Sc | -52 | 2 |
| Ti | -73 | 2 |
| V | -124 | 20 |
| Cr | -122 | 1 |
| Mn | -77 | 3 |
| Fe | -82 | 10 |
| Co | -80 | 7 |
| Ni | -21 | 10 |
| Ga | -64 | 0.4 |

## Table S8.2. Eight migration barriers in LaXO3 for several B-site cations.

Table S8.2. Eight migration barriers in LaXO$_3$ (compensation state indicated), moving the vacancy from atomic position 30 (an arbitrary choice for convenience) to the position indicated, and using a single image except where noted and a 4x4x4M kmesh. Barriers and range are measured in eV. The in-plane hop discussed in the main text and the out-of-plane hop discussed in the Supporting Information are marked in bold.

| Oxygen position | Sc uncompensated | Sc compensated | Ti uncomp. | V (3 images) uncomp. | Cr uncomp. | Mn (3 images) uncomp. | Fe uncomp. |
|---|---|---|---|---|---|---|---|
| In-plane hops | | | | | | | |
| 30 to 19 | 2.33 | 0.85 | 1.76 | 1.80 | 1.89 | 1.20 | 0.95 |
| 30 to 25 | 1.80 | 0.53 | 1.67 | 1.66 | 1.85 | 1.19 | 0.92 |
| **30 to 31** | **1.73** | **0.49** | **1.56** | **1.62** | **1.72** | **0.92** | **0.83** |
| 30 to 37 | 2.03 | 0.75 | 1.66 | 1.72 | 1.79 | 1.16 | 0.95 |
| Out-of-plane hops | | | | | | | |
| 30 to 17 | 1.97 | 0.53 | 1.61 | 1.60 | 1.86 | 1.19 | 0.86 |
| 30 to 20 | 1.96 | 0.46 | 1.61 | 1.64 | 1.79 | 0.98 | 0.83 |
| **30 to 29** | **1.96** | **0.46** | **1.61** | **1.64** | **1.72** | **0.98** | **0.81** |
| 30 to 32 | 1.97 | 0.53 | 1.61 | 1.60 | 1.84 | 1.09 | 0.86 |
| *Range (eV)* | *0.60* | *0.39* | *0.20* | *0.20* | *0.17* | *0.28* | *0.14* |

### Table S12.1. Comparing PV$_{mig}$ for constant pressure and E$_{mig}$ for constant volume

Table S12.1. Comparing $PV_{mig}$ for constant pressure and $E_{mig}$ for constant volume for the LaCrO$_3$ in-plane hop. The pressure given is that of the perfect strained cell. The migration volume was calculated using the Birch-Murnaghan equation procedure in ESI Section S12d *and the bulk modulus information from ESI Table S12.3*. The "Schichtel expected" pressure value is calculated from ESI Equation S12.1, given the LaCrO$_3$ prefactor from Table S12.2R. The $\Delta E_{mig}$ value is taken as the difference between the strained $E_{mig}$ and the zero-strain $E_{mig}$ (0.90 eV). The Schichtel expected $\Delta G_{mig}$ is calculated from ESI Equation S12.2, given both the LaCrO$_3$ prefactor from Table S12.2R and the migration volume in this table.

| Epitaxial strain | P (kbar) | $V_{mig}$ (Å³) | $PV_{mig}$ (eV) | Schichtel expected P (kbar) | $E_{mig}$ (eV) | $\Delta E_{mig}$ (eV) | Schichtel expected $\Delta G_{mig}$ (eV) |
|---|---|---|---|---|---|---|---|
| -0.02 | 43.89 | 6.4 | 0.175 | 46.11 | 1.065 | 0.162 | 0.184 |
| 0.02 | -45.08 | 7.4 | -0.208 | -46.11 | 0.726 | -0.177 | -0.213 |

## Table S12.2. Strained-bulk calculated Poisson's ratio, Young's modulus, and prefeactor, with errors.

Table S12.2. Strained-bulk calculated Poisson's ratio, Young's modulus, and prefactor, with errors.

| B-site cation | Poisson's ratio | Error in $\nu$ | $Y$ (eV/Å$^3$) | Error in $Y$ (eV/Å$^3$) | Prefactor (eV/Å$^3$) | Error in prefactor (eV/Å$^3$) |
|---|---|---|---|---|---|---|
| Sc | 0.290 | 0.001 | 1.23 | 0.02 | -1.16 | 0.02 |
| Ti | 0.292 | 0.003 | 1.37 | 0.03 | -1.29 | 0.03 |
| V  | 0.337 | 0.110 | 1.11 | 0.75 | -1.12 | 0.84 |
| Cr | 0.266 | 0.001 | 1.59 | 0.02 | -1.44 | 0.02 |
| Mn | 0.350 | 0.003 | 0.95 | 0.04 | -0.97 | 0.04 |
| Fe | 0.321 | 0.008 | 1.03 | 0.06 | -1.01 | 0.07 |
| Co | 0.311 | 0.022 | 1.17 | 0.15 | -1.13 | 0.17 |
| Ni | 0.386 | 0.015 | 0.76 | 0.21 | -0.82 | 0.23 |
| Ga | 0.321 | 0.000 | 1.14 | 0.02 | -1.12 | 0.02 |

## Table S12.3. Bulk modulus values and errors

Table S12.3. Bulk modulus values and errors

| B-site cation | $B_0$ (kbar) | error in $B_0$ (kbar) | $B_0'$ | $B_0$ (GPa) | $B_0$ (eV/Å$^3$) | error in $B_0$ (eV/Å$^3$) |
|---|---|---|---|---|---|---|
| Sc | 1572 | 8  | 4.12 | 157.2 | 0.981 | 0.005 |
| Ti | 1761 | 10 | 4.25 | 176.1 | 1.099 | 0.006 |
| V  | 1820 | 6  | 4.52 | 182.0 | 1.136 | 0.004 |
| Cr | 1805 | 9  | 4.09 | 180.5 | 1.127 | 0.006 |
| Mn | 1689 | 16 | 3.93 | 168.9 | 1.054 | 0.010 |
| Fe | 1539 | 17 | 3.37 | 153.9 | 0.961 | 0.010 |
| Co | 1652 | 31 | 2.46 | 165.2 | 1.031 | 0.020 |
| Ni | 1766 | 72 | 3.99 | 176.6 | 1.102 | 0.045 |
| Ga | 1692 | 10 | 4.28 | 169.2 | 1.056 | 0.006 |

## Table S12.4. Vacancy effects on bulk modulus and elastic model

Table S12.4. Vacancy effects on bulk modulus and elastic model. The bulk modulus was calculated using a series of static calculations at different volumes under the specified conditions. All values are in eV/Å$^3$.

| Condition | B=Mn | | | B=Cr | | |
|---|---|---|---|---|---|---|
| | $B_0$ | E | prefactor | $B_0$ | E | prefactor |
| Undefected bulk | 1.05 | 0.95 | -0.97 | 1.13 | 1.59 | -1.44 |
| Initial state, compensated vacancy | 1.22 | 1.10 | -1.13 | 1.31 | 1.84 | -1.67 |
| Transition state, compensated vacancy, out-of-plane hop | 1.14 | 1.03 | -1.05 | 1.21 | 1.71 | -1.55 |
| Transition state, compensated vacancy, in-plane hop | not calculated | not calculated | not calculated | 1.23 | 1.73 | -1.57 |
| Initial state, uncompensated vacancy | 1.08 | 0.97 | -1.00 | 1.10 | 1.55 | -1.41 |
| Transition state, uncompensated vacancy, out-of-plane hop | 1.02 | 0.92 | -0.95 | 1.07 | 1.50 | -1.36 |

## Table S12.5. Migration volumes, calculated with BM equation or allowing volume-only relaxation

Table S12.5. Migration volumes, calculated with the Birch-Murnaghan equation (first two numeric columns) and calculated by allowing a volume-only relaxation (last two columns)

| B-site cation | dVmig IP, Birch-Murn. | dVmig OOP, Birch-Murn. | dVmig IP, volume relaxation (not used) | dVmig OOP, volume relaxation (not used) |
|---|---|---|---|---|
| Sc | 3.5 | 4.4 | 3.1 | 3.9 |
| Ti | 5.2 | 6.1 | 3.9 | 4.7 |
| V | 10.6 | 8.8 | 12.2 | 8.6 |
| Cr | 7.1 | 7.6 | 5.8 | 6.3 |
| Mn | 5.7 | 6.3 | 4.7 | 5.3 |
| Fe | 11.7 | 12.0 | 8.3 | 8.2 |
| Co | 5.8 | 6.2 | 5.2 | 6.0 |
| Ni | 4.8 | 5.2 | 4.3 | 5.5 |
| Ga | 4.1 | 5.5 | 3.9 | 5.2 |

**Figures**

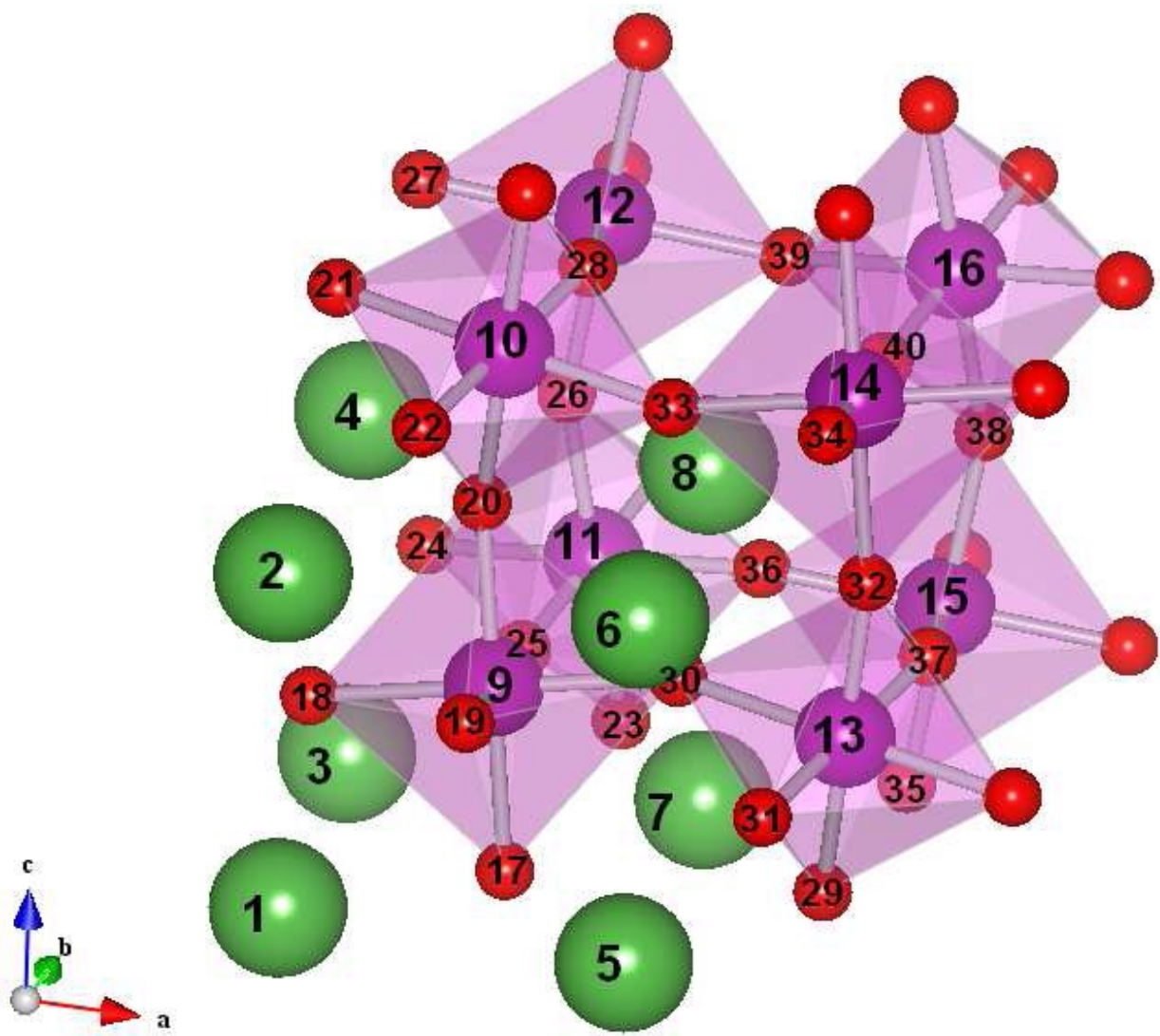

**Figure S2.1. Numbered atomic positions.**

Figure S2.1. Numbered atomic positions, taken from the relaxed LaMnO$_3$ bulk, pictured with atomic radii for clarity.

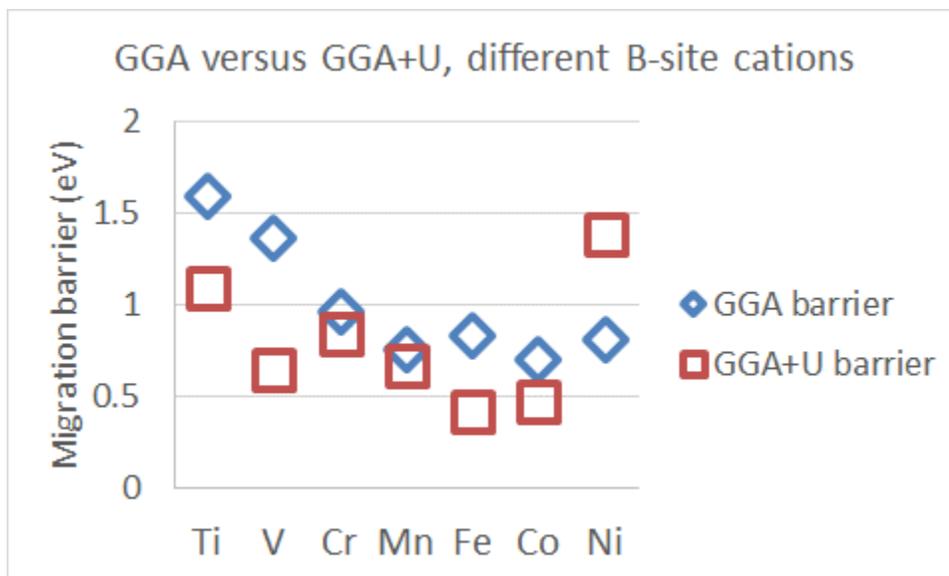

**Figure S5.1. GGA versus GGA+U no-strain migration barriers.**

Figure S5.1. GGA versus GGA+U no-strain migration barriers.

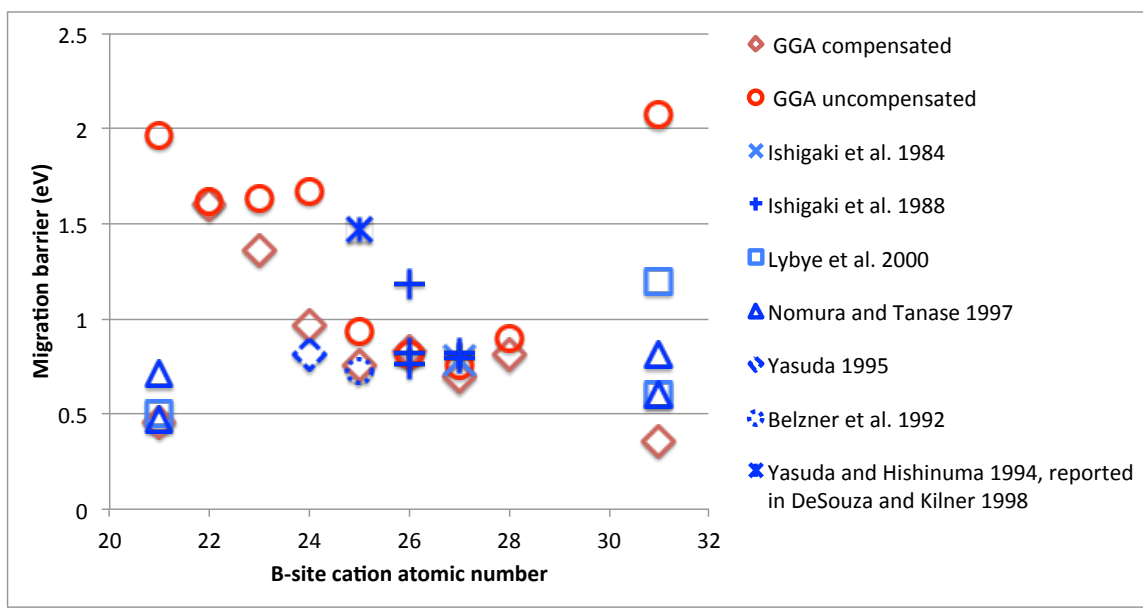

**Figure S6.1. Literature compared with LaXO3 barriers.**

Figure S6.1. Literature compared with LaXO3 uncompensated (reduced B-site cations) and compensated (all B-site cations nominally 3+ due to removal of extra electrons along with oxygen atoms) systems. Literature values are given in Table S6.3.

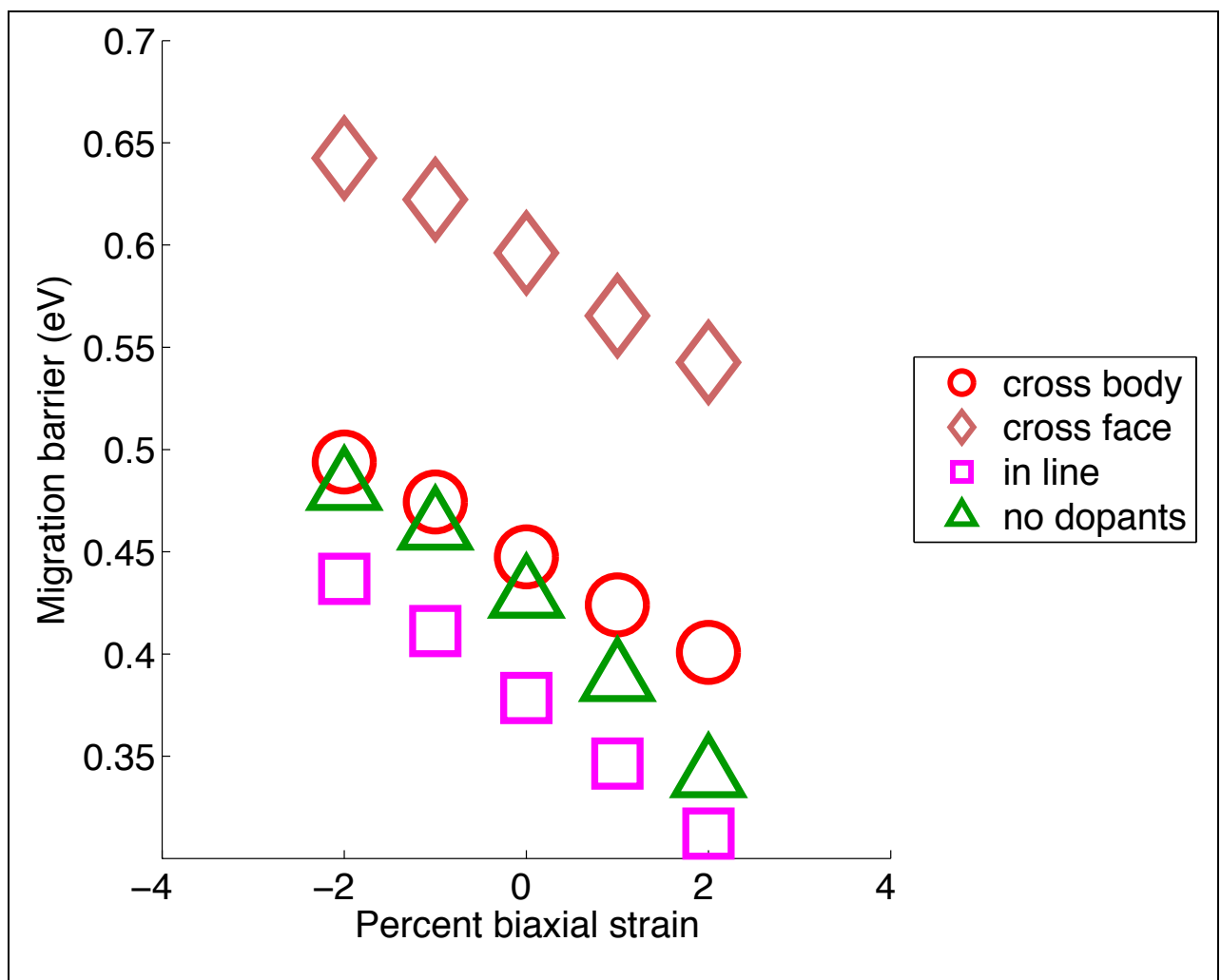

**Figure S7.1 LaScO$_3$ and La$_{0.75}$Sr$_{0.25}$ScO$_3$ migration barrier versus strain, in-plane hop**

Figure S7.1 LaScO$_3$ and La$_{0.75}$Sr$_{0.25}$ScO$_3$ migration barrier versus strain, in-plane hop from oxygen position o31 to o30 (see Figure S2.1). Cross-body diagonal dopant positions are a1 and a8. Cross-face diagonal dopant positions are a2 and a8. In-line dopant positions are a4 and a8.

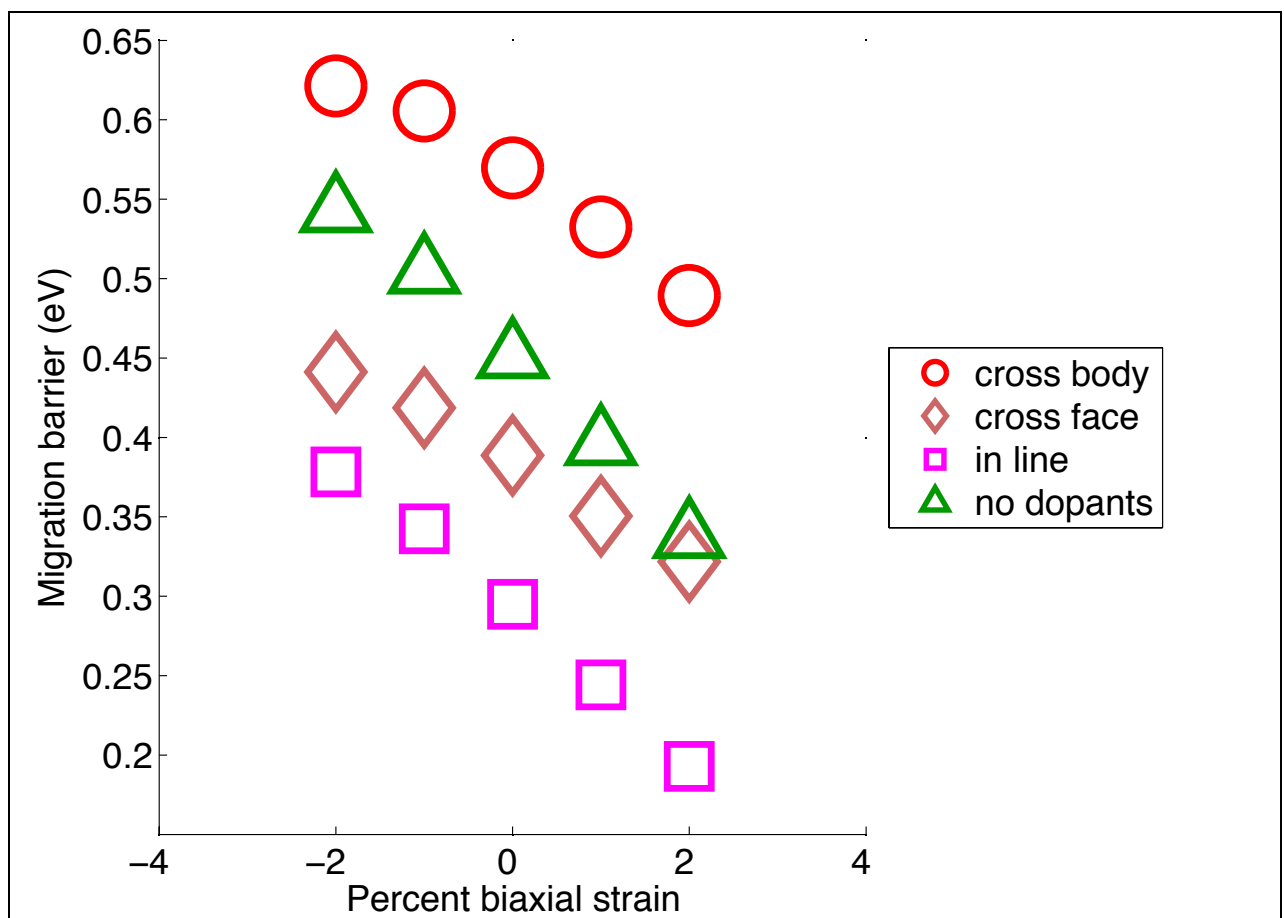

**Figure S7.2 LaScO$_3$ and La$_{0.75}$Sr$_{0.25}$ScO$_3$ migration barrier versus strain, out-of-plane hop**

Figure S7.2 LaScO$_3$ and La$_{0.75}$Sr$_{0.25}$ScO$_3$ migration barrier versus strain, out-of-plane hop from oxygen position o29 to o30 (see Figure S2.1). Cross-body diagonal dopant positions are a1 and a8. Cross-face diagonal dopant positions are a2 and a8. In-line dopant positions are a4 and a8.

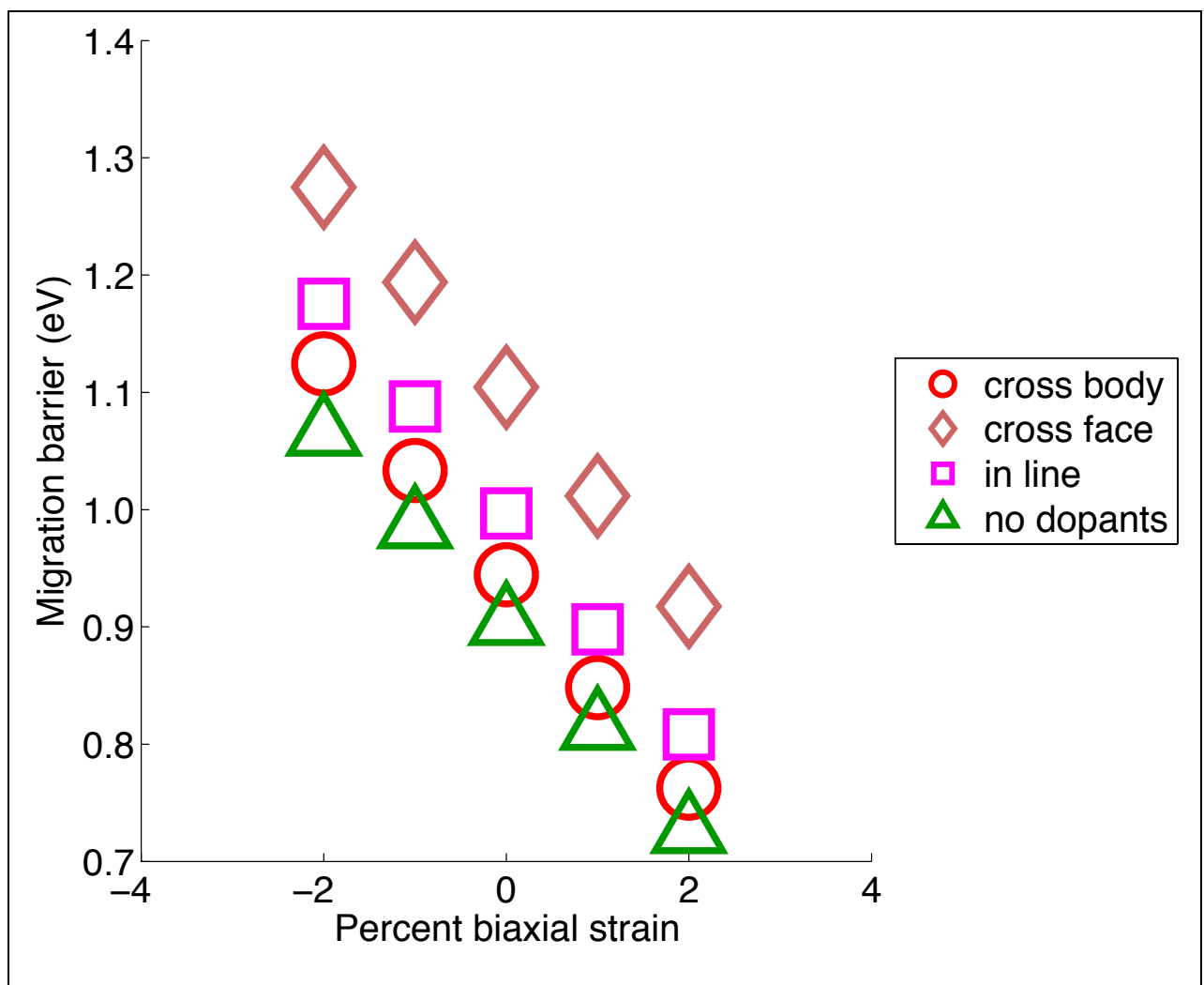

**Figure S7.3 LaCrO$_3$ and La$_{0.75}$Sr$_{0.25}$CrO$_3$ migration barrier versus strain, in-plane hop**

Figure S7.3 LaCrO$_3$ and La$_{0.75}$Sr$_{0.25}$CrO$_3$ migration barrier versus strain, in-plane hop from oxygen position o31 to o30 (see Figure S2.1). Cross-body diagonal dopant positions are a1 and a8. Cross-face diagonal dopant positions are a2 and a8. In-line dopant positions are a4 and a8.

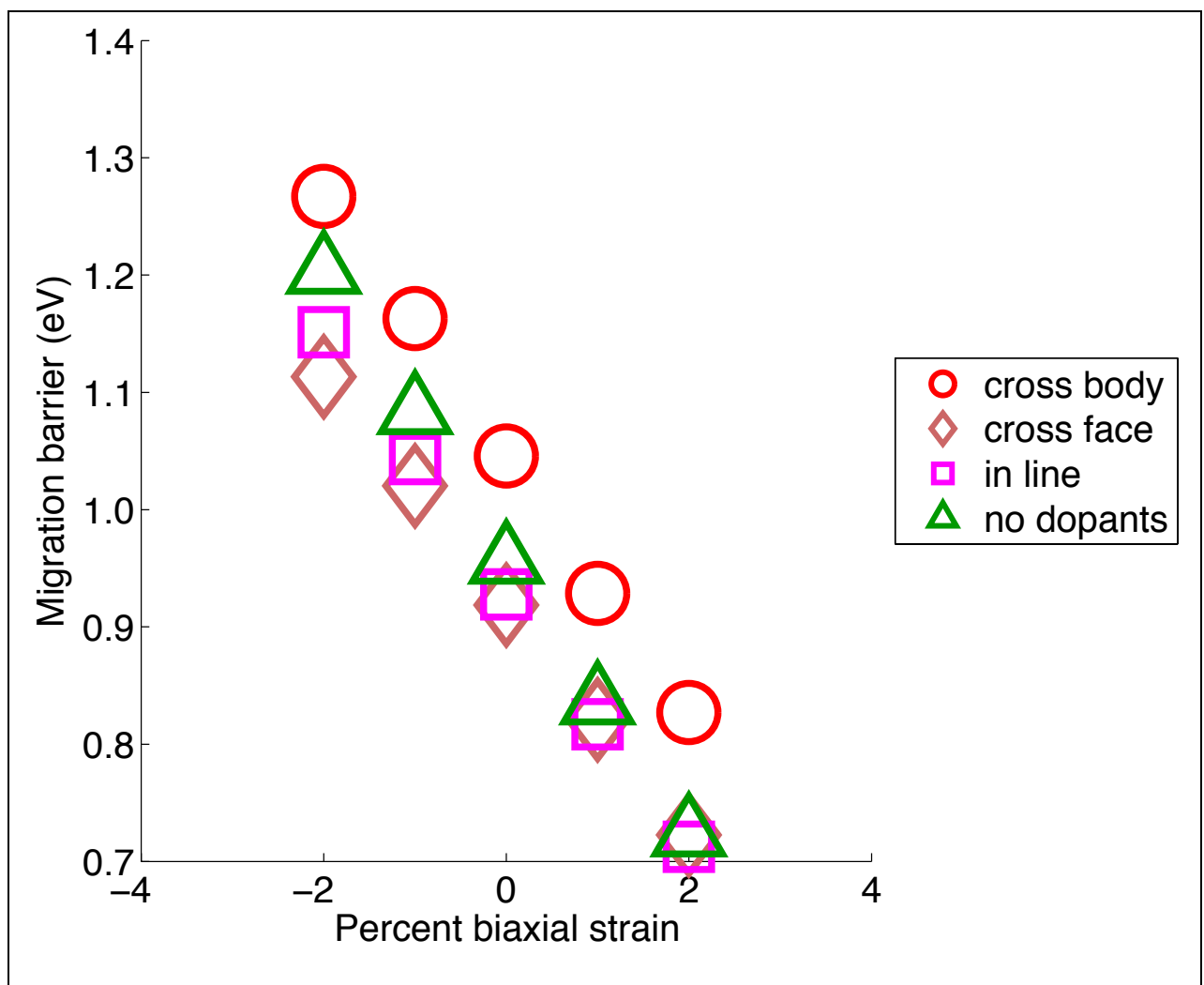

**Figure S7.4 LaCrO$_3$ and La$_{0.75}$Sr$_{0.25}$CrO$_3$ migration barrier versus strain, out-of-plane hop**

Figure S7.4 LaCrO$_3$ and La$_{0.75}$Sr$_{0.25}$CrO$_3$ migration barrier versus strain, out-of-plane hop from oxygen position o29 to o30 (see Figure S2.1). Cross-body diagonal dopant positions are a1 and a8. Cross-face diagonal dopant positions are a2 and a8. In-line dopant positions are a4 and a8.

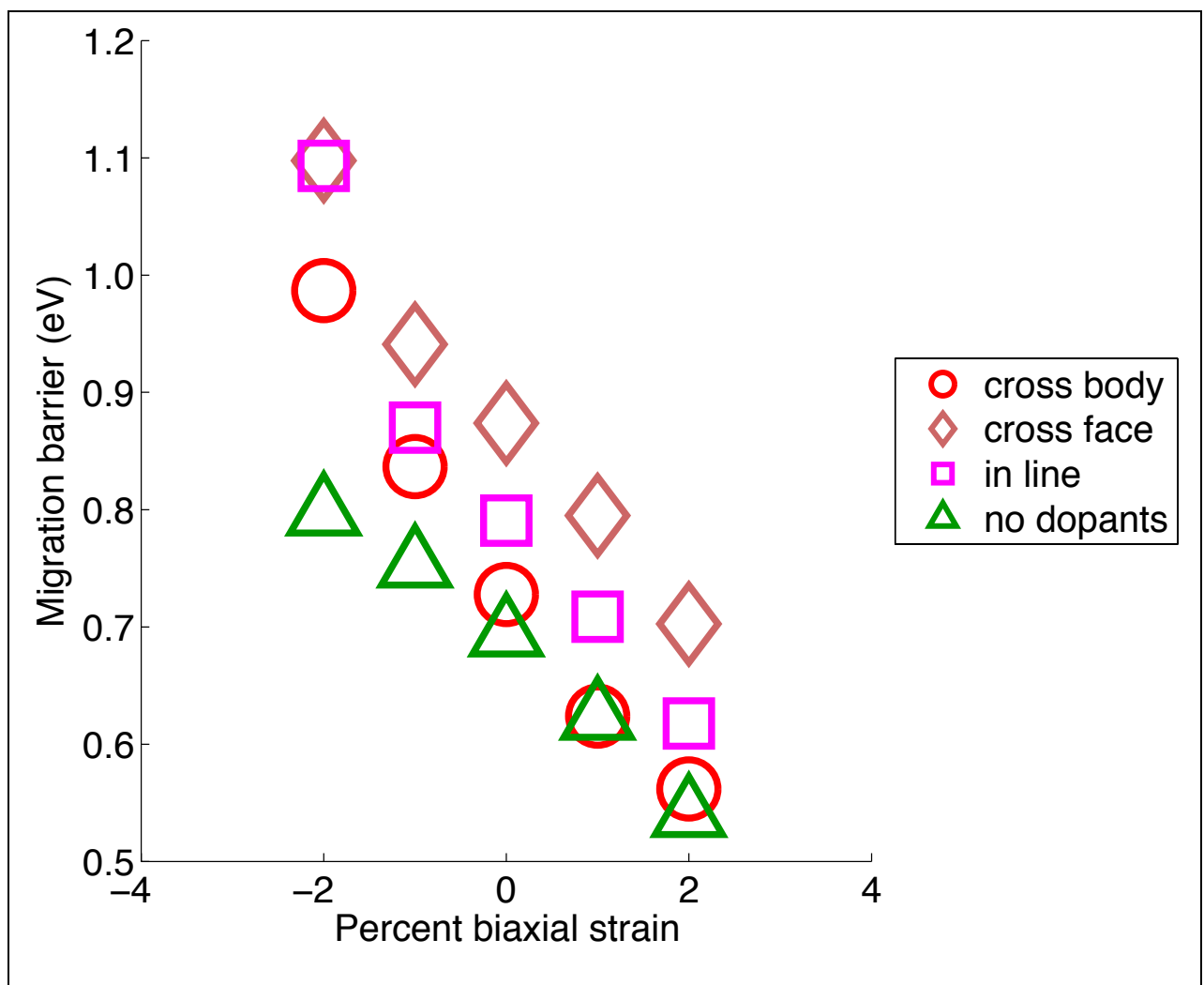

**Figure S7.5 LaMnO$_3$ and La$_{0.75}$Sr$_{0.25}$MnO$_3$ migration barrier versus strain, in-plane hop**

Figure S7.5 LaMnO$_3$ and La$_{0.75}$Sr$_{0.25}$MnO$_3$ migration barrier versus strain, in-plane hop from oxygen position o31 to o30 (see Figure S2.1). Cross-body diagonal dopant positions are a1 and a8. Cross-face diagonal dopant positions are a2 and a8. In-line dopant positions are a4 and a8.

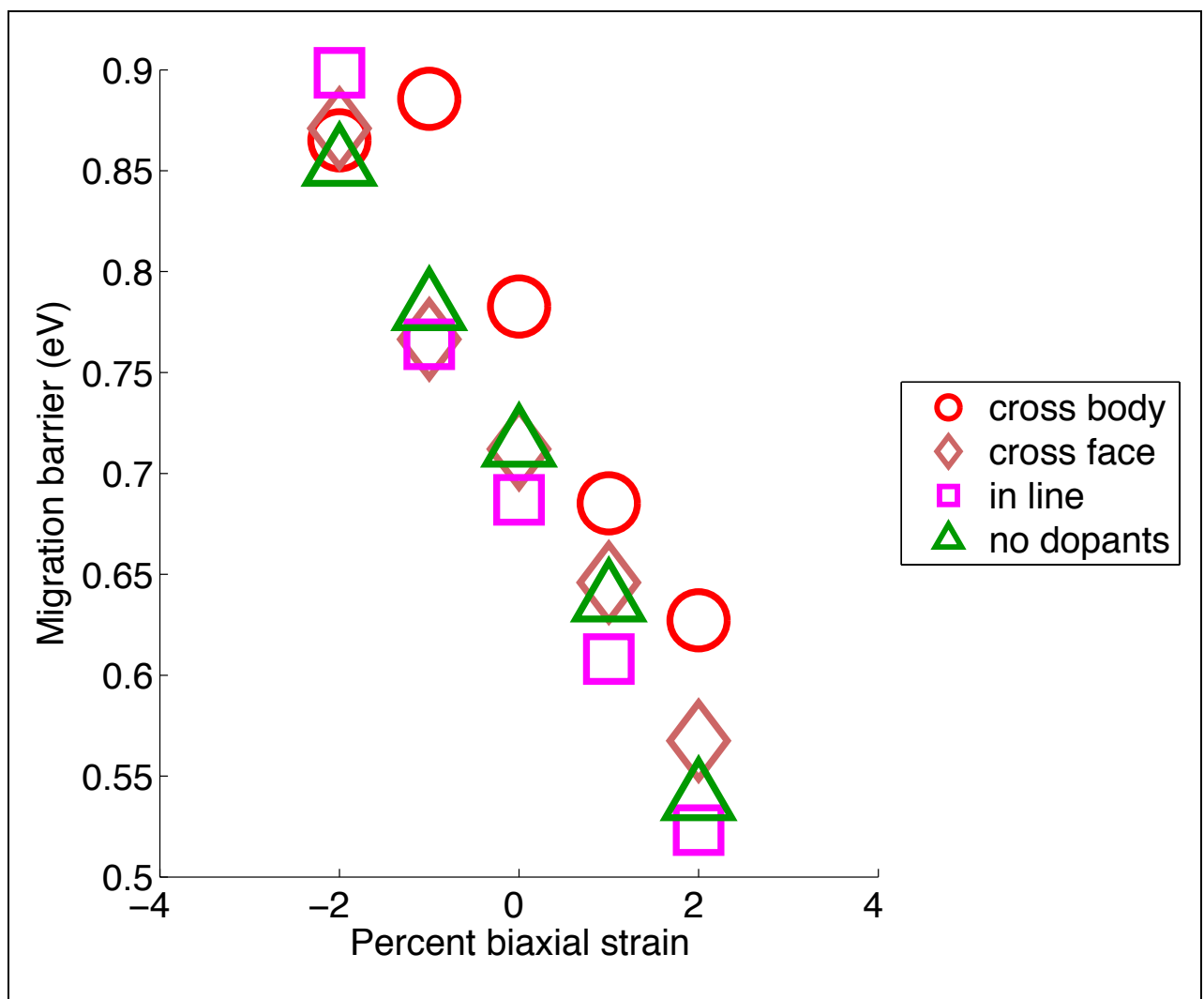

**Figure S7.6 LaMnO$_3$ and La$_{0.75}$Sr$_{0.25}$MnO$_3$ migration barrier versus strain, out-of-plane hop**

Figure S7.6 LaMnO$_3$ and La$_{0.75}$Sr$_{0.25}$MnO$_3$ migration barrier versus strain, out-of-plane hop from oxygen position o29 to o30 (see Figure S2.1). Cross-body diagonal dopant positions are a1 and a8. Cross-face diagonal dopant positions are a2 and a8. In-line dopant positions are a4 and a8.

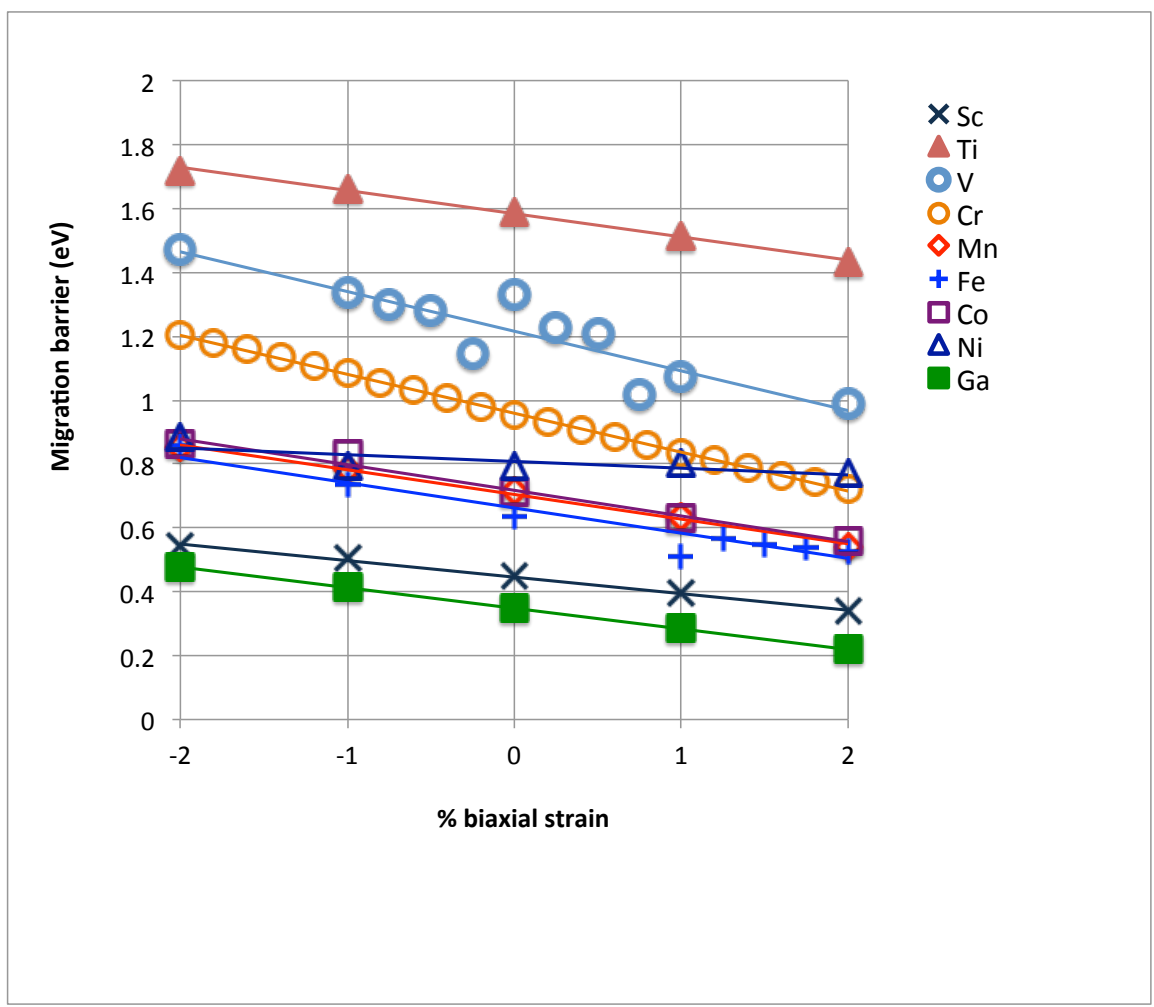

### Figure S8.1. Migration barrier versus strain, out-of-plane hop

Figure S8.1. Change in migration barrier versus biaxial strain for a selected out-of-plane hop for all systems (o29 to o30, see Figure S2.1). The legend indicates the B-site cation for $LaXO_3$ perovskites.

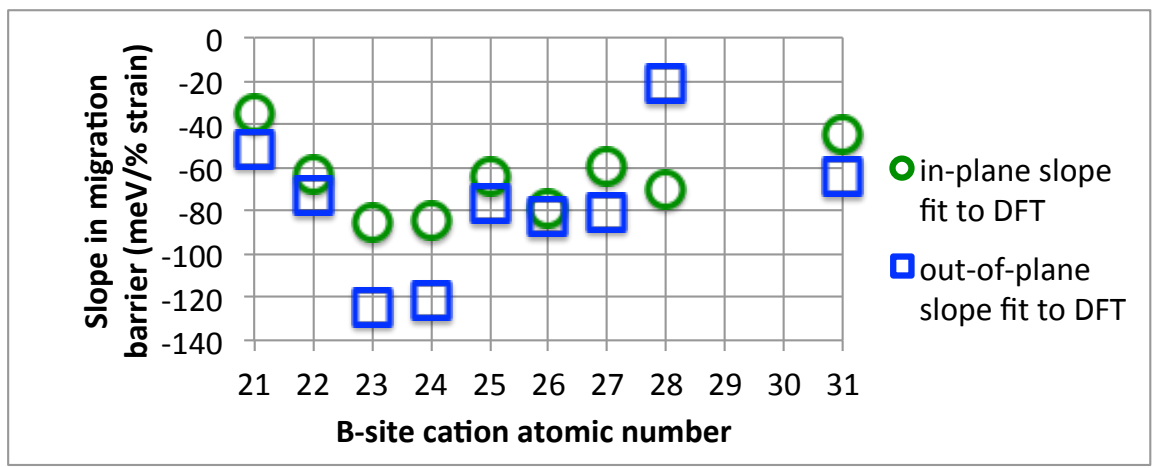

**Figure S8.2. Migration barrier versus strain slopes for in-plane and out-of-plane hops.**

Figure S8.2. Slopes in migration barrier for in-plane and out-of-plane hops across all systems, plotted by B-site cation atomic number. These slope values correspond to Table 1 and Table S8.1. No clear trend with B-site atomic number is evident.

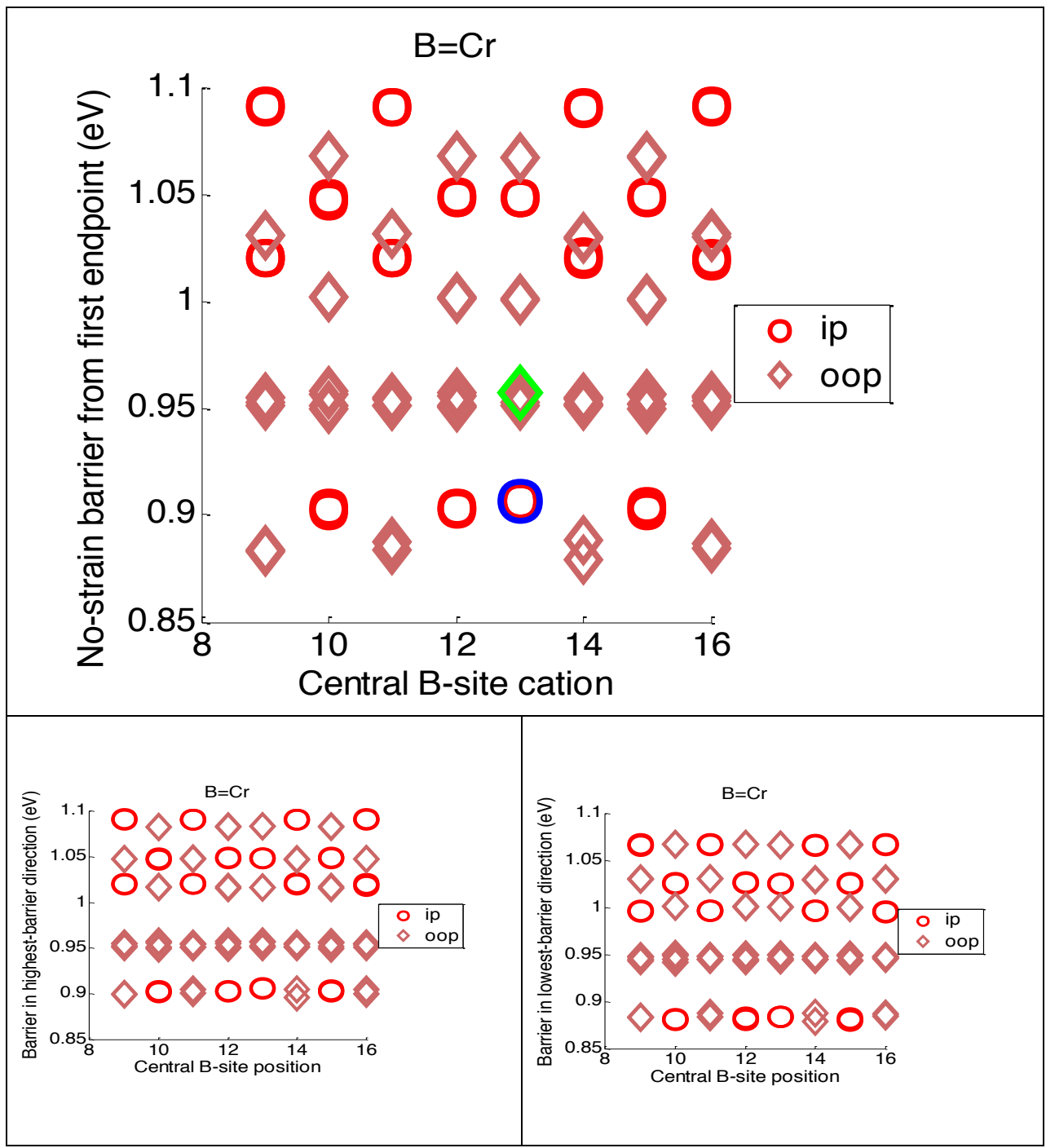

Figure S8.3. LaCrO3 barriers, all hops.

Figure S8.3. LaCrO$_3$ calculated barriers, all oxygen hops, all octahedra, with a total of 96 barriers (12 symmetry distinct) for each strain case. The no-strain case is shown here. Several barriers overlap, reducing the apparent number of points. The top plot shows the hop energy from the initial to the final endpoint, where the out-of-plane and in-plane hops calculated consistently for all systems are highlighted in green and blue, respectively. The smaller plots show the energies for hops in the maximum hop energy (bottom left plot) and minimum hop energy (bottom right plot) directions. The change in energy associated with hopping in the opposite direction is never more than 30 meV

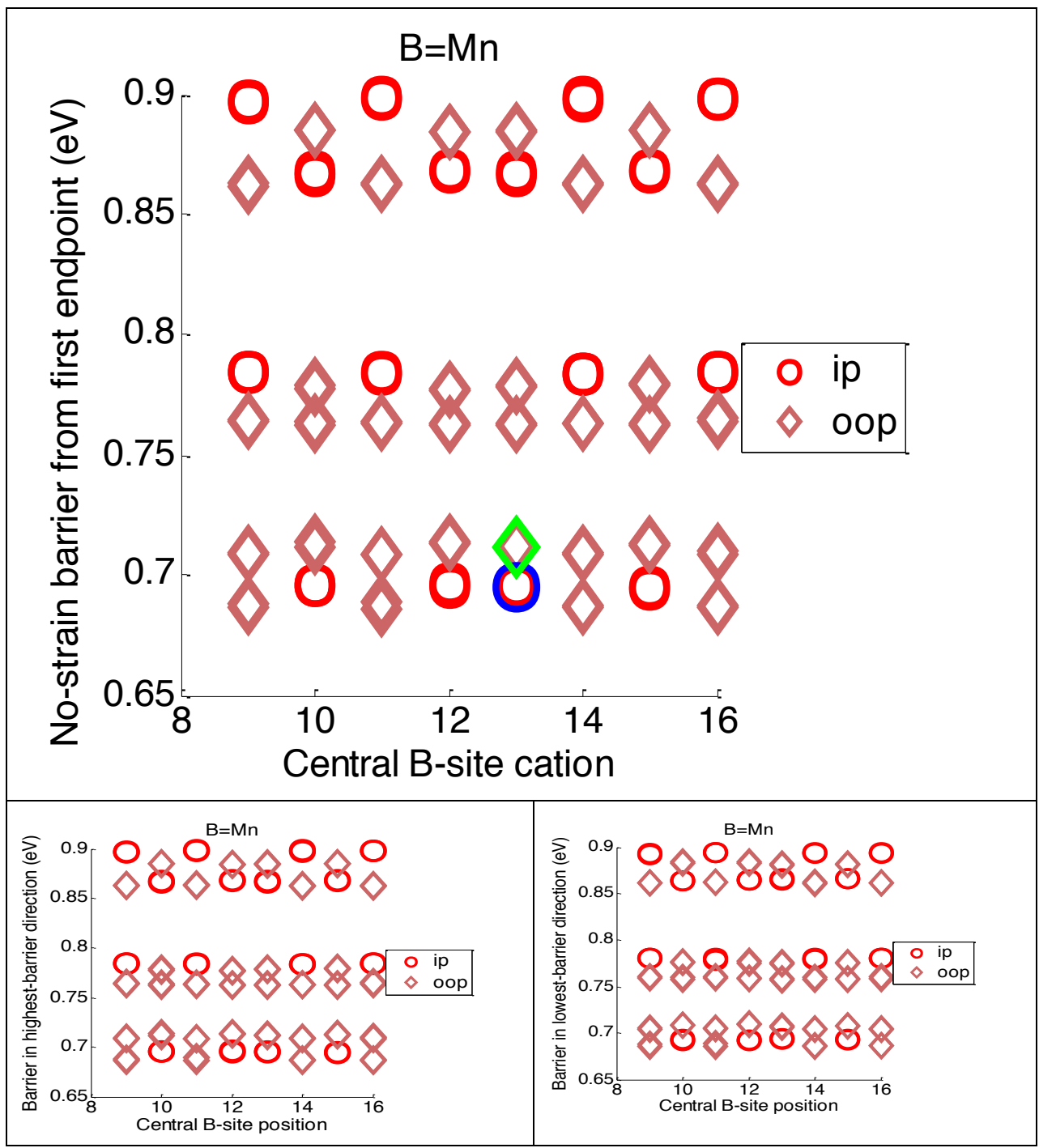

Figure S8.4. LaMnO3 barriers, all hops.

Figure S8.4. LaMnO₃ calculated barriers, all oxygen hops, all octahedra, with a total of 96 barriers (12 symmetry distinct) for each strain case. The no-strain case is shown here. Several barriers overlap, reducing the apparent number of points. The top plot shows the hop energy from the initial to the final endpoint, where the out-of-plane and in-plane hops calculated consistently for all systems are highlighted in green and blue, respectively. The smaller plots show the energies for hops in the maximum hop energy (bottom left plot) and minimum hop energy (bottom right plot) directions. The change in energy associated with hopping in the opposite direction is never more than 30 meV.

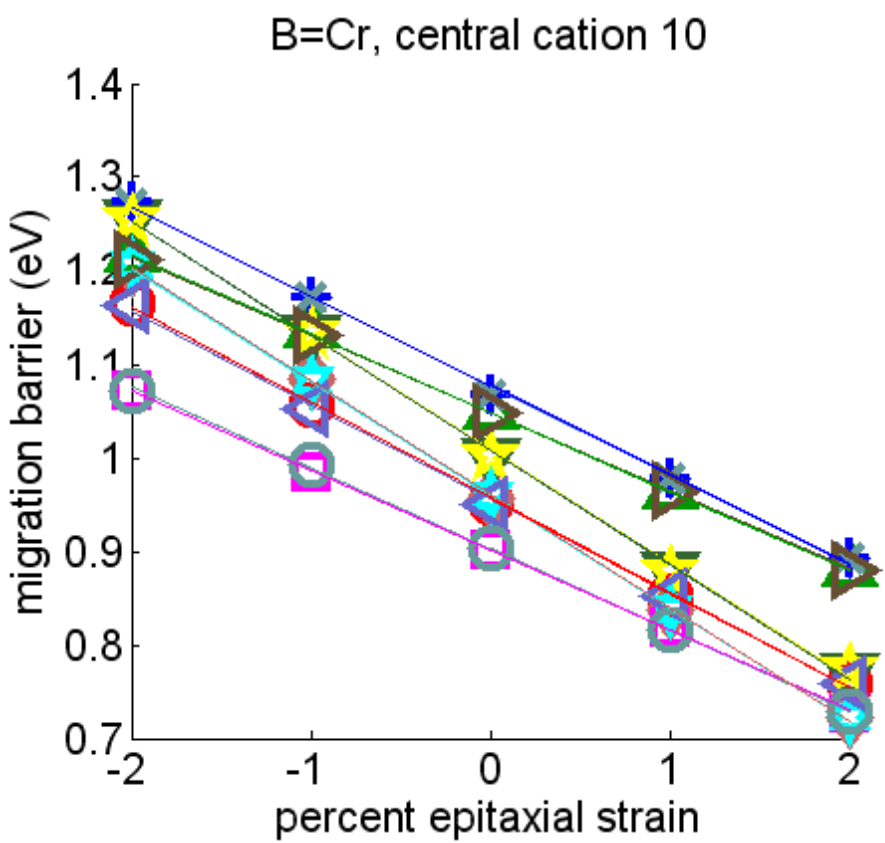

**Figure S8.5. Migration barrier versus strain for LaCrO3, central cation position 10.**

Figure S8.5. Migration barrier versus strain for LaCrO₃ with central cation position 10, for hops in the direction of initial endpoint to final endpoint, giving a representative example that all migration barriers decrease with increasing tensile strain.

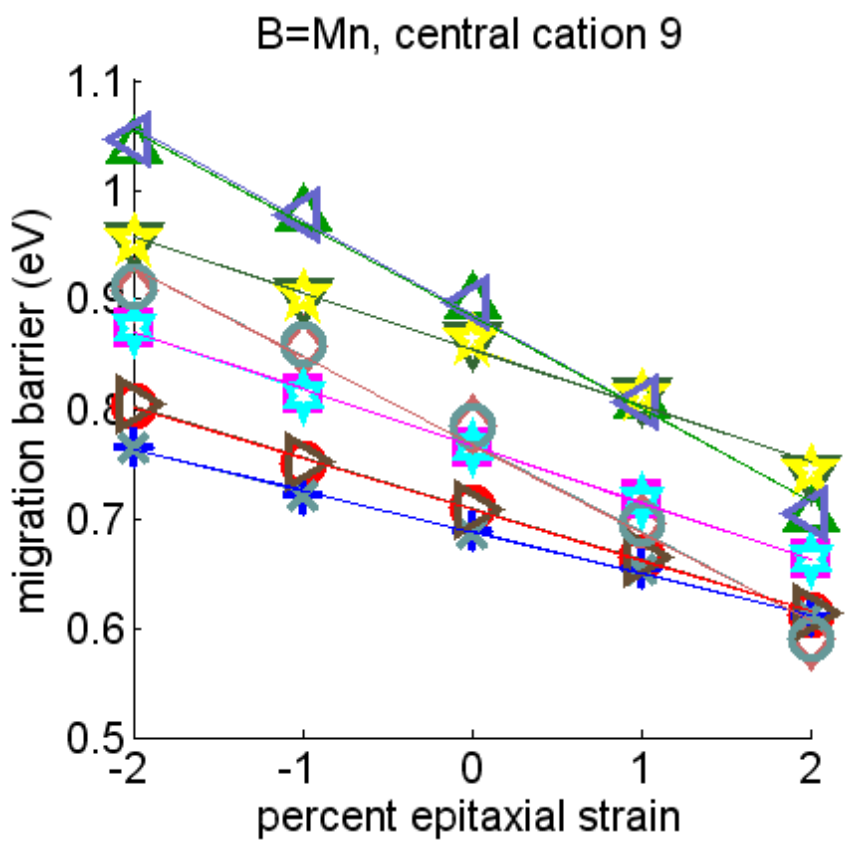

**Figure S8.6. Migration barrier versus strain for LaMnO3, central cation position 9**

Figure S8.6. Migration barrier versus strain for LaMnO$_3$ with central cation position 9, for hops in the direction of initial endpoint to final endpoint, giving a representative example that all migration barriers decrease with increasing tensile strain.

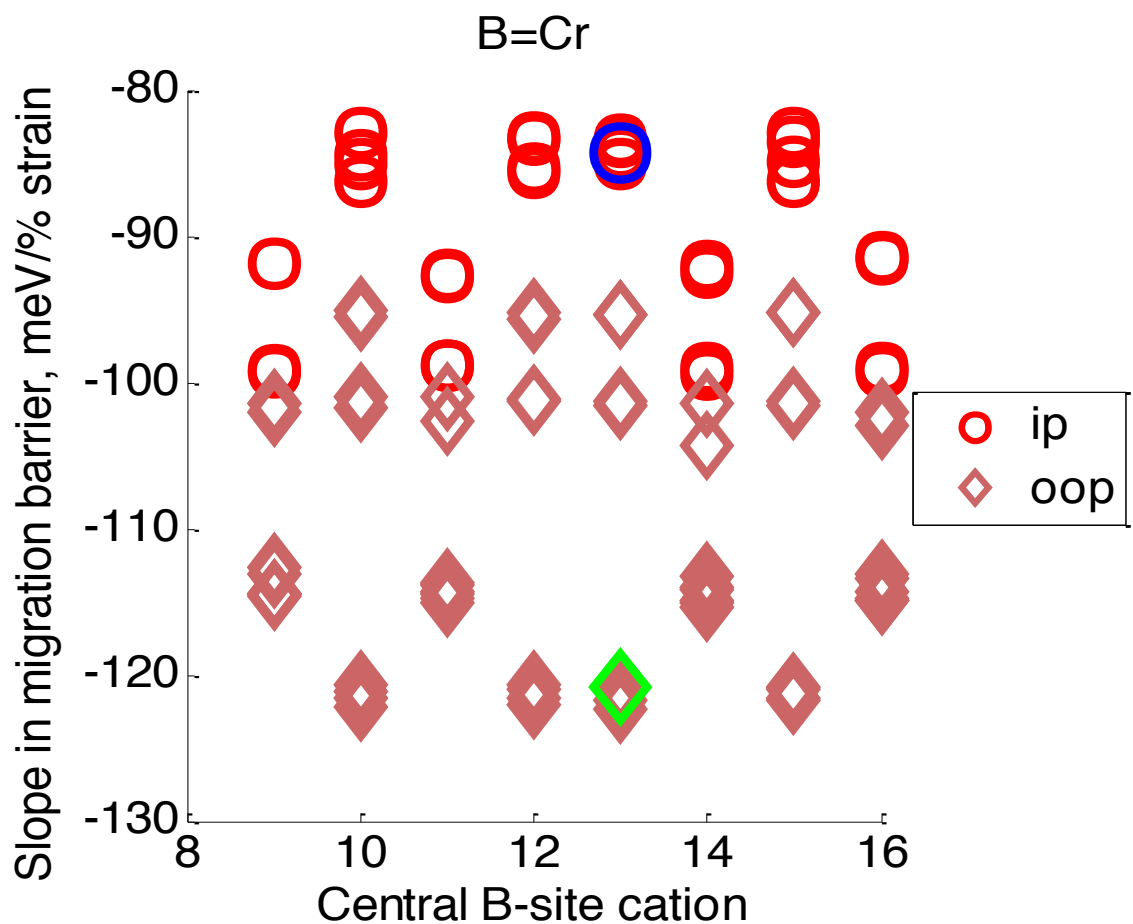

**Figure S8.7. Slopes in migration barrier for LaCrO₃, all hops.**

Figure S8.7. Slopes in migration barrier for LaCrO$_3$, all hops, in the direction of intial endpoint to final endpoint. The in-plane hop used for all systems and the out-of-plane hop used for all systems (described in Section S8) are highlighted in blue and green, respectively.

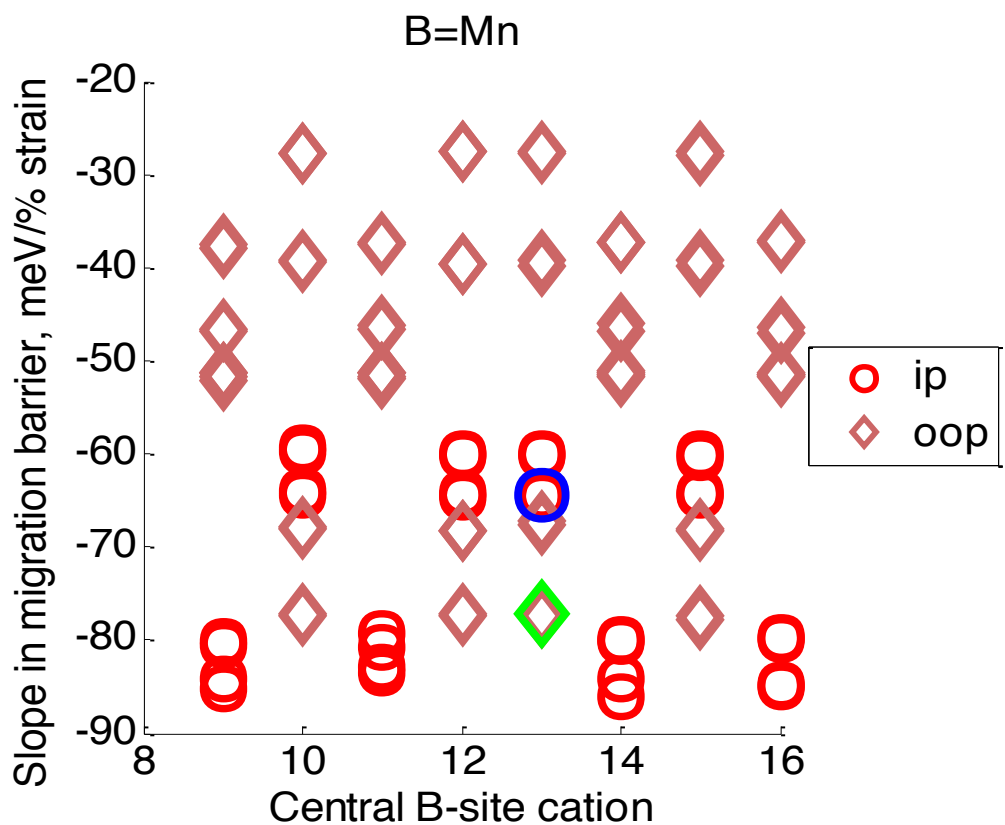

**Figure S8.8. Slopes in migration barrier for LaMnO3, all hops.**

Figure S8.8. Slopes in migration barrier for LaMnO$_3$, all hops, in the direction of initial endpoint to final endpoint. The in-plane hop used for all systems and the out-of-plane hop used for all systems (described in Section S8) are highlighted in blue and green, respectively.

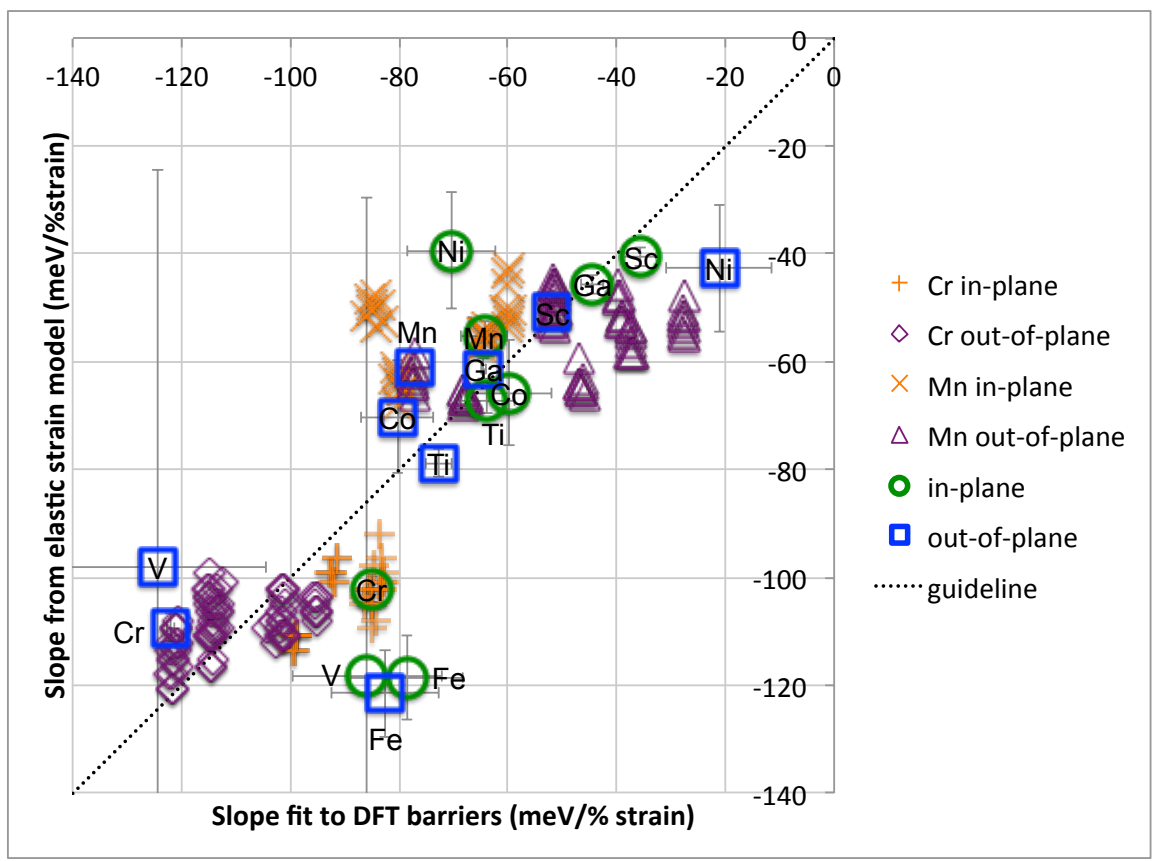

**Figure S8.9. Elastic strain model slopes versus slopes fit to DFT barriers, all LaCrO3 and LaMnO3 hops represented**

Figure S8.9. Elastic strain model slopes versus slopes fit to DFT barriers, with all LaCrO$_3$ and LaMnO$_3$ hops represented. The elastic strain model slope for each hop was calculated using that hop's no-strain Birch-Murnaghan calculated migration volume.

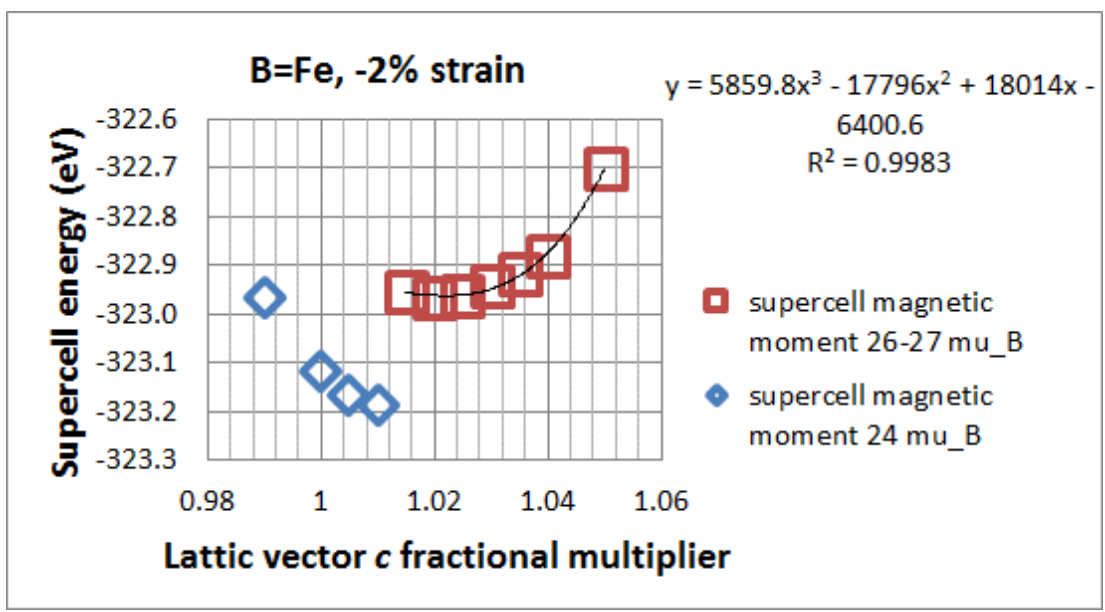

**Figure S10.1. Example of cubic fitting for lattice vector c fractional multiplier.**

Figure S10.1. Example of cubic fitting for lattice vector c fractional multiplier, for the B=Fe system at -2% biaxial strain. The volume-conserving multiplier would be around 1.04. The actual fit multiplier turns out to be 1.021. Note that there are two distinct magnetic moment curves.

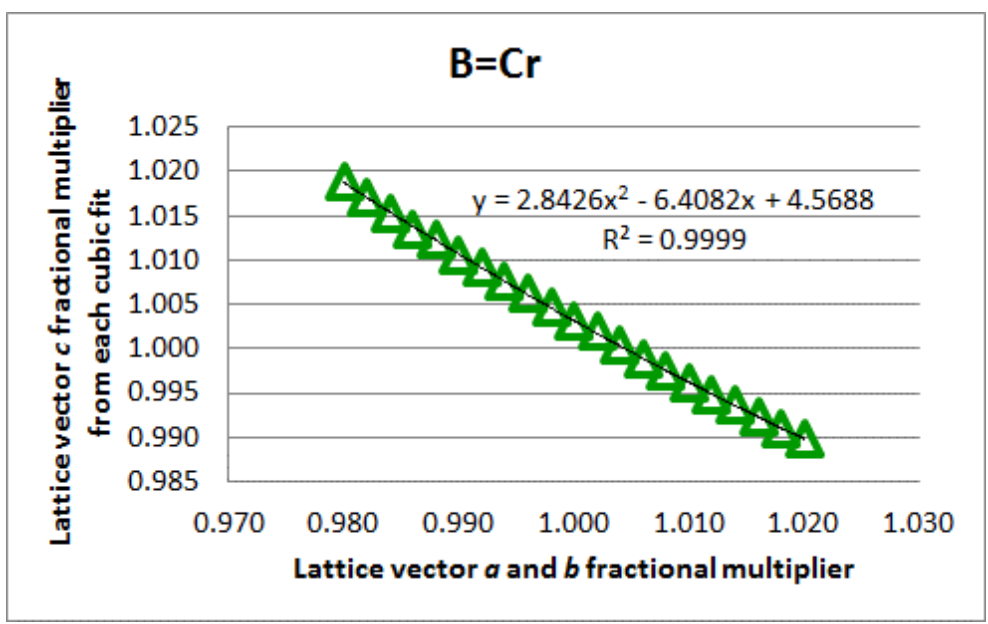

**Figure S10.2. Fine-gridding for B=Cr.**

Figure S10.2. Fine-gridding for the B=Cr system, showing a smooth curve. Each lattice vector *c* response multiplier was the result of a separate 7-point cubic fit at the given strain in lattice vectors *a* and *b*.

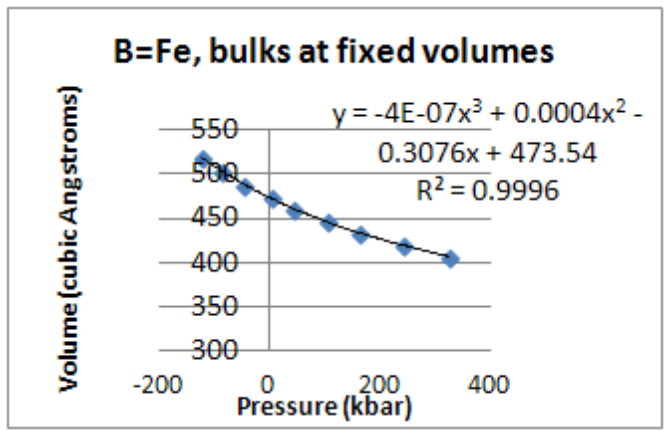

**Figure S12.1. Sample V(P) curve showing a cubic fit.**

Figure S12.1. Sample V(P) curve showing a cubic fit.

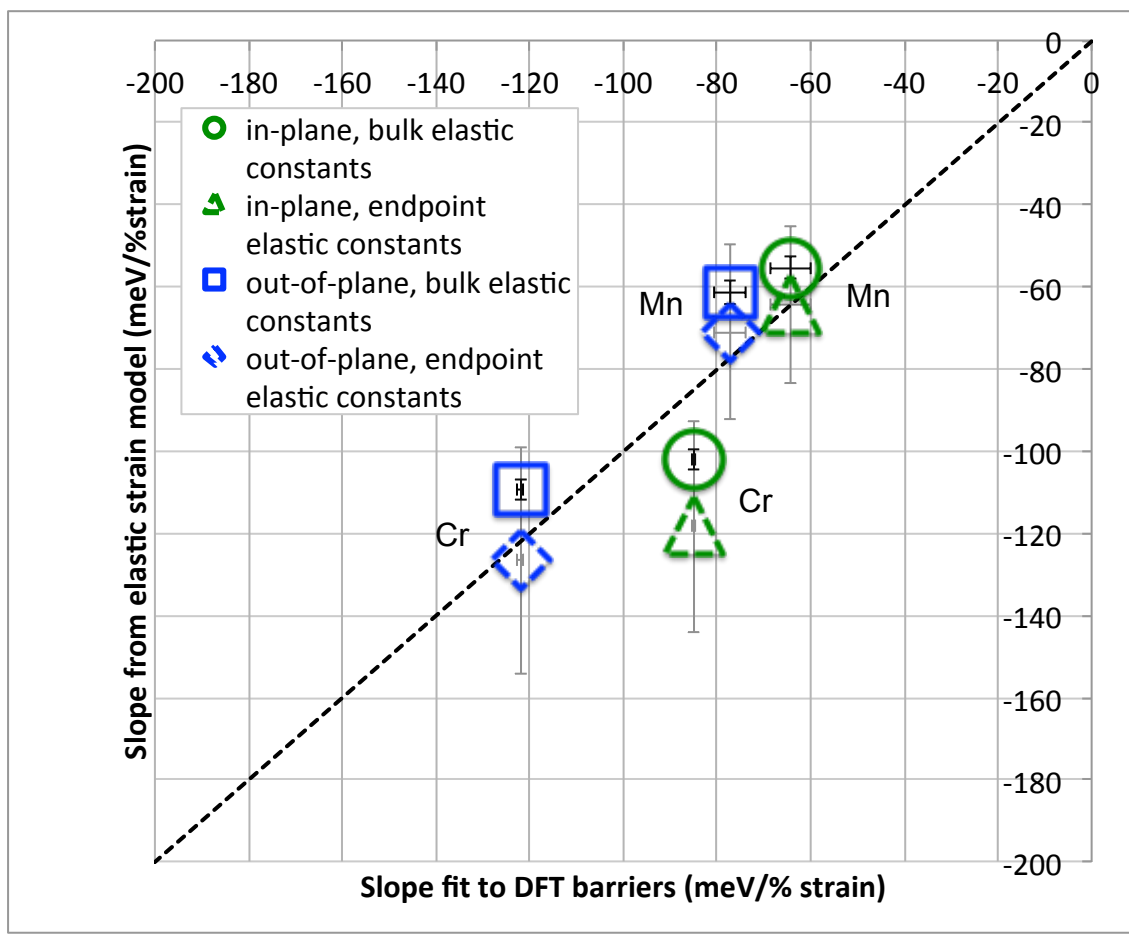

**Figure S12.2. Vacancy effects on elastic model**

Figure S12.2. Vacancy effects on the elastic model (compare with Figure 5 in the main paper). The largest shift effect is shown, which for the compensated system corresponds to elastic constants calculated at the initial state endpoint. Calculating the elastic constants with a defected supercell shifts the elastic model slope to be steeper, which may either improve or worsen its agreement with the slope fit to DFT barriers. However, the error in the model slope increases greatly due to increased error in fitting the elastic constants with the inclusion of the vacancy.

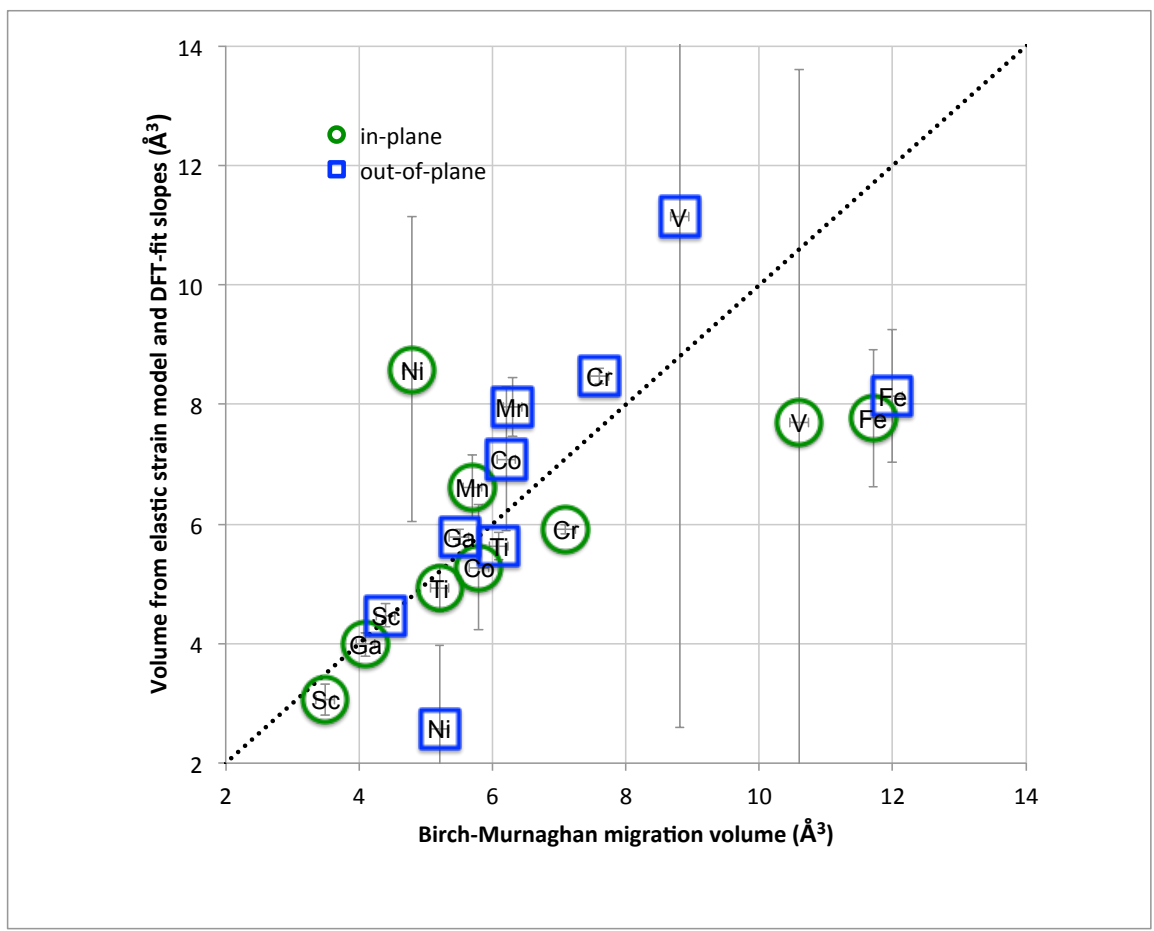

**Figure S12.3. Using elastic model to try to predict Vmig (BM)**

Figure S12.3. Elastic model migration volume versus Birch-Murnaghan formula migration volume. Data point is the center of each symbol. All error bars are symmetric.

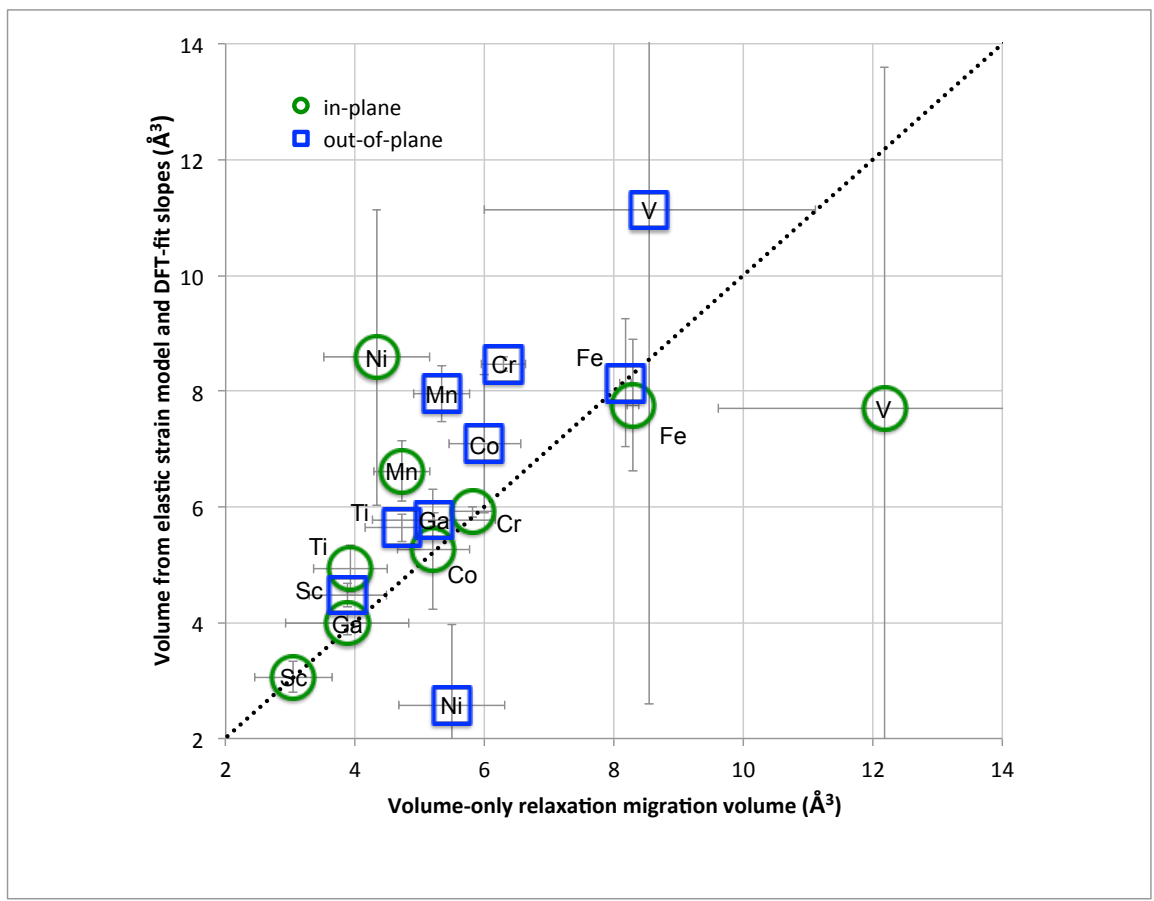

**Figure S12.4. Using elastic model to try to predict Vmig (volume-only relaxation)**

Figure S12.4. Elastic model migration volume versus volume-only relaxation migration volume. Data point is the center of each symbol. All error bars are symmetric. Error bars in the volume-only relaxation are the standard deviation between the in-plane and out-of-plane volumes.

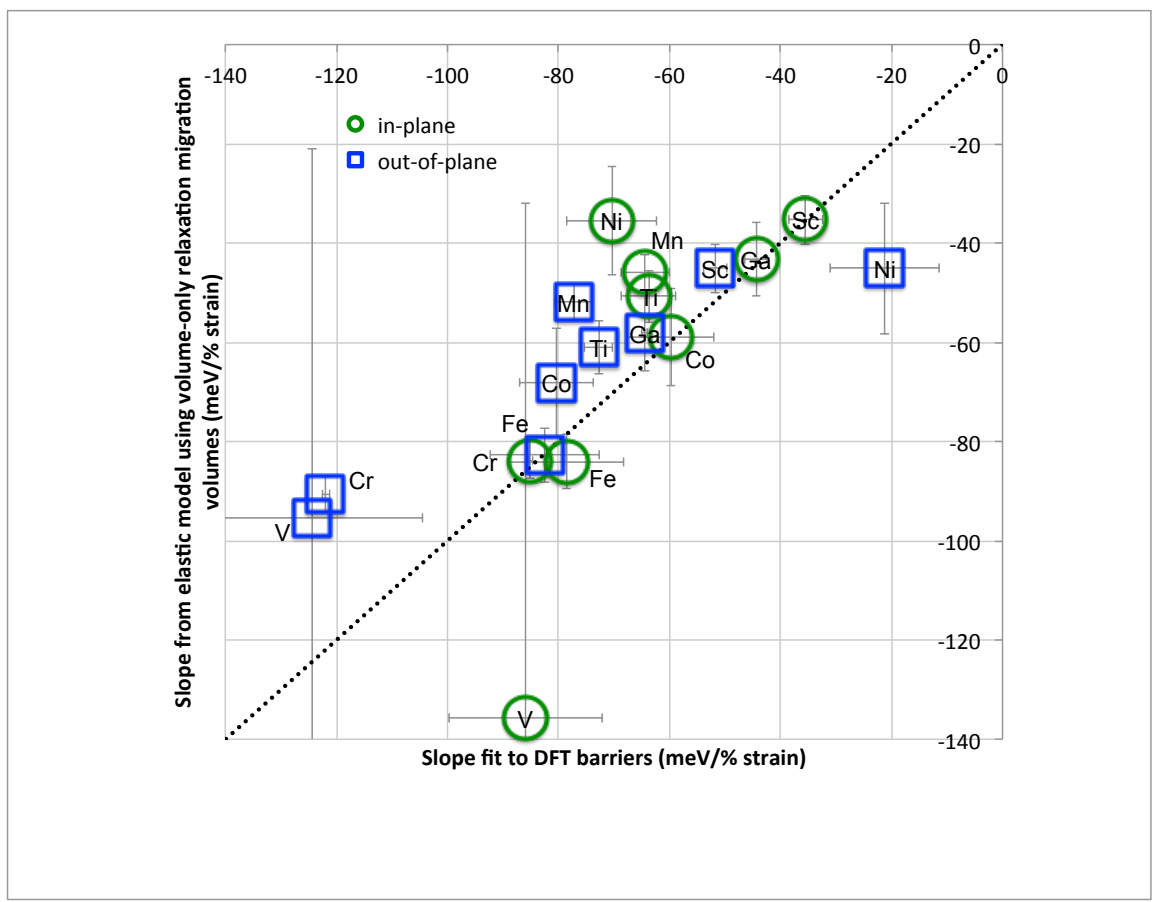

**Figure S12.5. Using elastic model to try to predict slopes, with Vmig from volume-only relaxation**

Figure S12.5. Elastic model-calculated slopes using volume-relaxation volumes, versus DFT-fit slopes. Data point is the center of each symbol. All error bars are symmetric.